\documentclass[twocolumn,english,pra]{revtex4}
\usepackage[T1]{fontenc}
\usepackage[latin9]{inputenc}
\usepackage{babel}
\usepackage{amsmath}
\usepackage{amssymb}
\usepackage{graphicx}
\usepackage{esint}
\usepackage{xcolor}
\usepackage[unicode=true,pdfusetitle,
 bookmarks=true,bookmarksnumbered=false,bookmarksopen=false,
 breaklinks=false,pdfborder={0 0 1},backref=false,colorlinks=false]
 {hyperref}
\usepackage{breakurl}

\def\be{\begin{equation}}
\def\ee{\end{equation}}
\def\beq{\begin{eqnarray}}
\def\eeq{\end{eqnarray}}
\def\ba#1{\begin{array}{#1}}
\def\ea{\end{array}}

\makeatletter
\@ifundefined{textcolor}{}
{%
 \definecolor{BLACK}{gray}{0}
 \definecolor{WHITE}{gray}{1}
 \definecolor{RED}{rgb}{1,0,0}
 \definecolor{GREEN}{rgb}{0,1,0}
 \definecolor{BLUE}{rgb}{0,0,1}
 \definecolor{CYAN}{cmyk}{1,0,0,0}
 \definecolor{MAGENTA}{cmyk}{0,1,0,0}
 \definecolor{YELLOW}{cmyk}{0,0,1,0}
}

\usepackage{multirow}

\makeatother

\begin{document}

\title{Semiclassical dynamics of a disordered two-dimensional Hubbard
model \\with long-range interactions}

\author{Adam S. Sajna$^{1, 2}$}
\author{Anatoli Polkovnikov$^1$}
\affiliation{\mbox{$^1$Department of Physics, Boston University, 590 Commonwealth Avenue, Boston, Massachusetts 02215, USA}
\mbox{$^2$Faculty of Physics, Adam Mickiewicz University, Uniwersytetu Pozna\'nskiego 2, 61-614 Pozna\'n, Poland}}
\begin{abstract}
Quench dynamics in a two-dimensional system of interacting fermions is analyzed within the semiclassical truncated Wigner approximation (TWA). The models with short-range and long-range interactions are considered. We show that in the latter case, the TWA is very accurate, becoming asymptotically exact in the infinite-range limit, provided that the semiclassical Hamiltonian is correctly identified. Within the TWA, different dynamical timescales of charges and spins can be clearly distinguished. Interestingly, for a weak and moderate disorder strength, we observe subdiffusive behavior of charges, while spins exhibit diffusive dynamics. At strong disorder, the quantum Fisher information shows logarithmic growth in time with a slower increase for charges than for spins. It is shown that in contrast to the short-range model, strong inhomogeneities such as domain walls in the initial state can significantly slow down thermalization dynamics, especially at weak disorder. This behavior can put additional challenges in designing cold-atom experimental protocols aimed to analyze possible many-body localization in such systems. While within this approach we cannot  make any definite statements about the existence of a many-body localized phase, we see a very fast crossover as a function of disorder strength from rapidly thermalizing to a slow glassy like regime both for the short-range and long-range models.

\end{abstract}

\pacs{34234}

\maketitle

\section{Introduction}

Understanding the dynamics of isolated interacting disordered many-body systems has recently became a forefront of both theoretical and experimental research~\citep{mblschreiber,Choi2016,Smith2016,PhysRevLett.116.140401,Zhang2017,Choi2017,Lschen2017,PhysRevLett.119.260401,PhysRevX.7.041047,Kucsko2018,Lukin256,1910.06024,Nandkishore2015,RevModPhys.91.021001,PhysRevLett.122.170403}.
Such systems have been explored both with respect to possible applications to quantum information~\citep{Nandkishore2015,1910.06024} and as generic models of possible ergodicity breaking in interacting systems~\citep{Luitz2017,RevModPhys.91.021001}. It is well-known that a competition of interaction and disorder leads to a peculiar dynamical behavior of the entanglement entropy and information propagation \citep{PhysRevB.77.064426,PhysRevLett.109.017202,PhysRevLett.110.260601,Smith2016,Lukin256,1910.06024,PhysRevB.99.241114}.
In particular, this dynamical behavior can be highly sensitive to the interaction range~\citep{PhysRevB.90.174204,PhysRevB.95.094205,pandeyMBL2019,PhysRevB.99.054204,PhysRevA.99.033610}.
Disordered systems with long-range interaction have been already realized in experiments with trapped ions \citep{Smith2016,Zhang2017}. There also exist solid state disordered materials with long-range Coulomb interactions \citep{pollak2013}. Electrons in such materials are strongly localized and charge carries cannot screen long-range interactions making their long-range nature play a very important role.  Coulomb interactions also remain unscreened in two-dimensional materials like suspended graphene \cite{RevModPhys.84.1067}. There is thus a very clear need for development of efficient theoretical methods which could simulate such systems in any dimension.

Conceptually, interplay of disorder and interactions can be understood within the framework of the Hubbard model \cite{Altland}. Originally introduced as a toy model to understand interacting systems, it has been experimentally realized in different spatial dimensions. In particular, a realization of disordered/quasiperiodic Hubbard model in one and two spatial dimensions has been reported in Refs.~ \citep{mblschreiber,PhysRevLett.116.140401,PhysRevLett.119.260401,PhysRevX.7.041047,PhysRevLett.122.170403}. In several recent works it has been argued that two-component fermions might not have the many-body localized phase in one spatial dimension due to the coexistence of spin and charge excitations~\citep{PhysRevB.94.241104,PhysRevB.97.064204,PhysRevLett.120.246602,
PhysRevB.98.115106,1910.11889, 1911.11711, PhysRevB.100.125132,PhysRevB.99.121110}. In particular, charge and spin degrees of freedom can exhibit different localization properties and affect long time dynamics~\citep{PhysRevB.98.014203,PhysRevB.99.115111}. 

In this paper we systematically analyze quantum dynamics in an interacting fermionic Hubbard model with long-range interactions using the fermionic version of the truncated Wigner approximation (fTWA)~ \citep{Davidson2017,PhysRevB.99.134301}. We focus on two-dimensional (2D) systems,  but also mention some results in the one-dimensional (1D) case, mostly to benchmark the approach against the exact diagonalization. We show that for the accuracy of the method it is crucial to choose the correct representation of the Weyl symbol of the Hamiltonian and of the observables. In particular, the fermionic number operator $\hat n_\alpha=\hat c_\alpha^\dagger \hat c_\alpha$, where $\alpha$ is some index labeling of the corresponding single-particle state, always satisfies the identity $\hat n_\alpha^2=\hat n_\alpha$. At the same time the Weyl symbols of $\hat n_\alpha$ and $\hat n_\alpha^2$ are different. Thus there is an ambiguity in defining the phase space representation of the corresponding operators. Within the exact analysis of the dynamics, this ambiguity is irrelevant, but within the semiclassical TWA approximation it plays a significant role. In this paper we remove the ambiguity by choosing the representation of the Hamiltonian, which leads to asymptotically exact fTWA dynamics when the range of interactions becomes infinite. It is shown that the choice of this particular representation also leads to a dramatic improvement in the accuracy of fTWA over a more naive representation if interactions decay as a power law.  We note that such an ambiguity might exists for other setups, e.g. spin $1/2$ systems, where the spin operators satisfy similar identities: $\hat s_{x,y,z}^2= \hat I/4$. Our work suggests that in those situations choosing the right representation of the Hamiltonian can significantly improve the accuracy of TWA.

Using this improved representation, fTWA was applied to analyze the charge and spin dynamics in the long-range interacting systems.  In particular, we study transport and time-dependent correlation functions and the role of disorder and interactions. It is found that within fTWA it is possible to clearly distinguish different dynamical time scales in transport of charge and spin degrees of freedom. For weak/moderate disorder strength, the charges exhibit subdiffusive dynamics, while
the spin dynamics remains nearly diffusive. This separation of time scales is related to the fact that both spin components are subject to the same disorder potential.
 For strong disorder both charge and spin are nearly localized undergoing very slow glassy dynamics. In this regime the quantum Fisher information (QFI) is shown to be a good indicator of the time scales associated with spin and charge sectors. Such QFI has been recently measured in a quantum simulator of a one-dimensional disordered spin system with long-range interactions~\citep{Smith2016}. We observed a logarithmic growth of QFI for both degrees of freedom. However, the growth rate for charges is slower than for spins, which is consistent with a stronger tendency of localization of the charge degrees of freedom. This difference disappears in the non-interacting limit, in which the system exhibits the Anderson localization reflected in a rapid saturation of  QFI~\citep{PhysRevB.99.054204,PhysRevB.99.241114}. Moreover, in contrast to the Hubbard model with short-range interactions, strong inhomogeneities in the initial state like that used in the short-range systems (Ref.~  \citep{Choi2016}), is found to significantly slow down the dynamics even at weak disorder. Thus in the long-range systems, extra care should be taken to choose the right initial state needed to check possible existence of many-body localization.

Most of the numerical studies are performed for 2D square lattices, which are intractable by exact methods. While we cannot definitely address all questions, in particular whether the system can be in a localized state beyond some disorder threshold, we can extract many quantitative and qualitative features of the dynamics in such systems showing the power of the fTWA approach to study cold atom systems and possibly even real materials.

The rest of the paper is organized as follows. In Sec. II  the semiclassical fTWA method and in particular its implementation in the Hubbard model are discussed. In Secs. III and IV, we analyze charge and spin transport in the presence of quenched disorder. In Sec. V, the impact of  different initial conditions on transport and thermalization time scales is revealed. In the last section we summarize our results. Additional technical details of the fTWA method are discussed in the Appendixes.

\section{fTWA implementation of the Hubbard model \label{sec: Phase-space-representation of HH}}

Semiclassical representation of fermionic dynamics within the fTWA in terms of phase space string-like variables was recently exploited in Refs. \citep{Davidson2017,PhysRevB.99.134301}. These string variables can be introduced through Weyl symbols of the following bilinear operators 
\be
\hat{E}_{\beta}^{\alpha}=\frac{1}{2}(\hat{c}_{\alpha}^{\dagger}\hat{c}_{\beta}-\hat{c}_{\beta}\hat{c}_{\alpha}^{\dagger}),\;
\hat{E}_{\alpha\beta}=\hat{c}_{\alpha}\hat{c}_{\beta},\; \hat{E}^{\alpha\beta}=\hat{c}_{\alpha}^{\dagger}\hat{c}_{\beta}^{\dagger}=-\hat{E}_{\alpha\beta}^{\dagger},
\label{eq:bilinear_operators}
\ee
where $\hat{c}_{\alpha}^{\dagger}$ and $\hat{c}_{\alpha}$, $\alpha=\{i,\sigma\}$ are the fermionic creation and annihilation operators where $i$ is the site position and $\sigma$ is the spin index. These bilinear operators generate $SO(2N)$ group and their corresponding  Weyl symbols are $\rho_{\alpha\beta}=\left(\hat{E}_{\beta}^{\alpha}\right)_{W}$, $\tau_{\alpha\beta}=\left(\hat{E}_{\alpha\beta}\right)_{W}$, $-\tau_{\alpha\beta}^{*}=\left(\hat{E}^{\alpha\beta}\right)_{W}$, which satisfy canonical Poisson bracket relations with the structure constants of this $SO(2N)$ group~\cite{Davidson2017}. In addition the subset of number conserving operators $\hat{E}_{\beta}^{\alpha}$ serve as generators of the $U(N)$ subgroup of $SO(2N)$. Using phase space representation of the operators and the Hamiltonian in terms of $\rho_{\alpha\beta}$ and $\tau_{\alpha\beta}$ one can define dynamics within fTWA, which is a straightforward generalization of the classical dynamics of coupled  rigid rotators (see also Appendix A).

In Ref. \citep{Davidson2017} it was shown that phase space (Weyl) representation of the interaction term in the Hamiltonians can lead to ambiguities. For example, two-particle interactions in the Hubbard model of the type
\be
U_{\alpha\beta\gamma\delta} \hat c_\alpha^\dagger \hat c_\beta^\dagger \hat c_\gamma \hat c_\delta
\ee
can be represented either through a product of the operators $\hat E^{\alpha}_\delta$ and $\hat E^{\beta}_\gamma$ (permutation of indexes $\alpha,\beta$ or $\gamma,\delta$ leads to an equivalent representation~\cite{Davidson2017}) or alternatively through a product of operators $\hat E^{\alpha\beta}$ and $\hat E_{\gamma\delta}$. In the first representation the Weyl symbol of the Hamiltonian is represented entirely through $\rho$ - variables, while in the second representation the Hamiltonian is generally expressed through both $\rho$ and $\tau$ variables. These two different representations are not equivalent and while in some situations the first $\rho$-representation gives accurate description of dynamics within the fTWA in other situations, like e.g. for the SYK model, the second $\rho,\tau$-representation leads to accurate (and even asymptotically exact) fTWA description~\cite{Davidson2017,PhysRevB.99.134301}. In a way a choice of representation in fTWA is similar to the choice of a particular decoupling in mean-field approximations. Here we show that even if we focus on the first $\rho$-representation there are still some ambiguities in rewriting the interaction term. We use this ambiguity to our advantage significantly improving accuracy of simulations of the semiclassical many-body dynamics in systems with long-range interactions.

This new ambiguity comes from noticing that the quantum operators $\hat{n}_{i\sigma}$ and $\hat{n}_{i\sigma}^{2}$,
where $\hat{n}_{i\sigma}=\hat{c}_{i\sigma}^{\dagger}\hat{c}_{i\sigma}$, are identical. However, the Weyl symbols for these two operators lead to different phase space representations of the corresponding terms. In particular:
\be
 \hat{n}_{i\sigma}\rightarrow\rho_{i\sigma i\sigma}+1/2,
\quad \hat{n}_{i\sigma}^{2}\rightarrow\left(\rho_{i\sigma i\sigma}+1/2\right)^{2}.
\ee
Generally within the semiclassical dynamics there is no conservation law of the single cite occupation number $\rho_{i\sigma i\sigma}$, because this conservation law does not originate from the corresponding Lie algebra, but rather from its particular (fundamental) representation. A simple way to see this inequivalence is to observe that if the Hamiltonian contains the corresponding $\hat n_{i\sigma}$ term then the second representation leads to nonlinear equations of motion within the fTWA, while the first representation keeps equations linear. In the following we explain that this ambiguity can be resolved by requiring that the fTWA becomes exact in the limit of infinite range interactions. It is found that the corresponding representation also significantly improves accuracy of the fTWA for algebraically decaying long-range interactions.

In this work we focus on the Hubbard Hamiltonian with long-range interactions:
\begin{eqnarray}
\hat{H}_{I} & = & -J\sum_{\langle ij\rangle\sigma}\left(\hat{c}_{i\sigma}^{\dagger}\hat{c}_{j\sigma}+h.c.\right)+\sum_{i\sigma}\Delta_{i}\hat{n}_{i\sigma}\nonumber \\
 &  & +\sum_{ij}U_{ij}\hat{n}_{i\uparrow}\hat{n}_{j\downarrow}+\sum_{i<j,\sigma}V_{ij\sigma}\hat{n}_{i\sigma}\hat{n}_{j\sigma},\label{eq: Hamiltonian - operator form}
\end{eqnarray}
where $J$ is the spin-independent hopping amplitude between neighboring sites, $\Delta_{i}$  is the on-site disorder potential with the strength uniformly distributed in the range $\Delta_{i}\in[-\Delta,\,\Delta]$, $U_{ij}$
($V_{ij\sigma}$) is the density-density interaction coupling between different (identical) spin components. We assume that these interparticle interactions are translationally invariant and depend only on the distance between the particles: $U_{ij}=U\left(\left|\mathbf{r}_{i}-\mathbf{r}_{j}\right|\right)$ and $V_{ij\sigma}=V_{\sigma}\left(\left|\mathbf{r}_{i}-\mathbf{r}_{j}\right|\right)$ ($\mathbf{r}_{i}$ is a real space vector corresponding to the location of the site $i$). Open boundary conditions are used.

While identical fermions can not interact on the same site due to the Pauli principle, we can formally add this self interaction without affecting the dynamics considering a different Hamiltonian instead
\begin{eqnarray}
\hat{H}_{II} & = & -J\sum_{\langle ij\rangle\sigma}\left(\hat{c}_{i\sigma}^{\dagger}\hat{c}_{j\sigma}+h.c.\right)+\sum_{i\sigma}\Delta_{i}\hat{n}_{i\sigma}\nonumber \\
 &  & +\sum_{ij}U_{ij}\hat{n}_{i\uparrow}\hat{n}_{j\downarrow}+\frac{1}{2}\sum_{ij,\sigma}V_{ij\sigma}\hat{n}_{i\sigma}\hat{n}_{j\sigma}\nonumber \\
 & = & \hat{H}_{I}+\underset{\frac{1}{2}V_{0\uparrow}\hat{N}_{\uparrow}+\frac{1}{2}V_{0\downarrow}\hat{N}_{\downarrow}}{\underbrace{{1\over 2}\sum_{\sigma}V_{0\sigma}\sum_{i}\hat{n}_{i\sigma}^{2}}},\label{eq: Hamiltonian - good}
\end{eqnarray}
where $V_{0\sigma}\equiv V_{\sigma}\left(\left|\mathbf{r}_{i}-\mathbf{r}_{i}\right|=0\right)$.  The difference between the two Hamiltonians  is proportional to terms with conserved number of fermions in each spin degree of freedom: 
\be
\hat H_{II}-\hat H_I=\frac{1}{2}V_{0\uparrow}\hat{N}_{\uparrow}+\frac{1}{2}V_{0\downarrow}\hat{N}_{\downarrow}, \quad \hat{N}_\sigma=\sum_{i}\hat{n}_{i\sigma},
\ee
which commutes $\hat H_{I}$ and hence both Hamiltonians lead to identical quantum dynamics. However, these two Hamiltonians lead to different semiclassical approximations. In the next section (Sec. \ref{sec: Benchmark-of-fTWA}) we show that fTWA based on $\hat{H}_{II}$ leads to much more accurate predictions. Intuitively this improvement follows from considering infinite range interactions where $U_{ij}, V_{ij\sigma}$ are independent of $\mathbf{r}_{i}-\mathbf{r}_{j}$. In this case it is easy to check (see Appendix B for details) that the interaction term commutes with the rest of the Hamiltonian and drops out from the equations of motion. However, within the fTWA the interaction term in this limit only drops if we use the Weyl representation of $\hat H_{II}$, but not $\hat H_I$. We checked that for the systems with short-range interactions both Weyl representations lead to similar results. 

In the rest of the paper we focus on the situation $U_{ij}=V_{ij\sigma}$ corresponding to an additional $SU(2)$ spin symmetry \citep{PhysRevB.59.12822,PhysRevB.95.054204}. We also consider long-range power law interactions such that
\begin{equation}
U_{ij}=V_{ij\sigma}=\frac{U}{\left|\mathbf{r}_{i}-\mathbf{r}_{j}\right|^{\alpha}}, \,\,\,i\neq j,
\end{equation}
and for on-site interactions $U_{ii}=V_{ii\sigma}=U$ is taken. As we already pointed in the infinite range case $\alpha=0$, the dynamics of the system becomes effectively noninteracting, i.e. equivalent to $U=0$, because in this case the interaction term simply reduces to the square of the total number of fermions with factor $U$.

\section{Benchmarking accuracy of the fTWA  in one dimension \label{sec: Benchmark-of-fTWA}}

Before proceeding with analyzing dynamics in two-dimensional systems we will check accuracy of fTWA and differences between the two semiclassical representations of the quantum Hamiltonian in smaller one-dimensional systems. In particular, we will analyze quench dynamics in the half filled one-dimensional lattice of eight
sites with open boundary conditions.  Such a system is amenable to exact diagonalization methods and hence can be used to test the semiclassical method. Following recent cold atom experiments~\citep{mblschreiber,Choi2016,PhysRevLett.116.140401,PhysRevLett.119.260401} we will study dynamics of the spin and charge imbalance after quenching from a charge/spin density wave state. This imbalance serves as a good indicator of ergodicity in the system. At strong disorder, where the imbalance does not decay in time due to localization a more sensitive probe distinguishing dynamics of interacting and non-interacting systems is the quantum Fisher information (QFI), which was recently measured for long-range interacting ions in the presence of  disorder \citep{Smith2016}. Like entanglement the QFI can distinguish between noninteracting Anderson
localization mechanism~\citep{PhysRev.109.1492} and possible many-particle localization (MBL) ~\citep{Basko2006,PhysRevB.82.174411,PhysRevLett.109.017202}. The details of implementation of the fTWA method are described in the Appendixes A and B.

\begin{figure}[tbh]
\includegraphics[scale=0.55]{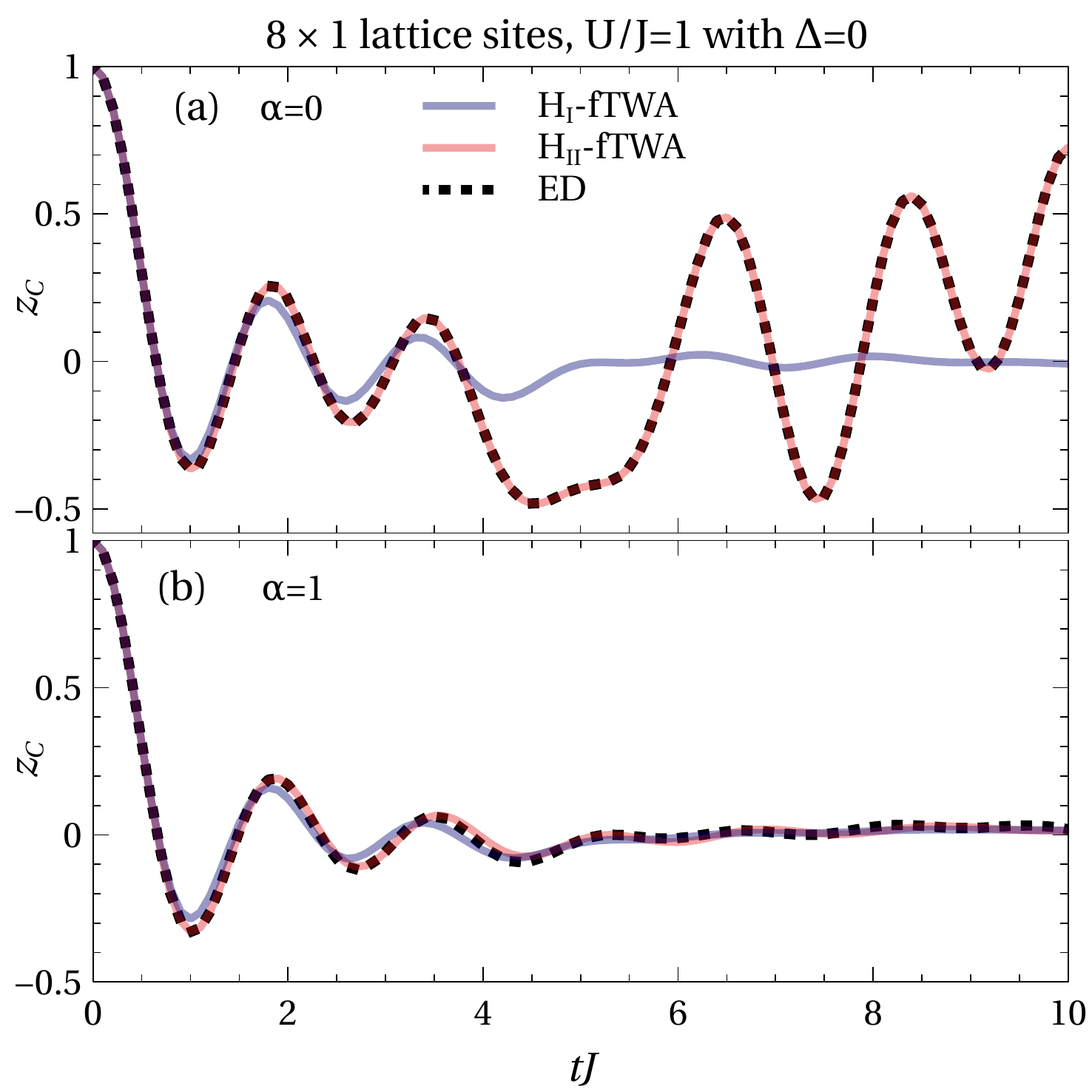}

\caption{(Color online) Time dependence of the charge imbalance $z_{C}$ (c.f. Eq.~\eqref{eq:zc_def}) for the initial CDW state. The black dashed line represents the exact diagonalization (ED) results, while the solid blue (dark gray) and the solid red (light gray) lines represent the results of the fTWA method based on the Hamiltonians $\hat{H}_{I}$ and $\hat{H}_{II}$ respectively. 
The top a) panel corresponds to the infinite range interactions ($\alpha=0$) and the bottom b) panel corresponds to the interactions decaying with the power $\alpha=1$. All simulations are obtained on a system consisting of 8-lattice sites at half filling with $U/J=1$ in the absence of disorder ($\Delta=0$). 
\label{fig: CDW - non-disorder}}
\end{figure}

\begin{figure*}[th]
\includegraphics[scale=0.4]{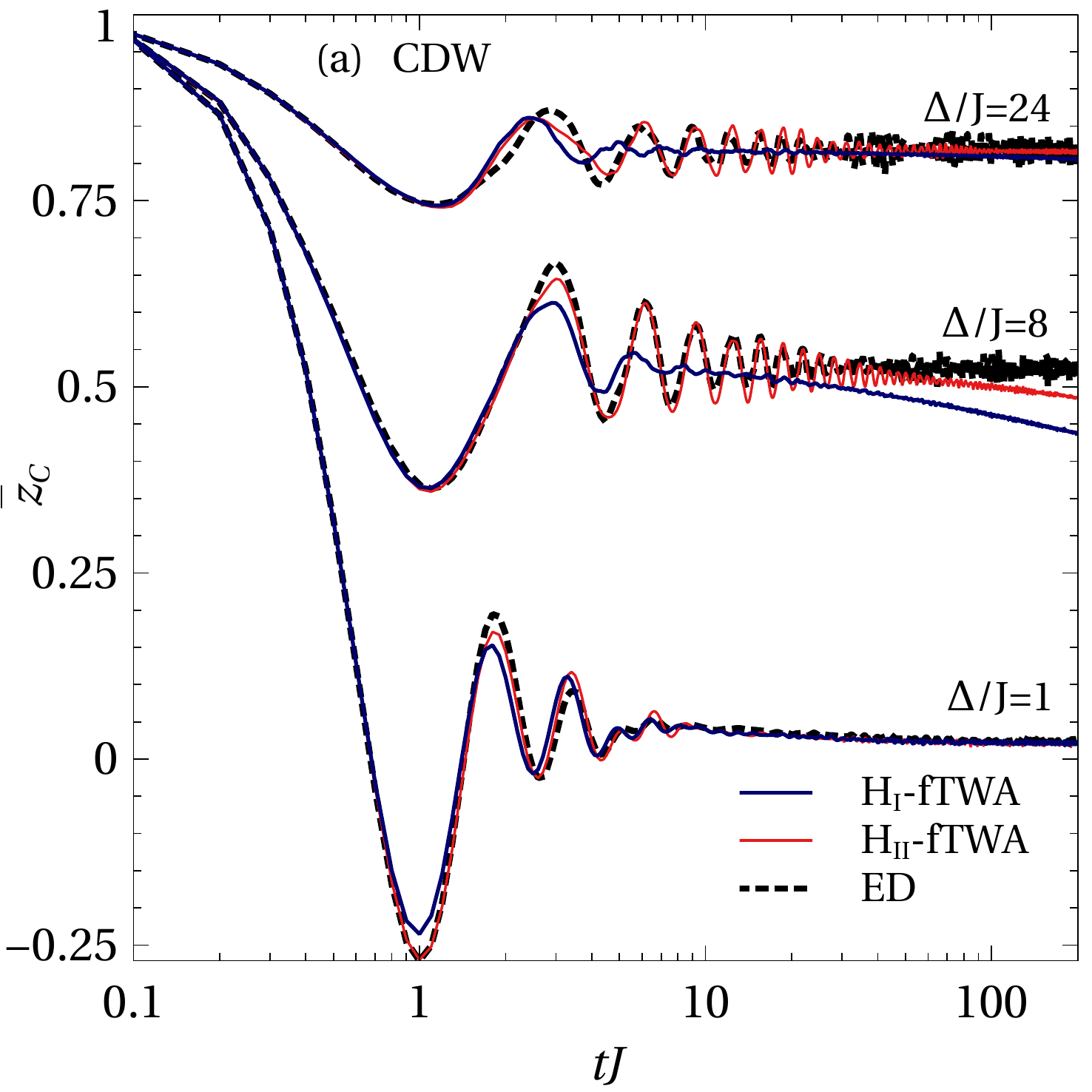}\includegraphics[scale=0.4]{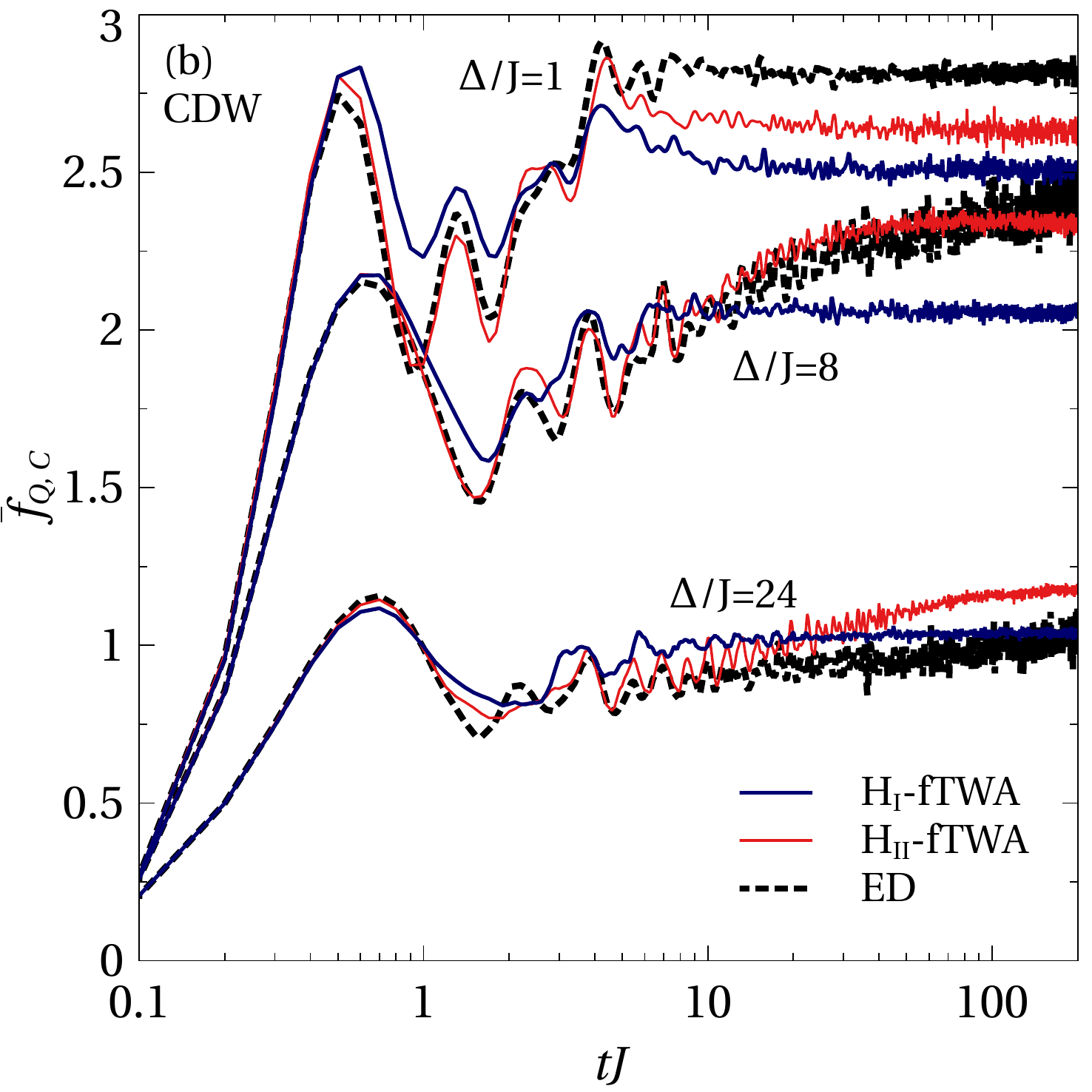}\includegraphics[scale=0.4]{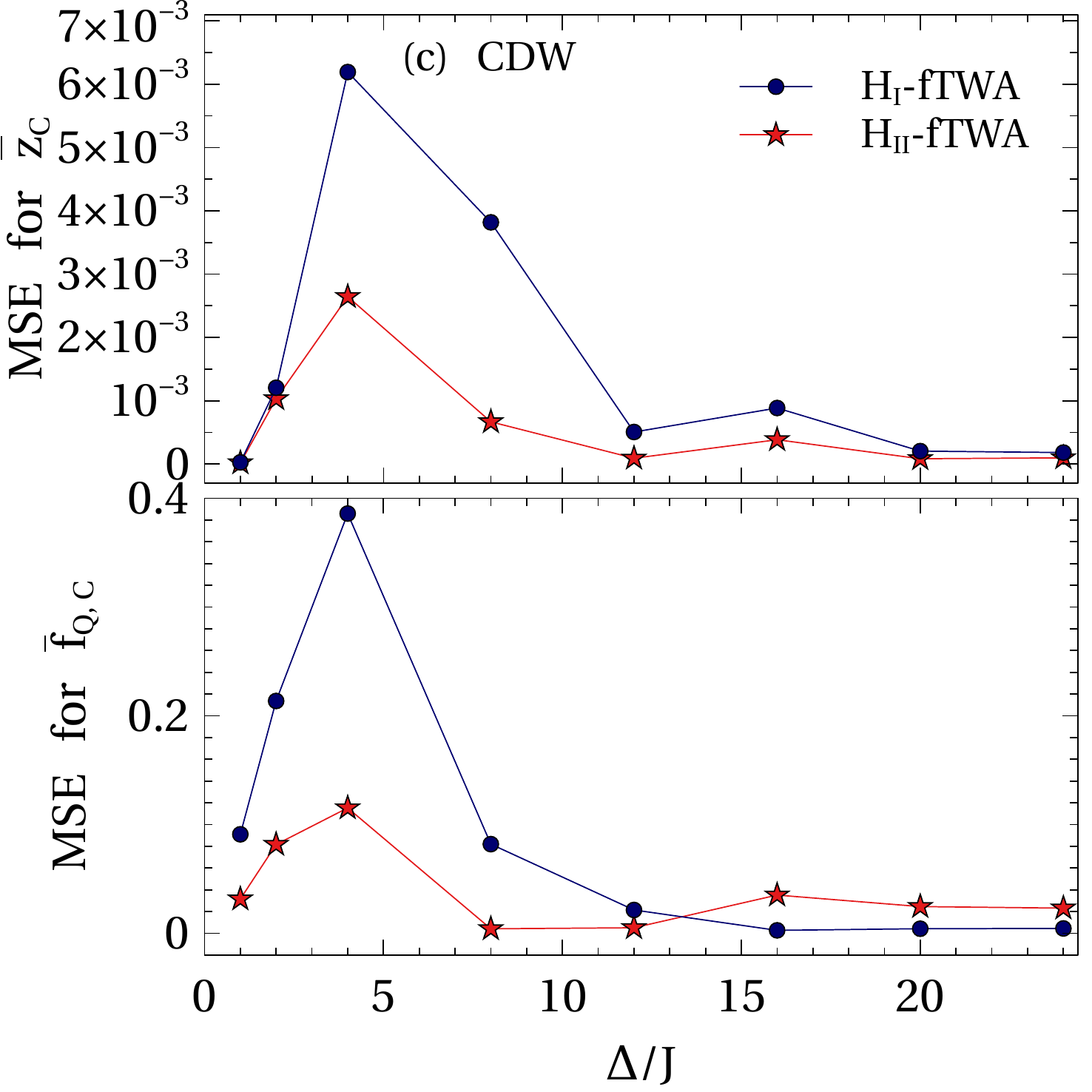}

\includegraphics[scale=0.4]{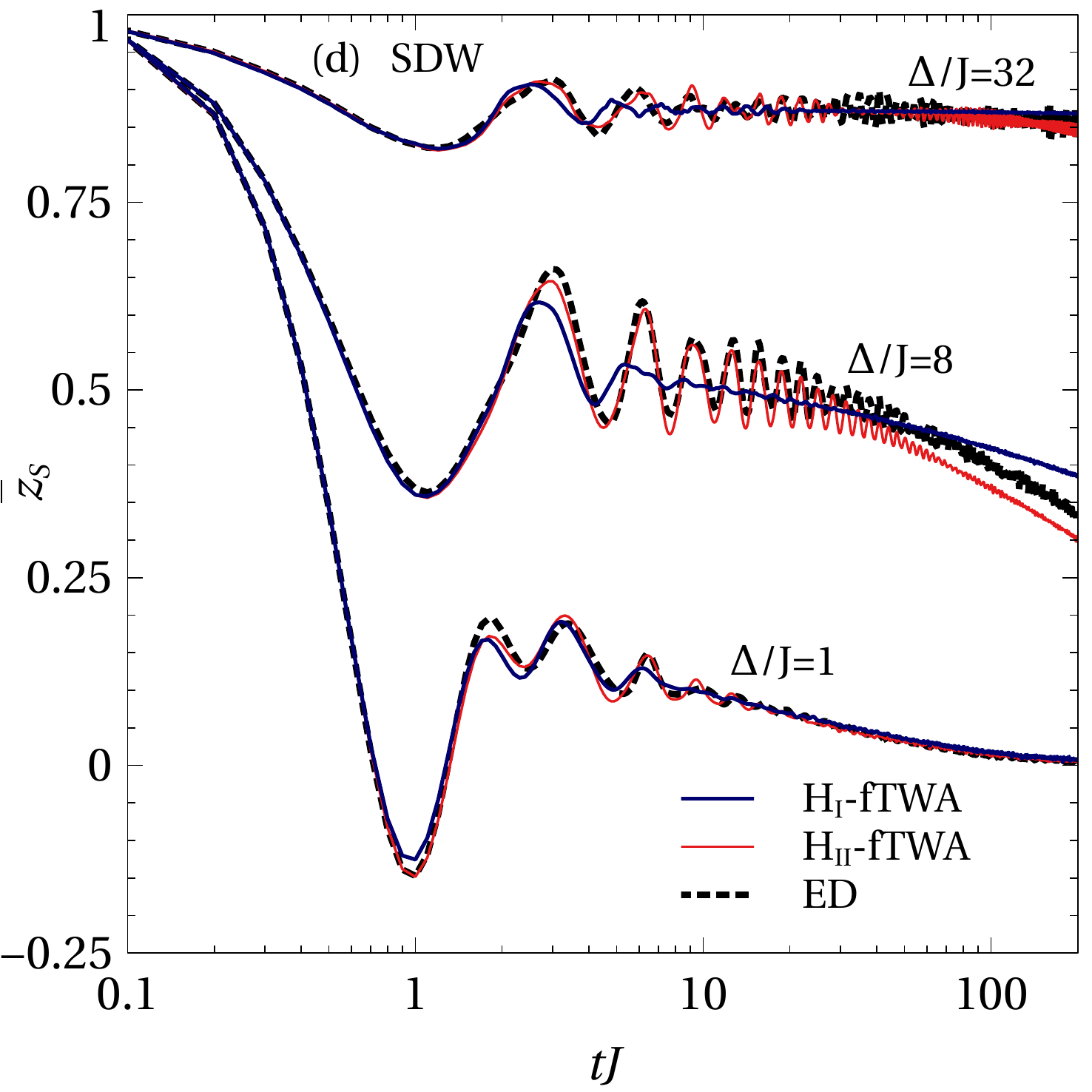}\includegraphics[scale=0.4]{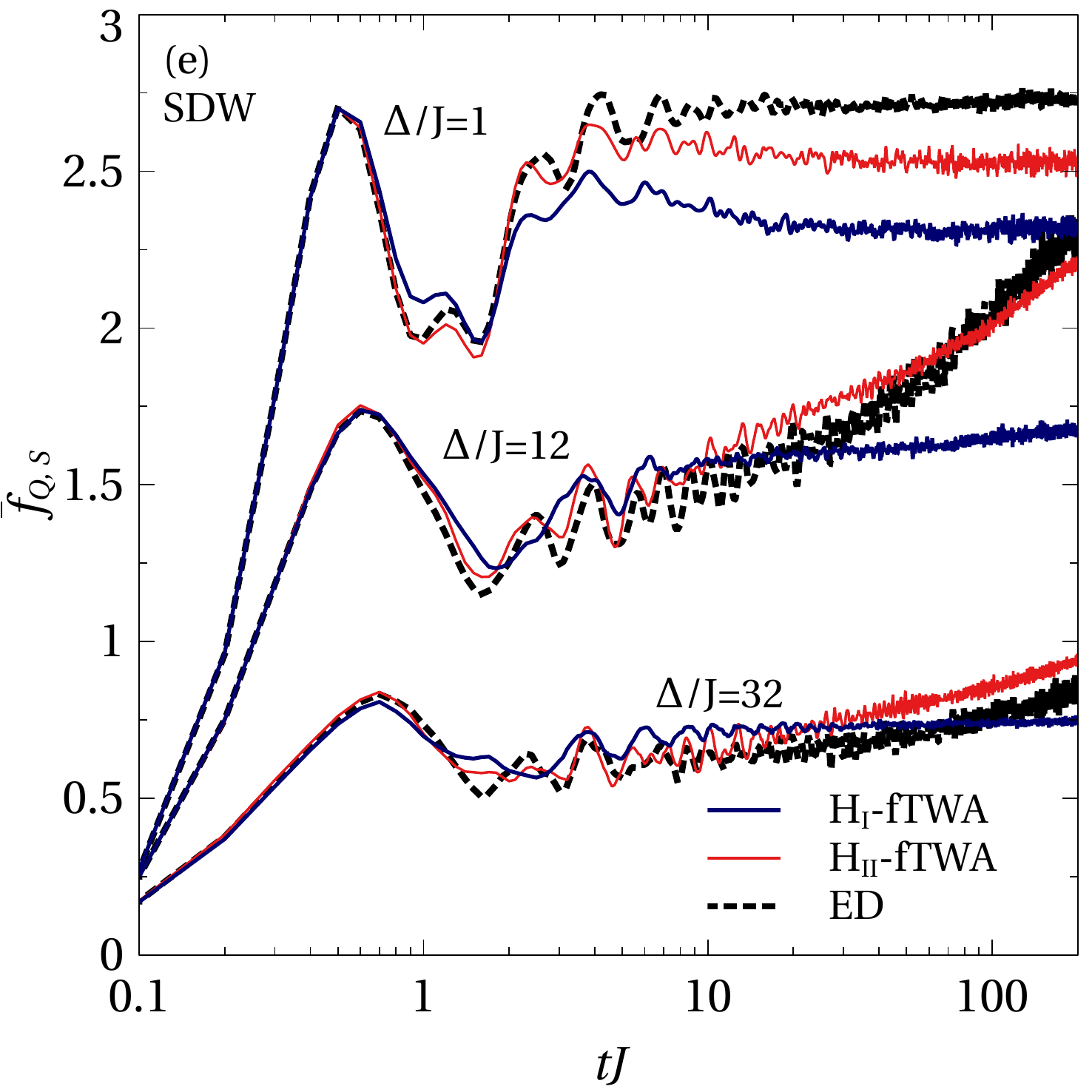}\includegraphics[scale=0.4]{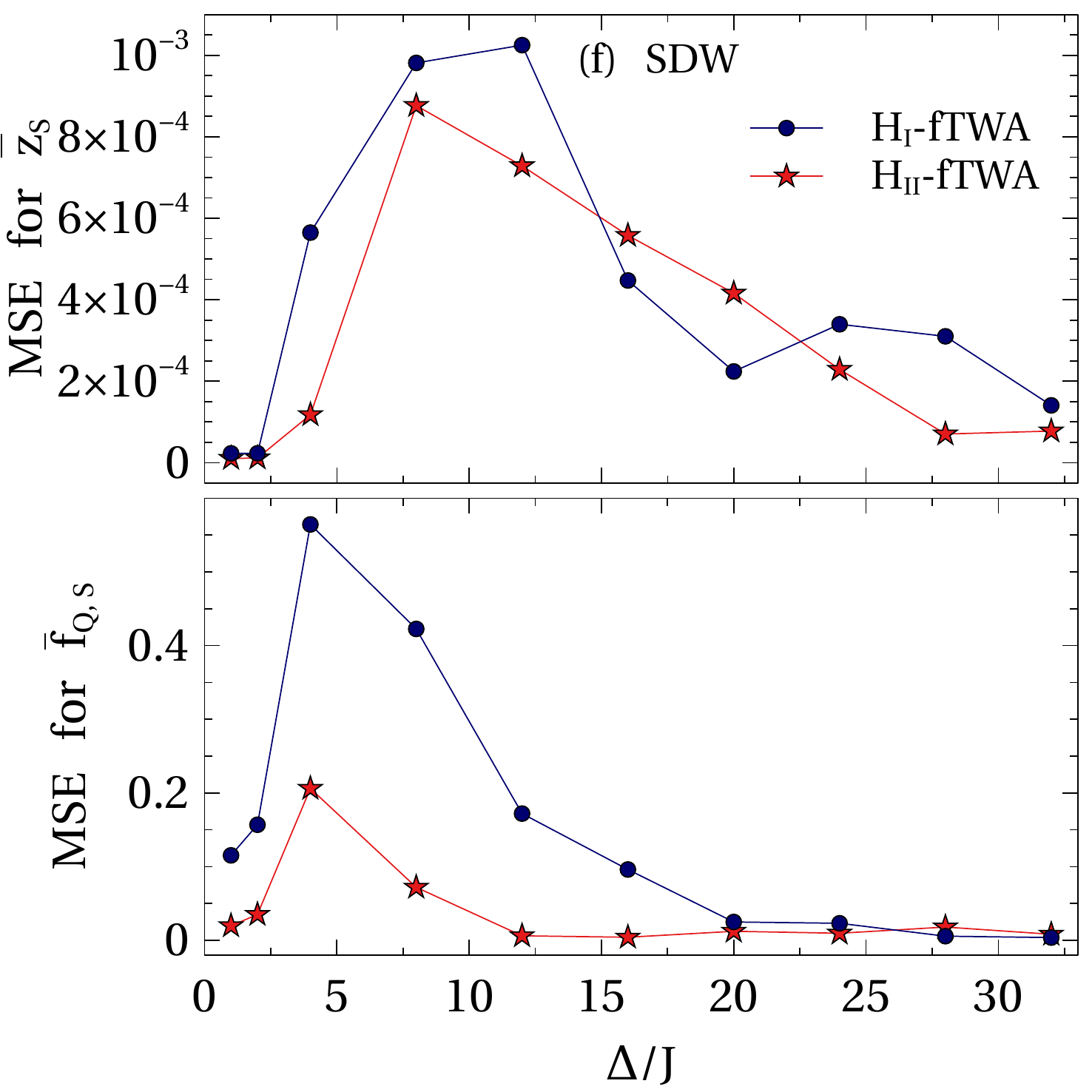}

\caption{(Color online) Time dependence of the imbalance $z_{C/S}$ (a and d) and of the QFI $f_{C/S}$ (b and e) for different disorder strengths $\Delta/J$.
Simulations are done for a 1D system of size 8 and open boundary conditions. The top/bottom rows correspond to the CDW/SDW initial states (see text for details). The solid blue (dark gray) and red (light gray) lines represent simulations done with $H_{I}-\textrm{fTWA}$ and $H_{II}-\textrm{fTWA}$, respectively; the black dashed line shows ED simulations. The panels (c) and (f) show the mean square error (MSE) of the two fTWA approximations. All the results  were averaged over $100$ different disorder realization. The remaining parameters of the Hamiltonian are  $U/J=1$, $\alpha=1$.
\label{fig: MSE}}
\end{figure*}

\begin{figure*}[t]
\includegraphics[scale=0.4]{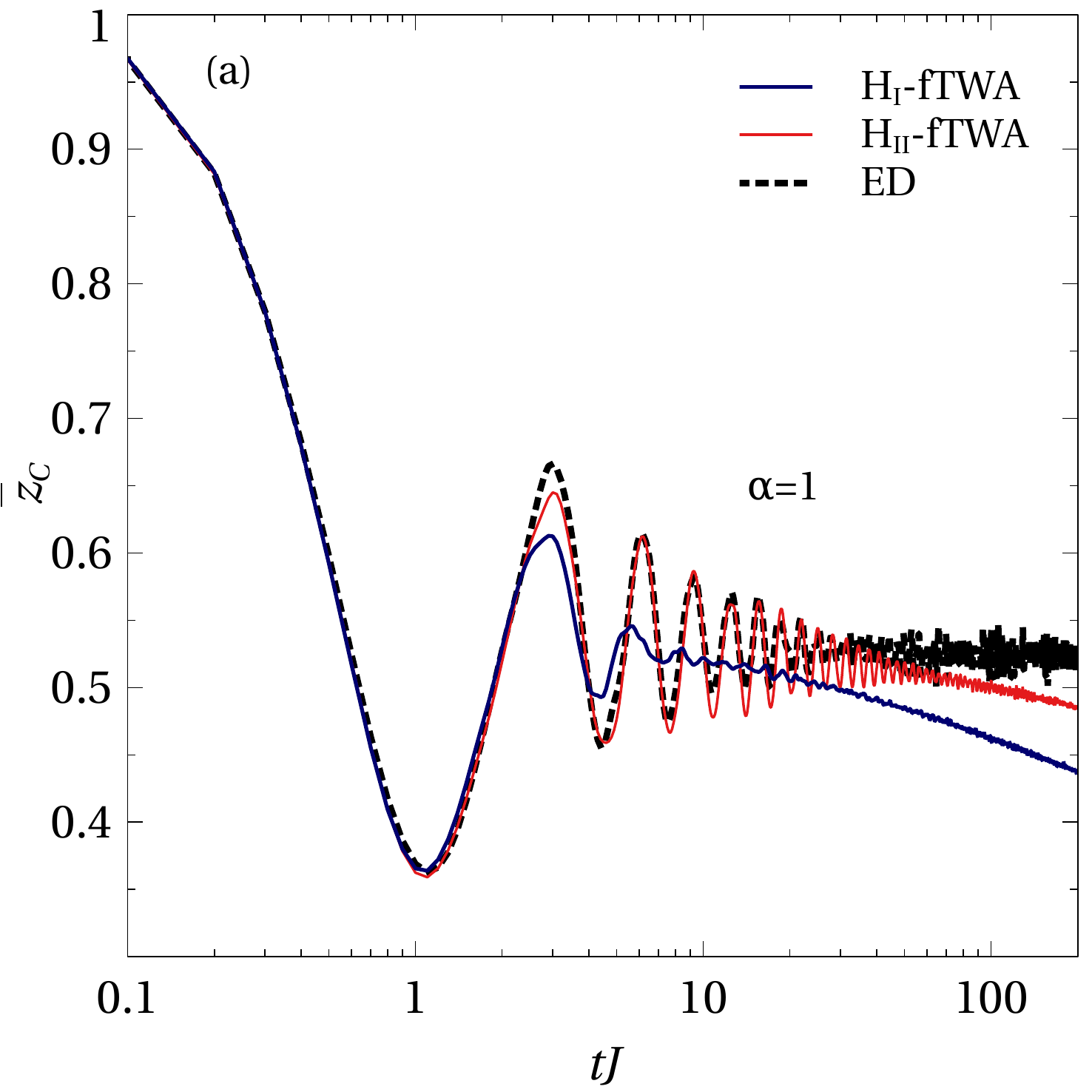}\includegraphics[scale=0.4]{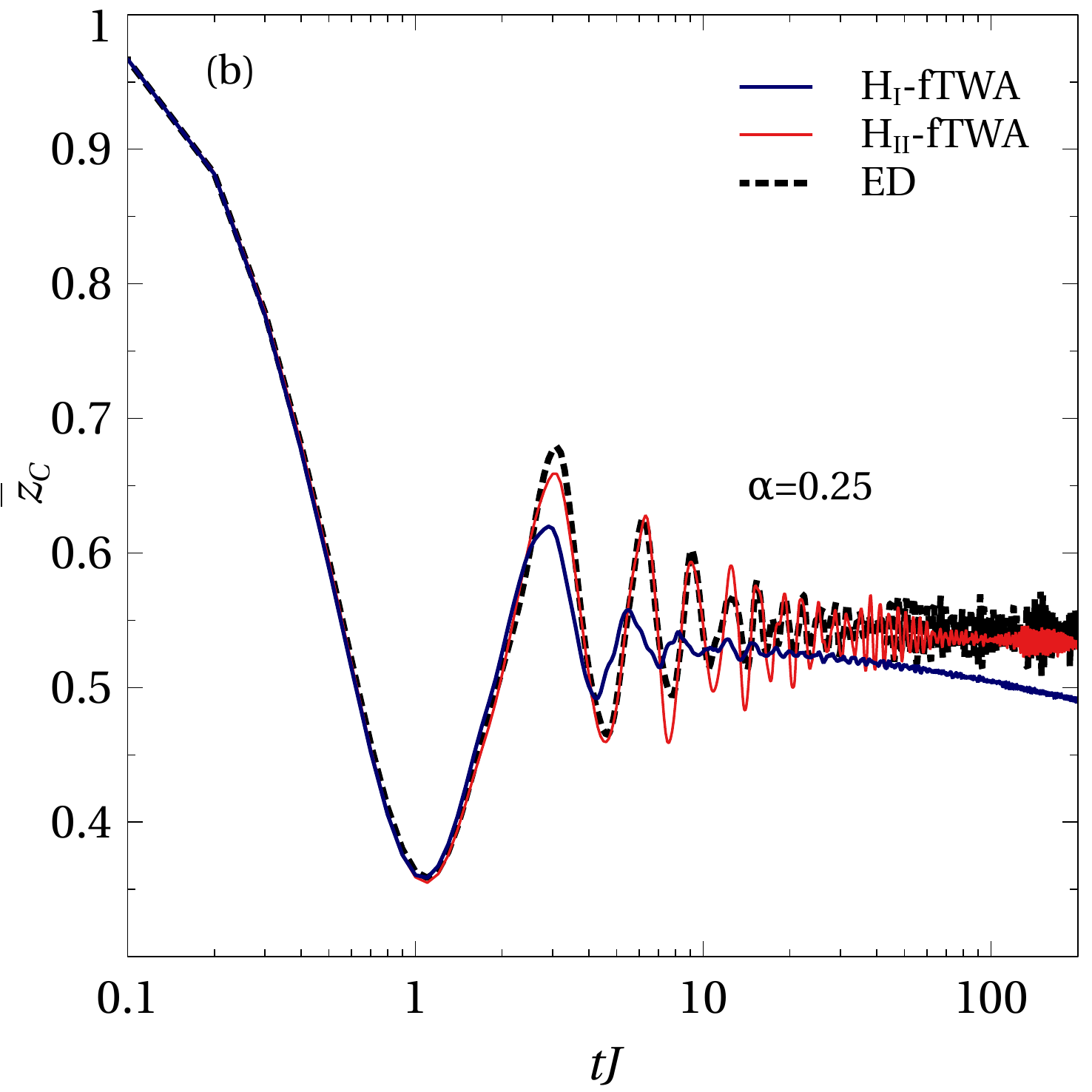}\includegraphics[scale=0.4]{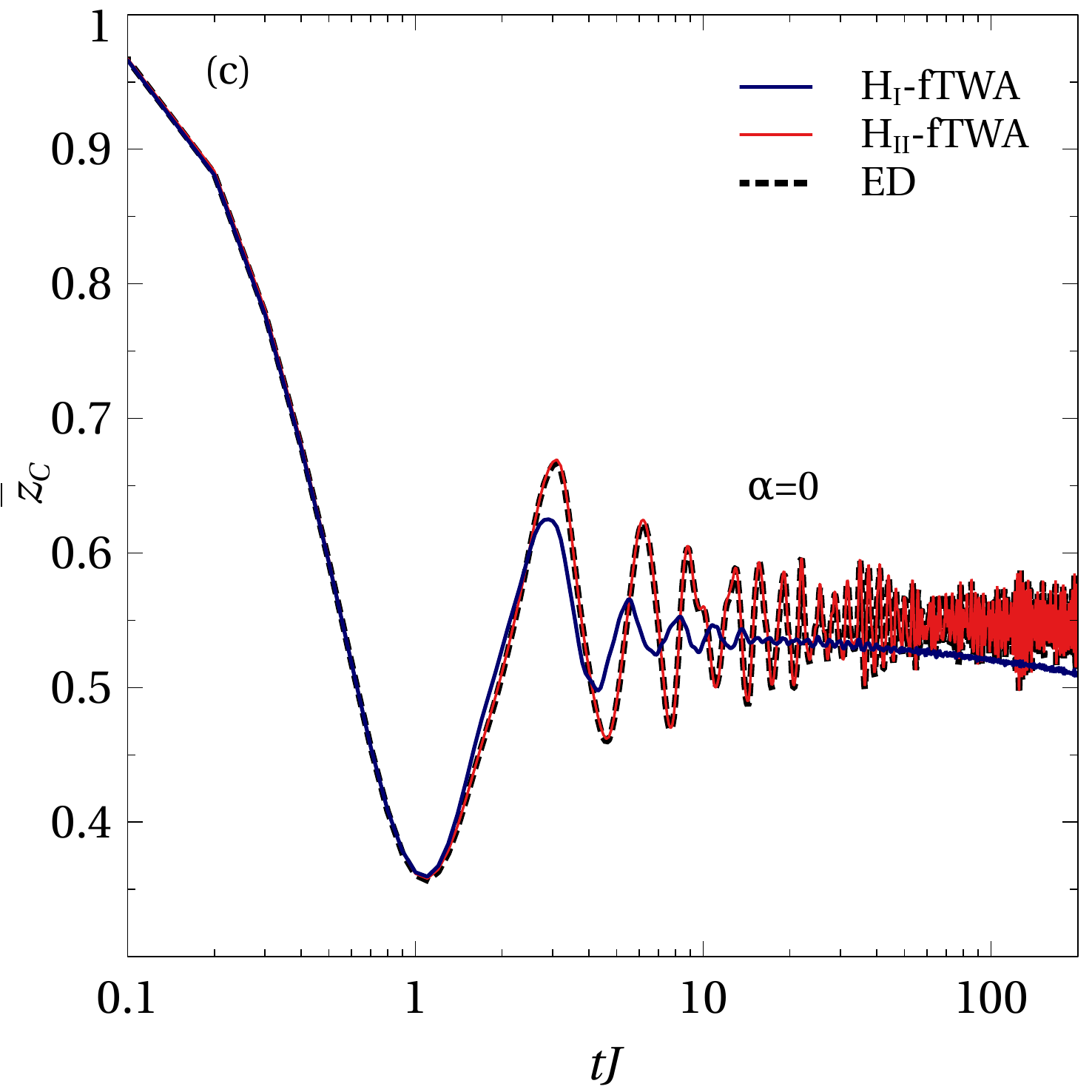}

\includegraphics[scale=0.4]{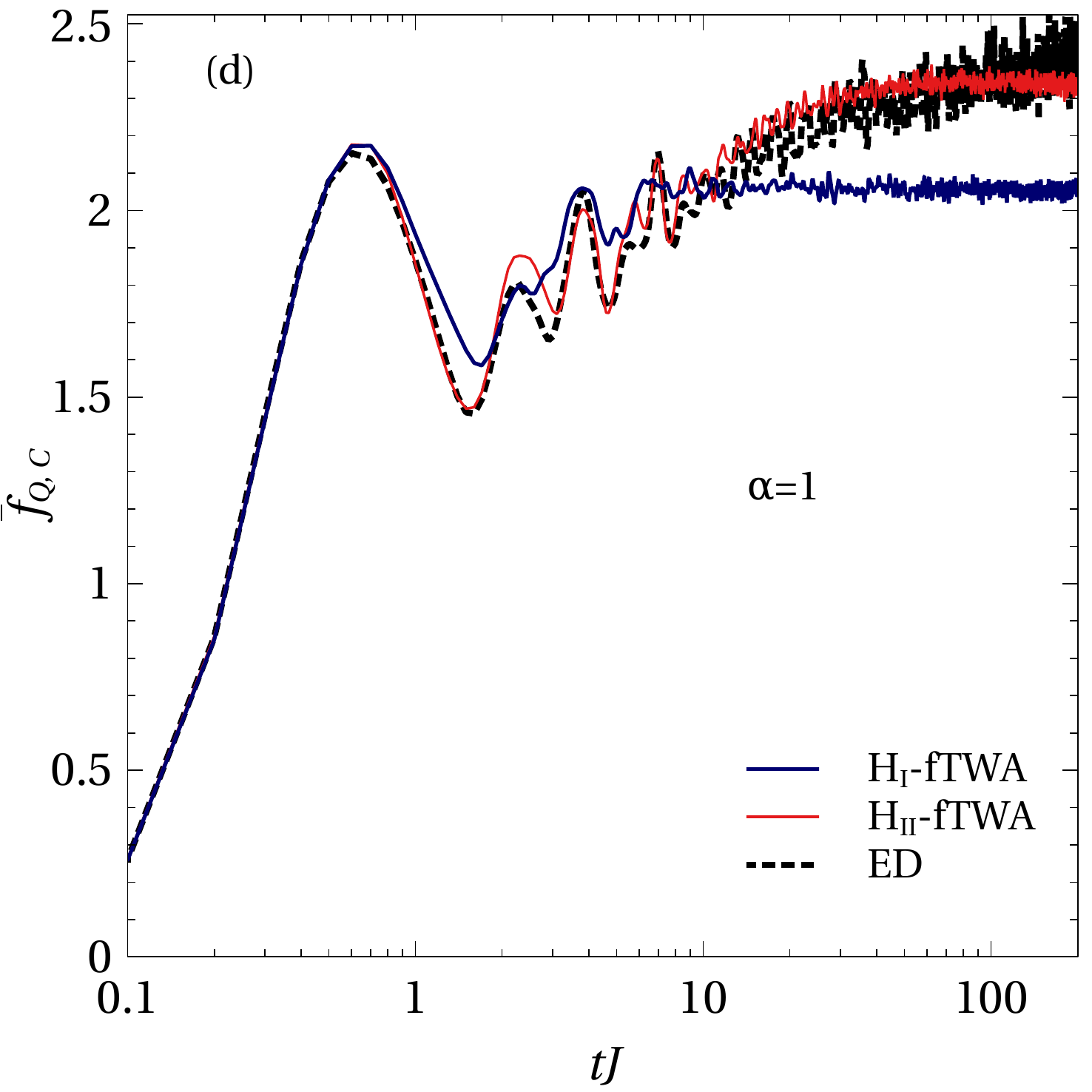}\includegraphics[scale=0.4]{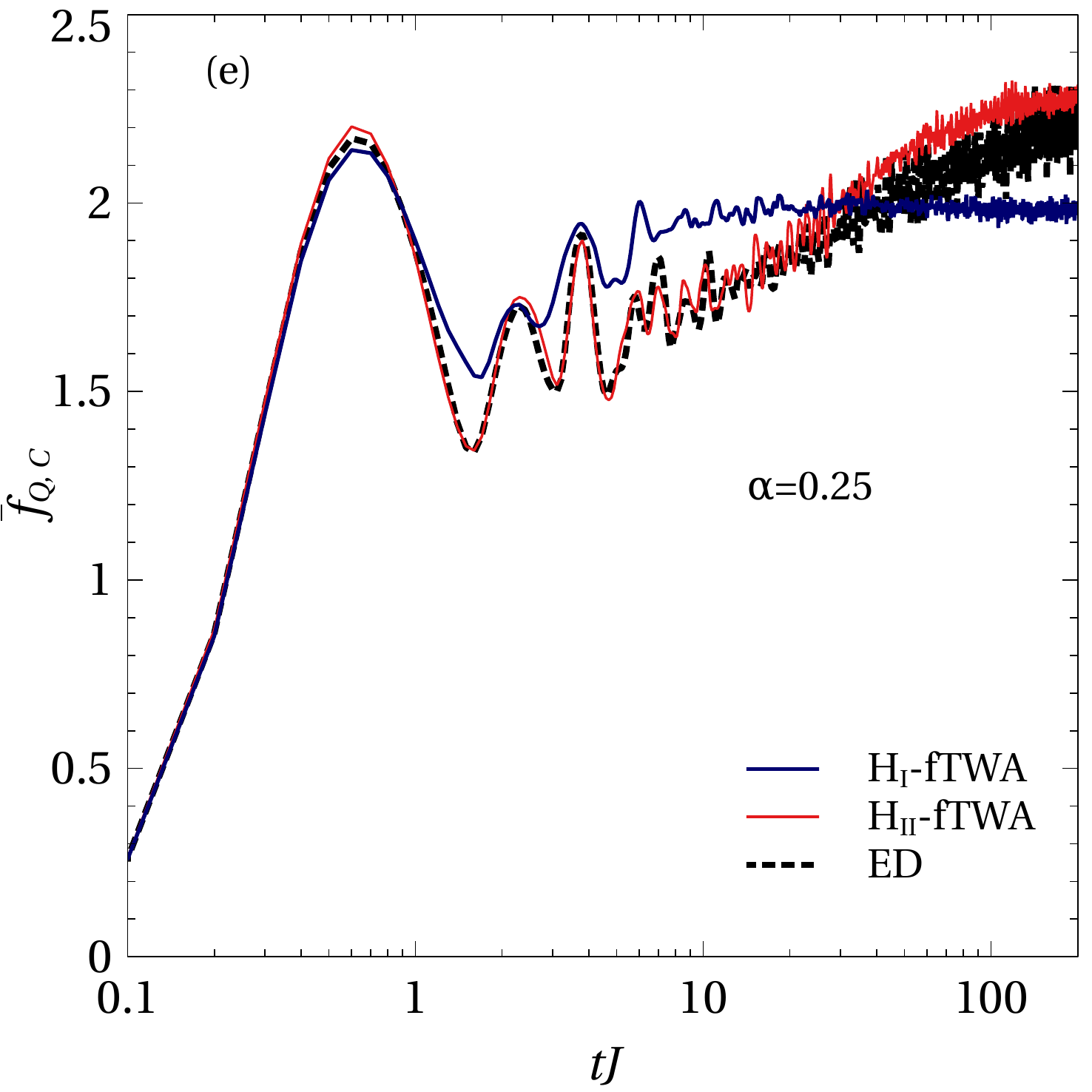}\includegraphics[scale=0.4]{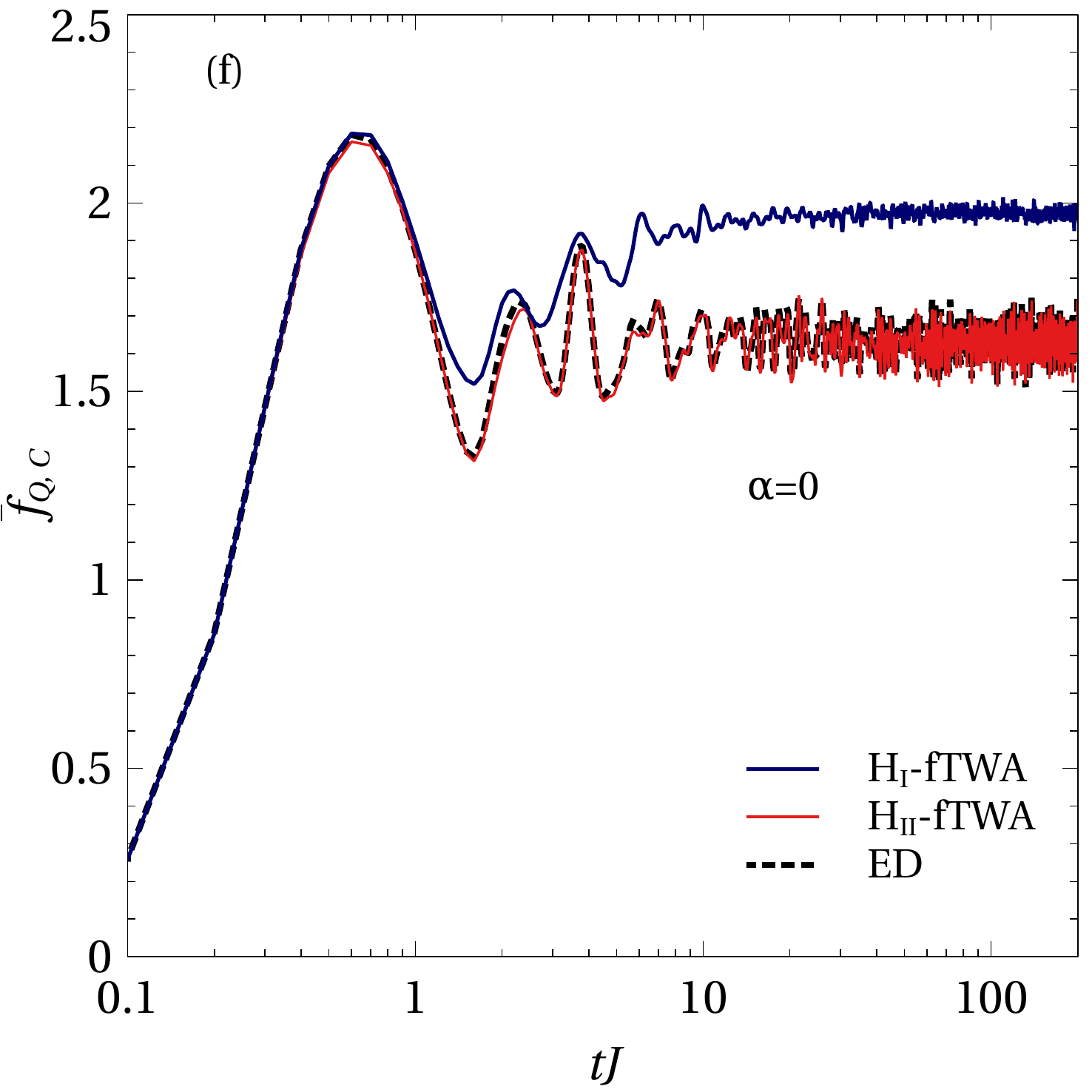}\caption{(Color online) Time dependence of the charge imbalance function $z_{C}(t)$ (a-c) and
QFI denisty $f_{Q,C}$ (d-f). Simulations are done for a 1D system with
8 sites and open boundary conditions. The system is initially prepared
in a CDW state (every even side is doubly occupied). The interaction strength
is $U/J=1$ with $\alpha=1$ for a and d, $\alpha=0.25$ for b and
e, $\alpha=0$ for c and f. Moreover, $H_{I}-\textrm{fTWA}$ data
are represented by the blue (dark gray) line, $H_{II}-\textrm{fTWA}$ by the red solid (light gray) line and ED by
the black dashed line. All the results were averaged over at least $100$ disorder
potentials with the strength $\Delta/J=8$. \label{fig: fTWA and ED}}
\end{figure*}

\begin{figure}[tbh]
\includegraphics[scale=0.55]{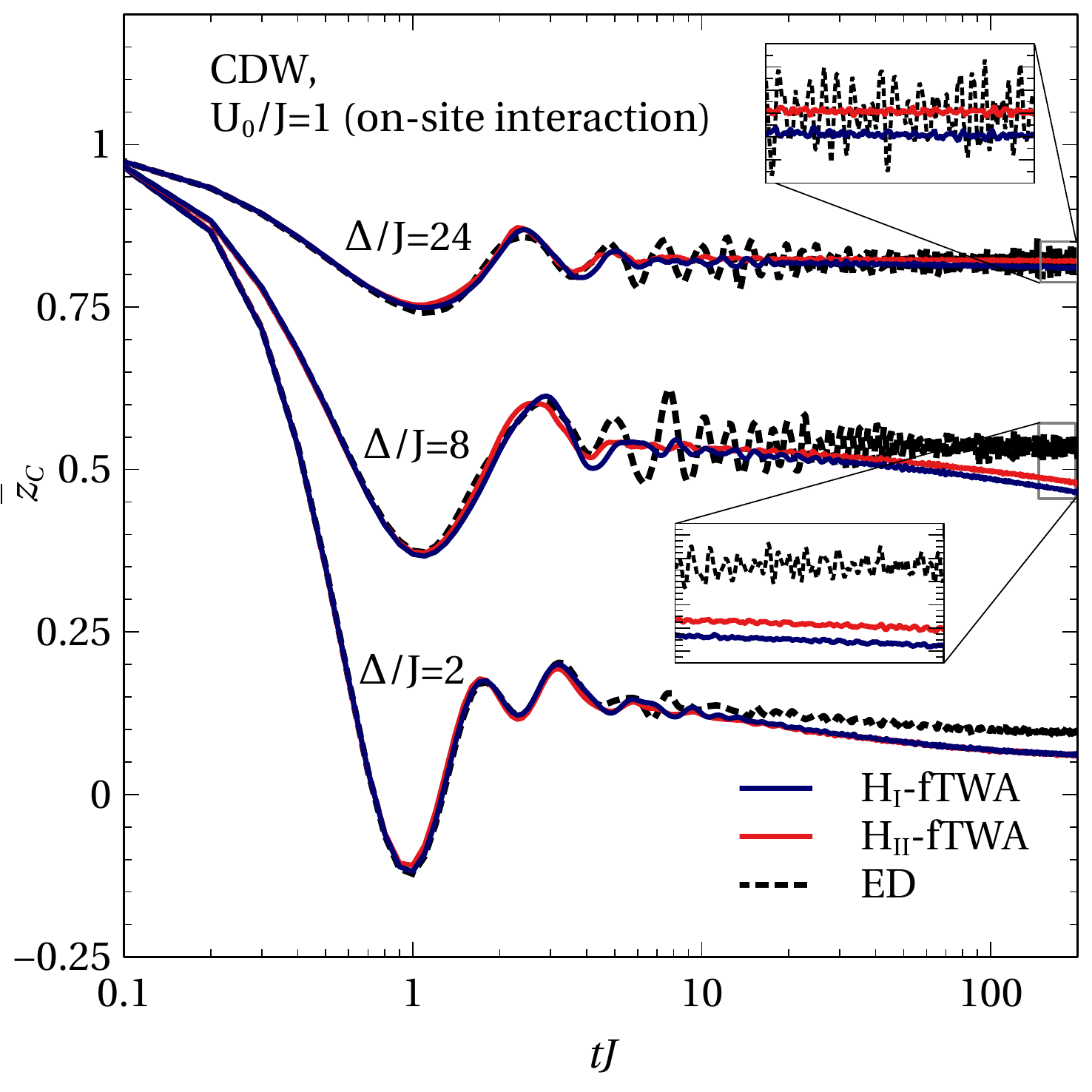}

\caption{(Color online) Time dependence of the charge imbalance function $z_{C}(t)$. Simulations are done for a 1D system of size
8 with open boundary conditions. The system is initially prepared
in the CDW state (every even site is doubly occupied) and the interactions are short-range (on-site i.e. $U_{ij}=V_{ij\sigma}=0$ for $i\neq j$ and $U_{ii}=V_{ii\sigma}=U_0\neq0$). The interaction strength is $U_{0}/J=1$. The $H_{I}-\textrm{fTWA}$ data
are represented by the blue (dark gray) lines, the $H_{II}-\textrm{fTWA}$ data - by the red (light gray) lines and the ED data - by
the black dashed lines. The results were averaged over at least $100$ disorder potentials with the strengths $\Delta/J=24, 8, 2$ (from top to bottom). 
\label{fig: CDW - short-range}}
\end{figure}

To study dynamics of charge and spin degrees of freedom we prepare
system in charge density (CDW) and spin density (SDW) wave, respectively.
For charge dynamics we consider a pure initial state 
\be
|\Psi_{init}^{C}\rangle=|0,\,\uparrow\downarrow,\,0,\,\uparrow\downarrow,\,...\rangle
\label{eq:psi_cdw}
\ee
and for spin dynamics we start from a different initial state
\be
|\Psi_{init}^{S}\rangle=|\downarrow,\,\uparrow,\,\downarrow,\,\uparrow,\,...\rangle.
\label{eq:psi_sdw}
\ee
As observables we choose the charge ($z_{C}(t)$) and the spin ($z_{S}(t)$) imbalances normalized to the total number of fermions $N$, which we define
\be
z_{C}(t)={1\over N}\left\langle \hat{Z}_{C}(t)\right\rangle ={1\over N}\sum_{i}(-1)^{i}\langle\Psi_{init}^{C}|\hat{n}_{i}(t)|\Psi_{init}^{C}\rangle
\label{eq:zc_def}
\ee
and
\be
z_{S}(t)={1\over N}\left\langle \hat{Z}_{S}(t)\right\rangle ={1\over N}\sum_{i}(-1)^{i}\langle\Psi_{init}^{S}|\hat{m}_{i}(t)|\Psi_{init}^{S}\rangle,
\label{eq:zs_def}
\ee
where  $\hat{n}_{i}=\hat{n}_{i\uparrow}+\hat{n}_{i\downarrow}$ and $\hat{m}_{i}=\hat{n}_{i\uparrow}-\hat{n}_{i\downarrow}$ are the on-site charge density and the on-site spin polarization respectively. In a two dimensional square lattice, analyzed in Sec.~\ref{sec:Charge-and-spin-in-2D}, we replace $(-1)^i$ with $(-1)^{i_x}$, where $i_x$ and $i_y$ are the integer site indexes along the $x$ and $y$-directions.

For a pure initial state we will also study the normalized QFI corresponding to the charge and spin imbalance operators $\hat Z_C$ and $\hat Z_S$~\citep{PhysRevLett.72.3439,PhysRevA.85.022321, PhysRevA.85.022322,Smith2016,Hauke2016,PhysRevB.99.054204,PhysRevB.99.241114}:
\begin{equation}
f_{Q,C/S}(t)={4\over N}\left[\left\langle \hat{Z}_{C/S}^{2}(t)\right\rangle -\left\langle \hat{Z}_{C/S}(t)\right\rangle ^{2}\right].\label{eq: QFI charges - definition}
\end{equation}
We stress that it is the presence of quantum noise, which is essential to get the nontrivial QFI in the fTWA method. In the mean-field approximation, which is equivalent to the fTWA if we suppress all the noise to zero, $f_{Q,C/S}(t)\equiv 0$ because there are no fluctuations and hence  $\langle \hat{Z}_{C/S}^{2}(t)\rangle=\langle \hat{Z}_{C/S}(t)\rangle^2$. On top of quantum averaging we will also perform averaging of the observables over different disorder realizations, which we denote by an over-line and show numerical results for $\bar{z}_{C/S}$, $\bar{f}_{Q,C/S}$.

First let us analyze the charge imbalance $z_{C}(t)$ starting from the initial CDW state. In Fig. \ref{fig: CDW - non-disorder}, we show comparison of ED and fTWA dynamics for a non-disordered lattice ($\Delta=0$). The top a) and the bottom b) plots correspond to the infinite interaction range ($\alpha=0$) and the interactions proportional to the inverse distance ($\alpha=1$). As we discussed  in Sec. \ref{sec: Phase-space-representation of HH} fTWA based on
$\hat{H}_{II}$ ($H_{II}$-fTWA) is exact for $\alpha=0$ for any system size. In comparison, the  fTWA based on $\hat{H}_{I}$ ($H_{I}$-fTWA) is only accurate for relatively short times. In the case of power law interactions the system is nonintegrable and generally exhibits thermalization (Fig. \ref{fig: CDW - non-disorder} b). In this case both representations of the fTWA give similar accurate predictions of the charge imbalance decay with  $H_{II}$-fTWA
still over-performing the $H_{I}$-fTWA. We checked that the situation is similar for the spin imbalance $z_{S}(t)$.

As we increase the disorder strength the improvement of the $H_{II}$-fTWA over the $H_{I}$-fTWA gets even more significant especially at longer times. This is illustrated  in Fig. \ref{fig: MSE}, where we plot $z_{C/S}$ and $f_{Q,C/S}$ for a fixed power $\alpha=1$. Here the upper panels represent the results for the charges and the lower panels correspond to spins. The right c) and f) panels show the mean square error (MSE) of the fTWA simulations as a function of the disorder strengths. The MSE is defined as
\be
\frac{1}{M}\sum_{i=1}^{M}(\bar{z}_{C/S}^{\textrm{ED}}(t_{i})-\bar{z}_{C/S}^{\textrm{fTWA}}(t_{i}))^{2}
\ee
for the charge/spin imbalance and 
\be
\frac{1}{M}\sum_{i=1}^{M}(\bar{f}_{Q,\, C/S}^{\textrm{ED}}(t_{i})-\bar{f}_{Q,\, C/S}^{\textrm{fTWA}}(t_{i}))^{2}
\ee
for the QFI. Here $M$ is the total number of time simulation steps within the time interval $t_{i}J\in[0,200]$. From Figs. \ref{fig: MSE} c) and f), we see that the fTWA gives satisfactory predictions both for short and long time dynamics in the limits of weak ($\Delta/J\approx1$) and strong ($\Delta/J\gg1$) disorder potentials. In the intermediate disorder regime the fTWA introduces a significant error. This situation is qualitatively similar to the one for spin systems~\citep{PhysRevA.96.033604, WURTZ2018341}. At these intermediate disorder strengths the semiclassical dynamics clearly leads to faster thermalization than exact quantum dynamics. A possible explanation for why classical systems thermalize faster was given in Ref.~\cite{Oganesyan_2009}. There the authors argued that it is discreteness of quantum levels, which further suppresses slow classical transport through chaotic resonances. While it is unclear how these considerations extend to fTWA, which deals with nonlocal bilinears, qualitatively the situation is very similar. We point that fTWA shows stronger tendency to localization than e.g. the cluster TWA. It is possible that accuracy of fTWA can be further improved by choosing a more efficient operator basis, e.g. the basis of l-bits~\cite{PhysRevB.98.184201}, which is obtained by a local unitary transformation of the local fermion basis. Unitary transformations do not change the commutation relations of the basis operators and hence the dressed operators still form a closed algebra and can be used to construct dressed versions of fTWA. This possibility needs further investigation, which is beyond the scope of the current manuscript.

It is also interesting to point out that both the ED and the fTWA (c.f. Fig. \ref{fig: MSE}) show that spin degrees of freedom tend to thermalize faster than charges. Similar tendency was also observed in several recent papers~\citep{PhysRevB.94.241104,PhysRevB.97.064204,PhysRevLett.120.246602, PhysRevB.98.115106,PhysRevB.98.014203,1910.11889, PhysRevB.100.125132,PhysRevB.99.121110,PhysRevB.99.115111}. The reason behind the asymmetry between spin and charge degrees of freedom is that the Hamiltonian~\eqref{eq: Hamiltonian - good} introduces only disorder in the charge sector allowing spins to delocalize much faster. In order to localize spins one can introduce additional disorder in the spin channel~\cite{PhysRevB.94.241104,PhysRevB.100.125132}. We checked that this is indeed correct in the Sec. \ref{sec:Charge-and-spin-in-2D}, where charge and spin dynamics in the two-dimensional setup is discussed.

From the MSE it is seen also that the $H_{II}$-fTWA is generally more accurate in predicting both the imbalance and the QFI especially in the crossover region (c.f. Fig. \ref{fig: MSE} c and f). Also the $H_{II}$-fTWA is able to capture initial the transient imbalance oscillations up to longer times ($tJ\approx10$) as compared to the $H_{I}$-fTWA, which agrees with the ED only up to $tJ\approx3$.

Next, we analyze accuracy of the fTWA as we vary the exponent $\alpha$. As we mentioned in Sec. \ref{sec: Phase-space-representation of HH}, $H_{II}$-fTWA should approach the exact results for $\alpha\rightarrow0$.
We only consider behavior of the charge imbalance $z_C(t)$ and $f_C(t)$ (the behavior of the spin imbalance is qualitatively similar), which are shown in Fig. \ref{fig: fTWA and ED} for a fixed quenched disorder with the strength $\Delta/J=8$. At this strong disorder the charge transport is suppressed but yet the imbalance changes significantly compared to its initial value. Within the ED the charge imbalance $z_C(t)$ is nearly identical for all three considered values of $\alpha=0, 0.25, 1$, while the QFI information clearly distinguishes the infinite range $\alpha=0$ regime from the other two. In all these three cases the $H_{II}$-fTWA gives more accurate results than the $H_{I}$-fTWA. As expected the $H_{II}$-fTWA becomes asymptotically exact as the exponent $\alpha$ approaches zero; in particular for $\alpha=0.25$ the charge imbalance $z_C(t)$ is nearly exactly reproduced by the $H_{II}$-fTWA. These observations are also consistent with the analytical considerations presented in Appendix B. Differences between the two fTWA simulations and the ED are even more pronounced for the QFI (Fig. \ref{fig: fTWA and ED} d-f). In particular, the $H_{I}$-fTWA is not able to predict the long-time behavior of the QFI for all values of $\alpha$ while the $H_{II}$-fTWA, yields significantly more accurate results. 

At the end of this section it is worth to add that the $H_{II}$-fTWA also slightly improves predictions for the long time dynamics of the imbalance in the Hubbard model with short-range (on-site)  interactions (i.e. $U_{ij}=V_{ij\sigma}=0$ for $i\neq j$ and $U_{ii}=V_{ii\sigma}=U_0\neq 0$). We checked this for the charge imbalance function, see Fig.~\ref{fig: CDW - short-range}. The improvement is observed for the higher disorder strengths but it is not so pronounced as in the long-range case. Perhaps the lack of significant improvement of fTWA in the short range model is expected as  both fTWA representations become exact in the noninteracting limit $U_0=0$ and none of them is favored over the other when interactions become large. On the contrary for the long-range model $H_{II}$-fTWA is significantly favored over $H_{I}$-fTWA by the proximity to the infinite-range model ($\alpha=0$), where $H_{II}$-fTWA is exact, while $H_{I}$-fTWA is not (see Appendix B for details).

\begin{figure*}[t]
\includegraphics[scale=0.55]{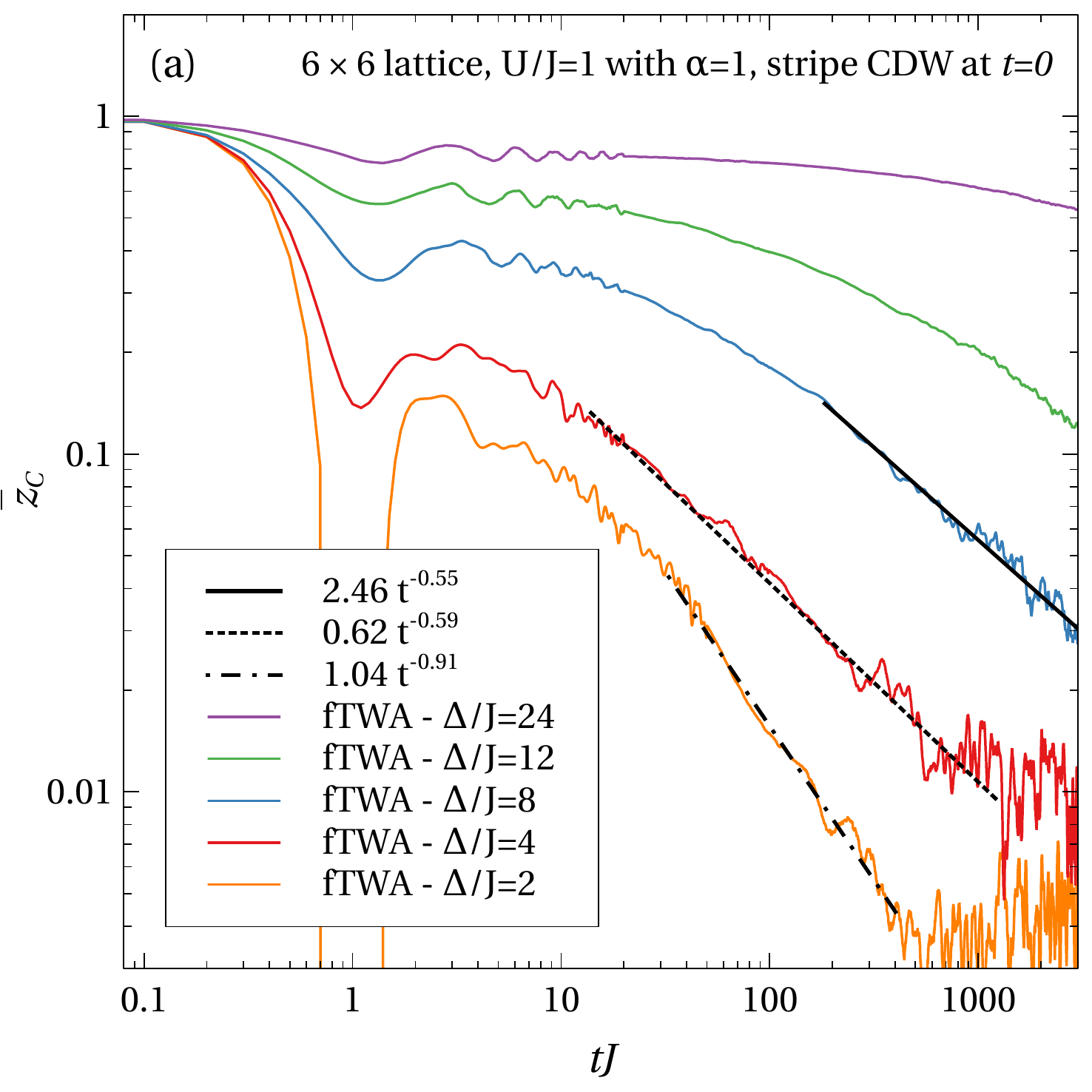}\includegraphics[scale=0.55]{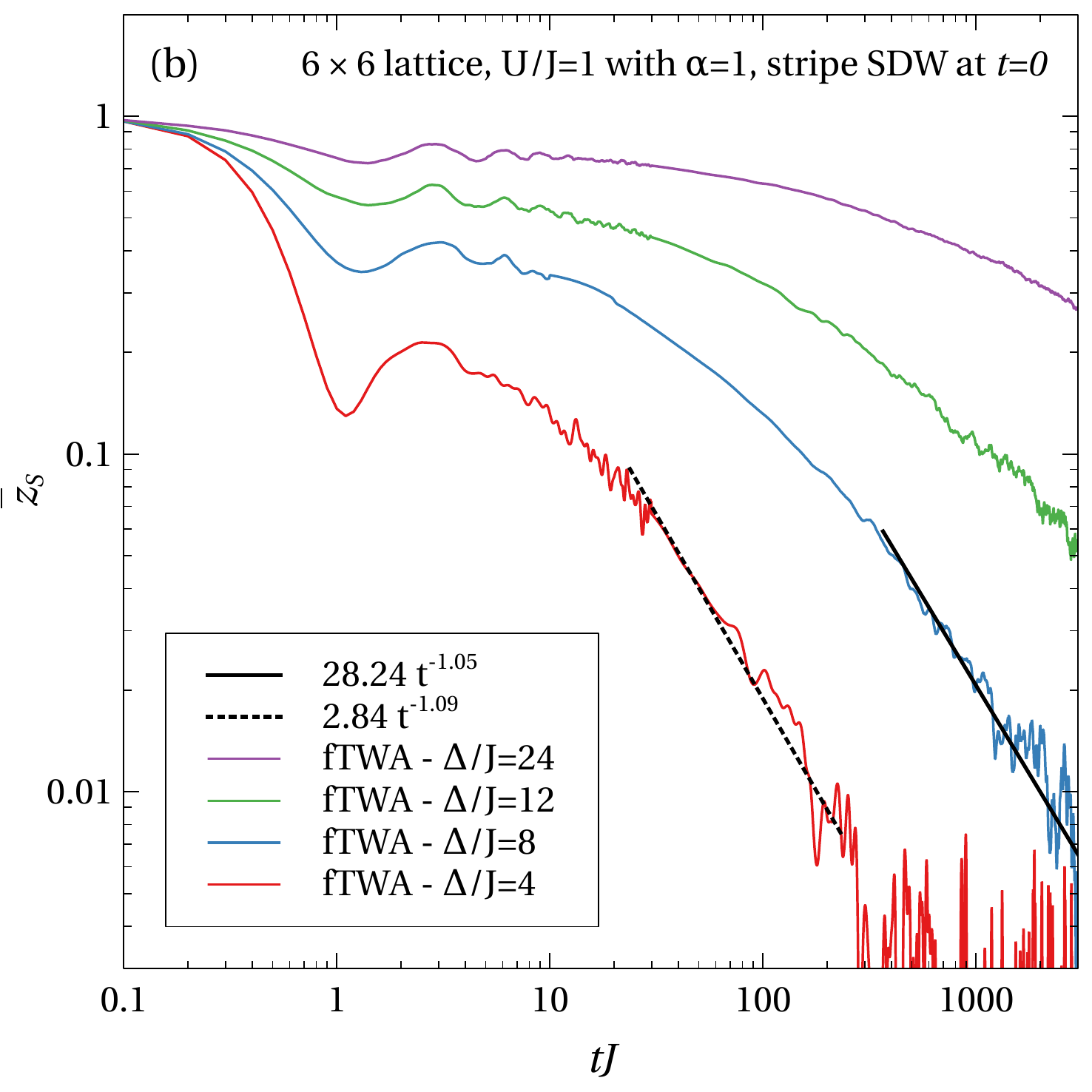}

\includegraphics[scale=0.55]{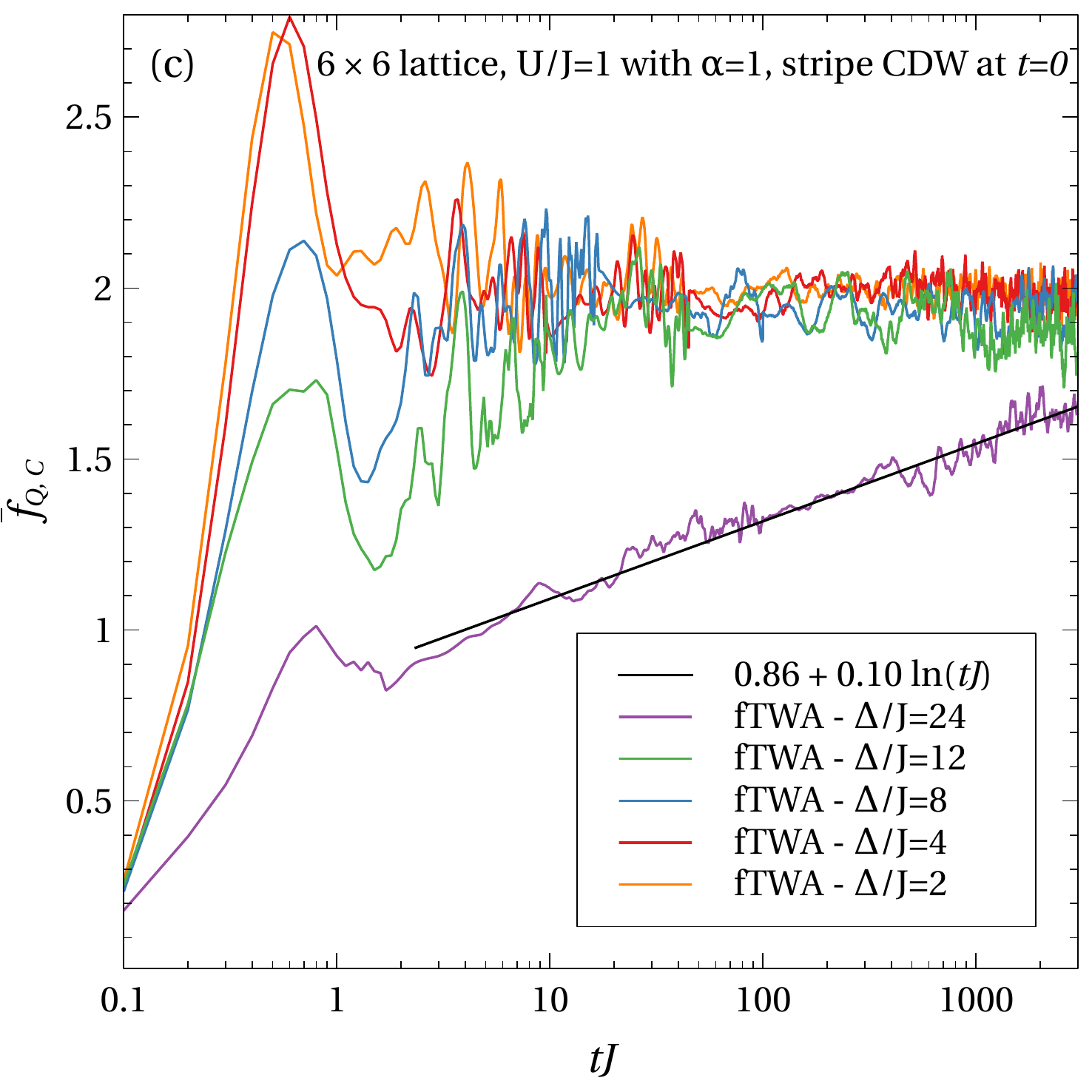}\includegraphics[scale=0.55]{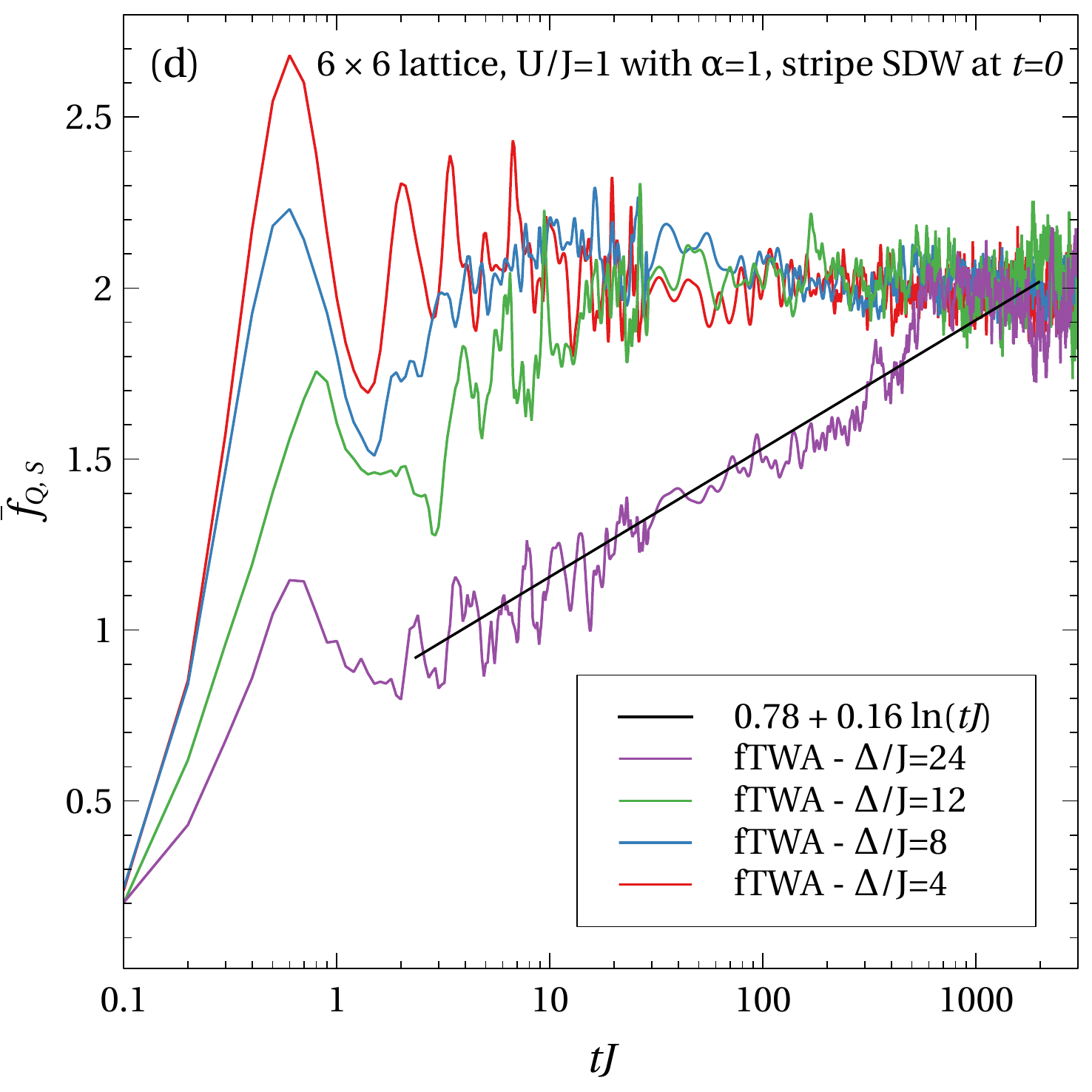}

\caption{(Color online) Time dependence of the disorder averaged charge/spin imbalance $\bar z_C(t)/\bar z_S(t)$ (plots a/b) and the corresponding QFI (plots c/d) for a $6\times6$ square lattice with the same on-site disorder for both spin components. Left plots (a and c) represent dynamics starting from the stripe CDW initial state and the right plots (b and d) represent dynamics starting from the stripe SDW initial state. Different colors represent different disorder strengths:  $\Delta/J=2,4,8,12,24$ (from bottom to top (a, b) and from top to bottom (c, d)). The straight black lines show the best algebraic (a, b) and the logarithmic (c, d) fits to the long time behavior of the corresponding imbalances and the QFI, respectively, and serve as a guide to an eye. All the shown results were averaged over at least $20$ different disorder realizations. The remaining parameters of the Hamiltonian are  $U/J=1$, $\alpha=1$.
 \label{fig: 2D - Z and QFI}}
 \end{figure*}

\begin{figure*}[t]
\includegraphics[scale=0.55]{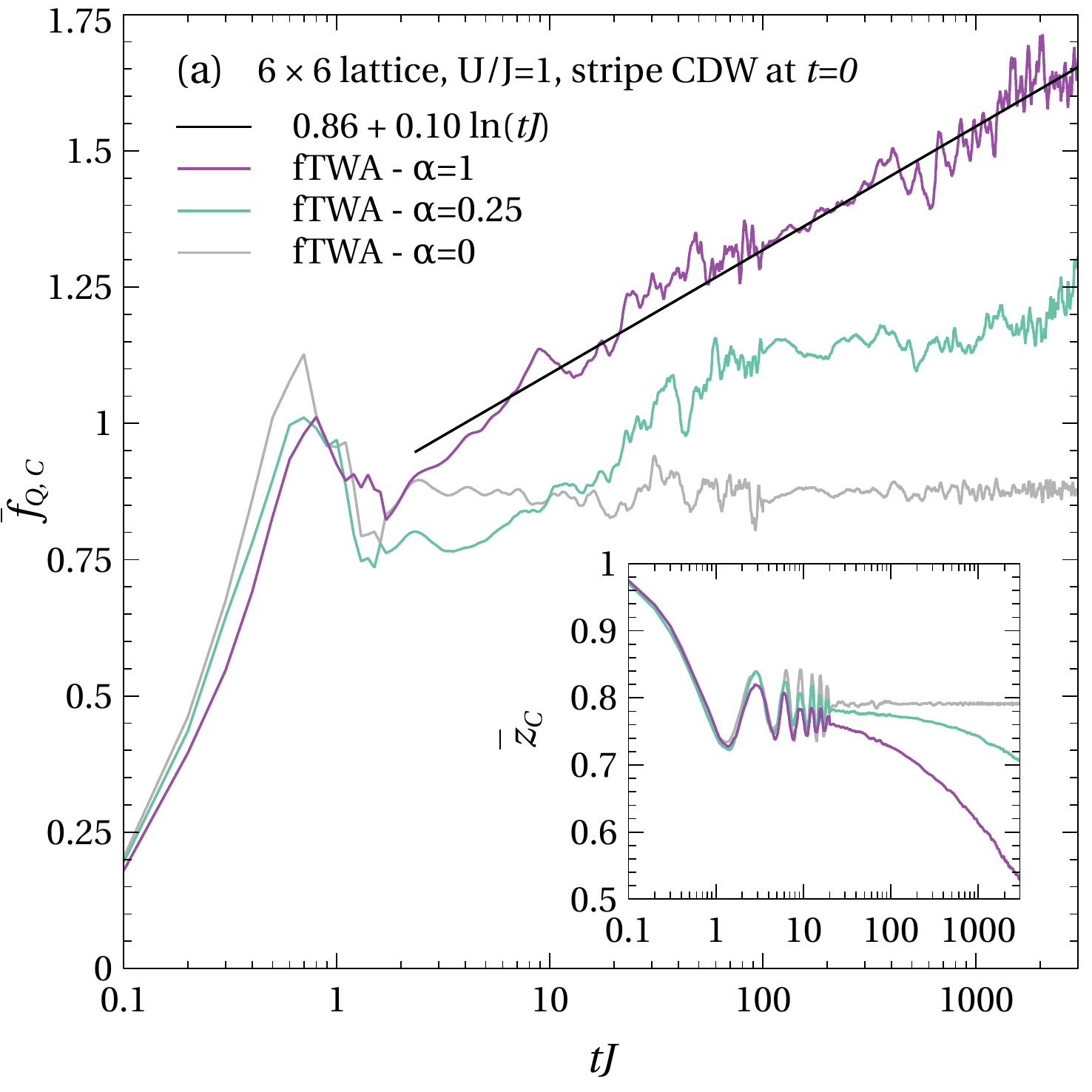}\includegraphics[scale=0.55]{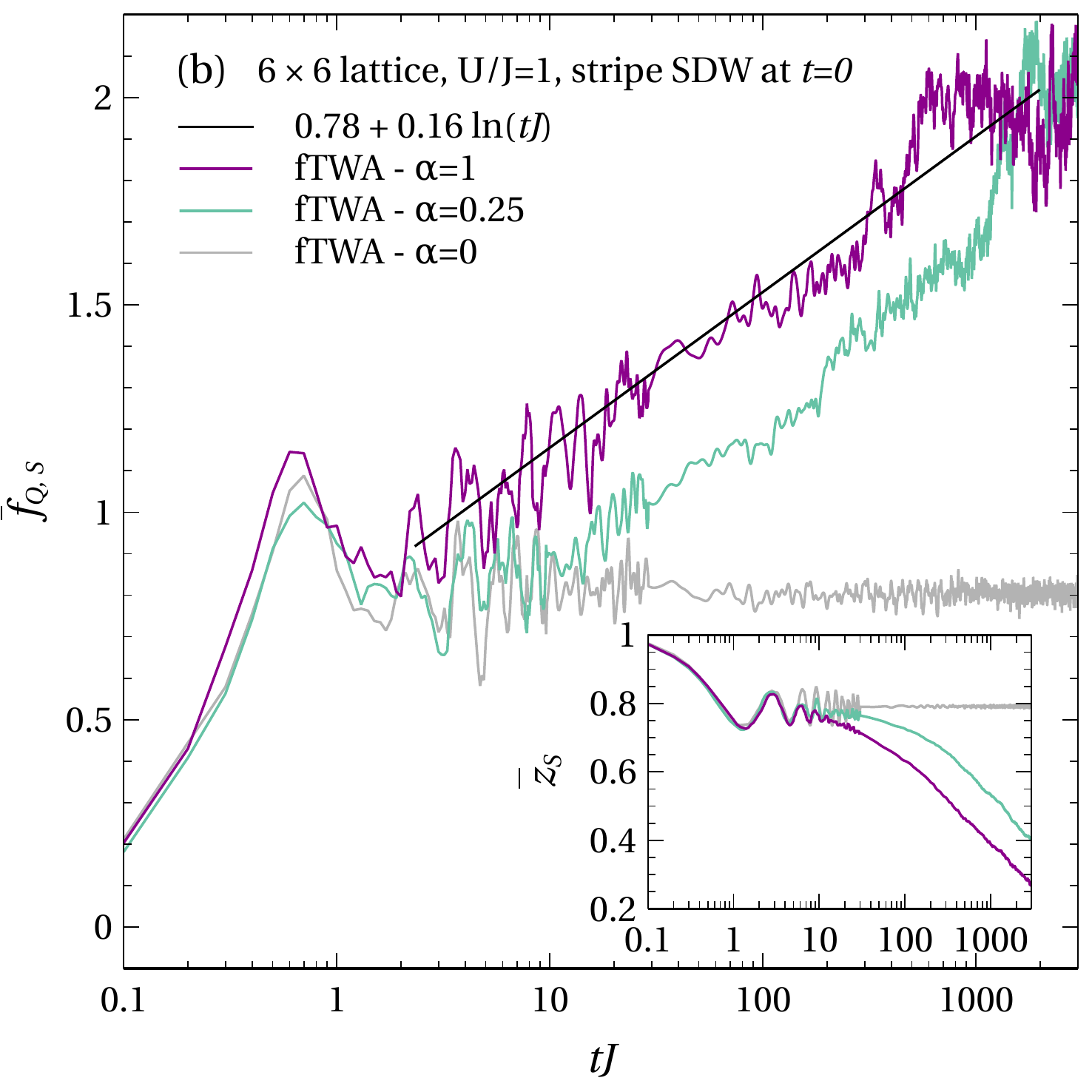}

\caption{(Color online) Time dependence of the QFI density for $6\times6$ lattice and the different values of $\alpha=0, 0.25, 1$ (from bottom to top). The disorder and interaction strength are $\Delta/J=24$, $U/J=1$. The other parameters are the same as in Fig.~\ref{fig: 2D - Z and QFI}. Left (right) plots represent the systems prepared in the stripe CDW (SDW) initial states. The black lines are the fits to the logarithmic time-dependence. The insets show time decay of the charge (spin) imbalance for $\alpha=0, 0.25, 1$ (from top to bottom).
\label{fig: 2D - QFI on alpha}}
\end{figure*}

\begin{figure}[th]
\includegraphics[scale=0.55]{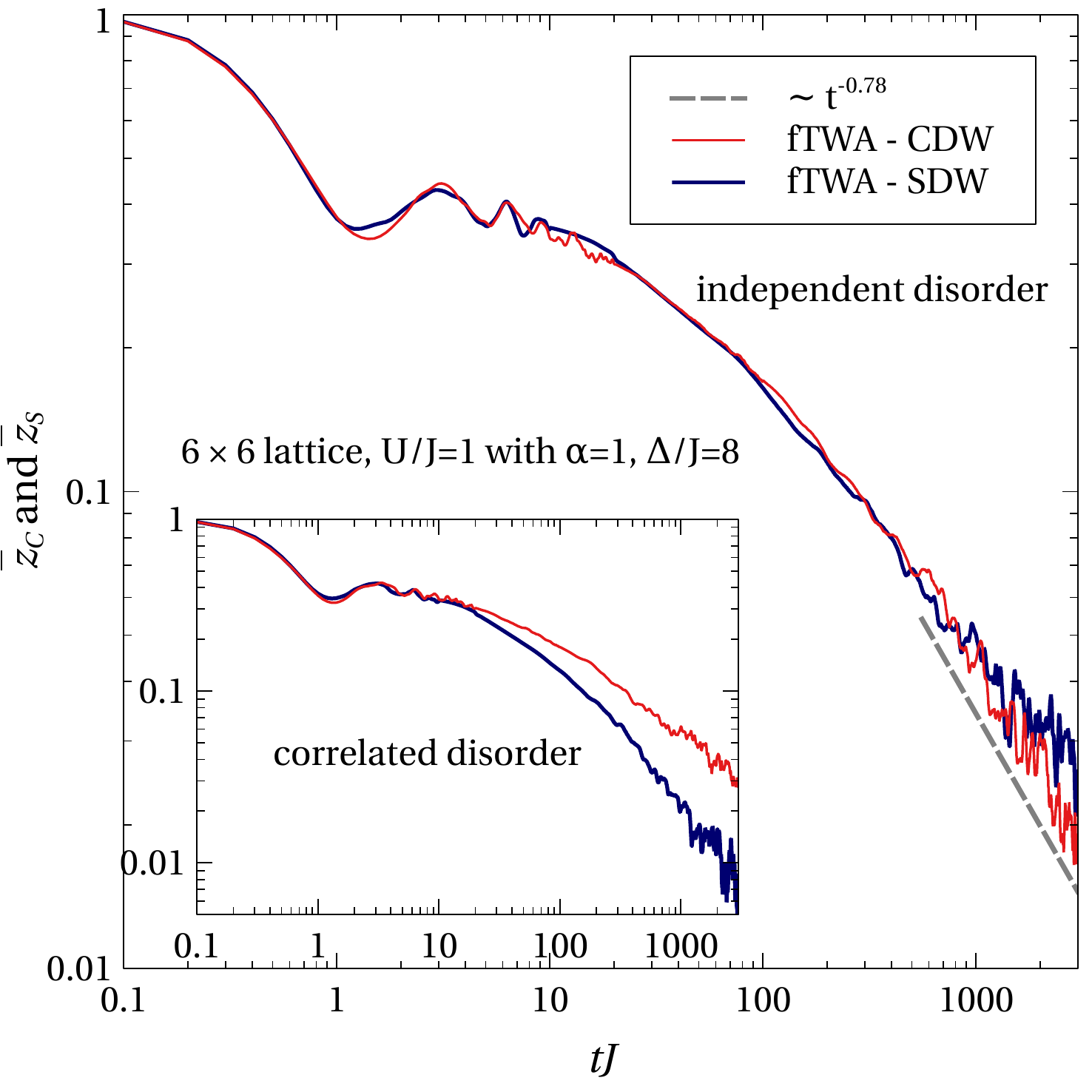}

\caption{(Color online) Decay of charge and spin imbalance for the uncorrelated disorder potential for each spin component.
Red (light gray) and blue (dark gray) line represent the system initially prepared in the stripe CDW and SDW state, respectively. The gray dashed line with
algebraic time dependence is a guide for an eye for the best fit of the long time imbalance to a power law. The disorder strength is $\Delta/J=8$ and the remaining parameters are the same as in Fig.~\ref{fig: 2D - Z and QFI}.
The inset shows the same results but for a spin-independent disorder potential~\label{fig: 2D - Z with uncorrelated disorder}}
\end{figure}

\section{Charge and spin dynamics in two dimensions\label{sec:Charge-and-spin-in-2D}}

We now proceed to analyzing the two-dimensional Hubbard Hamiltonian, where the ED is limited to very small system sizes such that any extrapolation to the thermodynamic limit is nearly impossible. In what follows we present 
the results of numerical simulations for a $6\times6$ square lattice using the  $H_{II}$-fTWA. Note that while these system sizes are far beyond reach of the ED, they are still relatively small and computationally demanding even within the fTWA approach. The reason is that the fTWA implementation requires solving a system of coupled non-linear differential equations with the number of degrees of freedom scaling as the square of the number of sites in the system (i.e. with the fourth power of the linear system size). It is highly plausible that one can go to larger system sizes by introducing further approximations into solving these nonlinear equations such as an effective media approximation beyond certain distance, but the corresponding analysis lies beyond the scope of our work. As we will see even for such system sizes we can effectively suppress finite size effects and make statements about the thermodynamic limit. We focus on the same observables as in the previous section, namely, on the charge and spin imbalances $z_{C/S}(t)$ and the corresponding QFI $f_{C/S}(t)$.

\begin{figure}[th]
\includegraphics[scale=0.55]{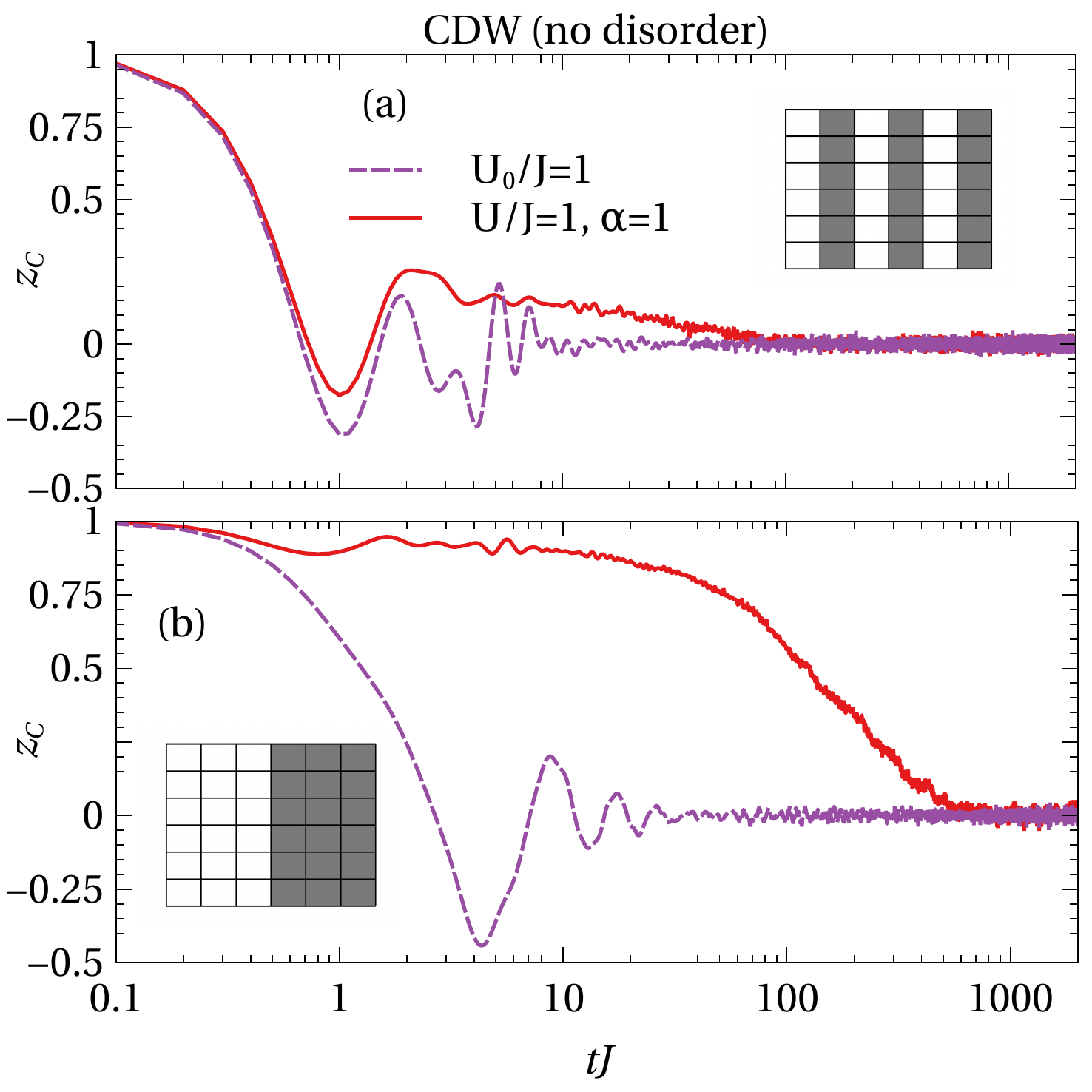}

\includegraphics[scale=0.55]{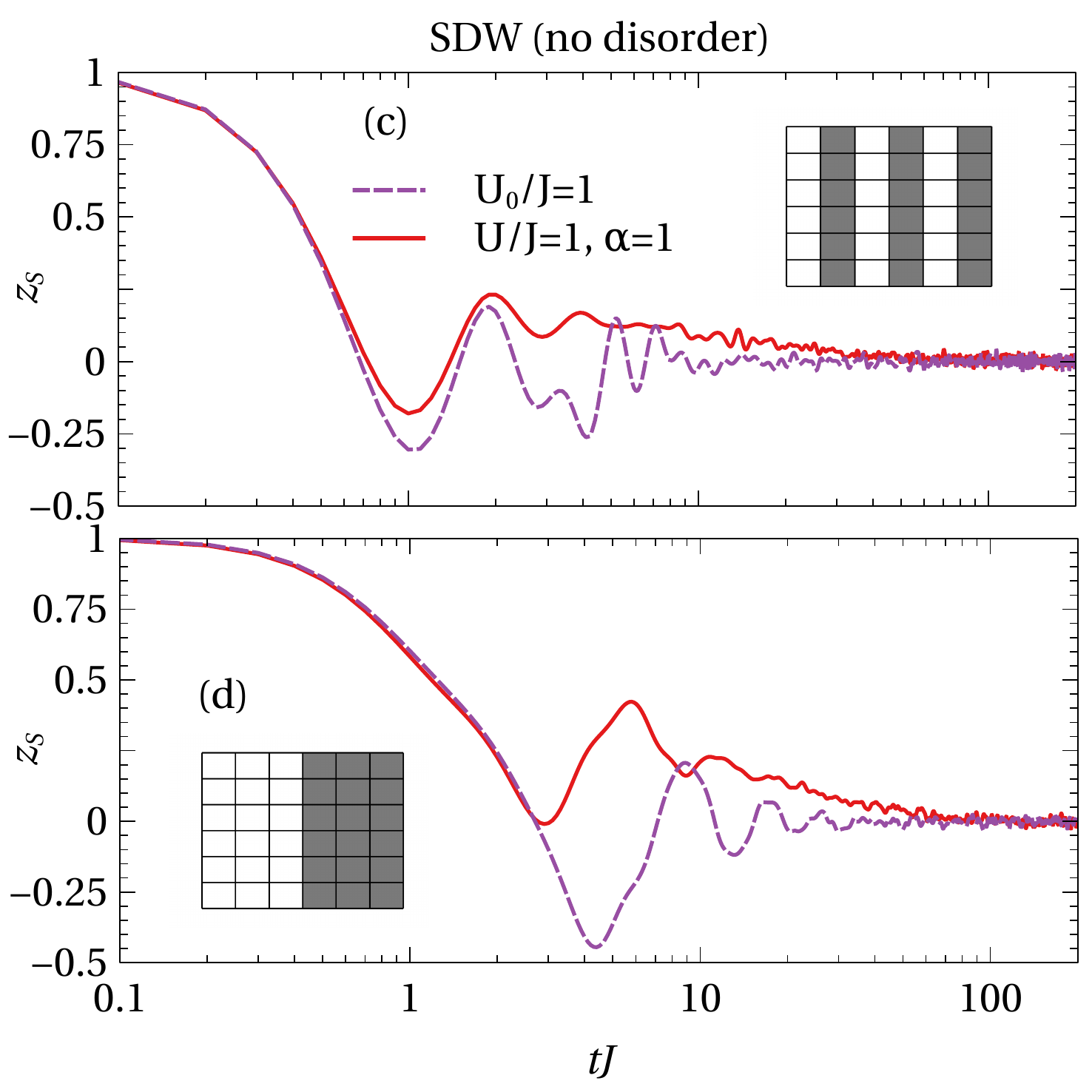}

\caption{Imbalance dynamics for short ($U_{0}/J=1)$ and long-range ($U/J=1$,
$\alpha=1$) interactions with different initial conditions. Stripes
in initial conditions are width $1$ in Fig. a and c, and $3$ in
Fig. b and d (see insets). Simulations are performed for a non-disordered
$6\times6$ lattice. Stripes contain doublons (spins up) or empty
(spins down) sites when system start from the CDW (SDW) initial condition.
Doublons or spins up states in initial conditions are denoted as a
shadow regions in the insets. 
\label{fig: no disorder 6x6}}
\end{figure}

We consider the initial striped CDW or SDW configurations, where the stripes are oriented along the $y$-axis and have a fixed period two along the $x$-axis (see the insets in panels a) and c) in Fig.~\ref{fig: no disorder 6x6}). These initial states are obtained from those used earlier in one dimensional systems (c.f. Eqs.~\eqref{eq:psi_cdw} and \eqref{eq:psi_sdw}) by adding more rows of lattice sites. We define the charge and spin imbalance operators as 
\be
\hat{Z}_{C}(t)=\sum_{i_{x}i_{y}}(-1)^{i_{x}}\hat{n}_{i_{x}i_{y}}
\ee
and 
\be
\hat{Z}_{S}(t)=\sum_{i_{x}i_{y}}(-1)^{i_{x}}\hat{m}_{i_{x}i_{y}}\,\,,
\ee
where $i_{x},\, i_{y}$ are $x$ and $y$ coordinates of the lattice site $i=(i_{x},\, i_{y})$ and $\hat{n}_{i}=\hat{n}_{i\uparrow}+\hat{n}_{i\downarrow}$,
$\hat{m}_{i}=\hat{n}_{i\uparrow}-\hat{n}_{i\downarrow}$. As before we use the notations $\bar z_C(t)$ and $\bar z_S(t)$ for the disorder averaged expectation values of these operators normalized by the total number of sites. Similarly we define the corresponding charge and spin QFI according to Eq. \ref{eq: QFI charges - definition}. All the expectation values are calculated either with the initial CDW or the initial SDW configurations. At half filling these configurations correspond to $z_{C/S}(t=0)=1$. 

In Fig.~\ref{fig: 2D - Z and QFI} we plot the results of simulations of the long-time dynamics of the imbalances and the QFI. These plots correspond to the interaction strength $U/J=1$ and the exponent  $\alpha=1$. Similarly to the 1D results discussed in Sec. \ref{sec: Benchmark-of-fTWA}, we observe the decay of charge and spin imbalances for any disorder strength. As expected at higher disorder the decay of the charge/spin imbalances is more suppressed. Interestingly, essentially at all values of the disorder potential charge transport exhibits subdiffusive behavior:
$\bar{z}_{C}\backsim t^{-\beta}$, $0<\beta<1$, while the spin dynamics is nearly always diffusive: $\bar{z}_{S}\approx t^{-1}$ with stronger disorder resulting only in longer approach to the asymptotic diffusive regime. Thus our results clearly demonstrate stronger and qualitatively different transport suppression in the charge channel. Qualitatively the situation is similar to that in one dimension discussed in the previous section (c.f. Fig.~\ref{fig: MSE}) and in the literature~\citep{PhysRevB.94.241104, PhysRevB.97.064204, PhysRevLett.120.246602, PhysRevB.98.115106, PhysRevB.98.014203,1910.11889, PhysRevB.100.125132, PhysRevB.99.121110, PhysRevB.99.115111} but the difference between the two is more pronounced. Our findings are also consistent with findings of Ref.~\cite{BarLev2016}, where sub-diffusive dynamics was observed in a short-range Hubbard model at the infinite temperature limit using self-consistent perturbation theory.

As in the one-dimensional case the QFI serves as a good indicator of information spreading due to interactions even when the charge and spin degrees of freedom are nearly localized clearly distinguishing the interacting system from the Anderson insulator \citep{Smith2016, PhysRevB.99.054204, PhysRevB.99.241114}. In Figs. \ref{fig: 2D - Z and QFI} (c) and (d) we show the charge and spin QFI density for different disorder strengths $\Delta/J$ and for $U/J=1$, $\alpha=1$. In both the charge and the spin sectors we observe a very fast (at $tJ \approx 1$) saturation of the of the QFI at low disorder ($\Delta/J\leqslant4$). With increasing disorder dynamics of the QFI slows down approaching the logarithmic in time growth at strong disorder (see e.g. QFI data for $\Delta/J = 24$ in Figs. \ref{fig: 2D - Z and QFI} (c) and (d)). At the same value of disorder the spin QFI grows at a faster rate than the charge QFI (c.f. black lines in Figs. \ref{fig: 2D - Z and QFI} c and d) and reaches the saturation value earlier at $tJ\approx 600$. From these simulations we can conclude that the information propagation is faster in the spin channel consistent with the faster imbalance decay there.

Next we analyze the QFI propagation through the system varying the range of interactions. In Fig. \ref{fig: 2D - QFI on alpha} we show the corresponding time dependences of the QFI for $\alpha=0,\,0.25$ and $1$. Interestingly, even for $\alpha=0.25$, i.e. for interactions which decay in space very slowly, we still observe very pronounced logarithmic growth of the QFI both for charges and for spins. Interestingly the anisotropy of the decay times between charge and spin sectors gets larger at smaller values of $\alpha$. These results suggest that as the interaction range increases the charge sector is localized especially strongly. We note that the fTWA is expected to be nearly exact for $\alpha=0.25$ (see, Sec. \ref{sec: Benchmark-of-fTWA}).  As anticipated, after transient behavior the QFI growth rate for spins and charges decreases with lowering $\alpha$. This happens because the system is quickly approaching a non-interacting limit ($\alpha=0$) for which its ability to store new information is lower, i.e. the system becomes less complex. Exactly at $\alpha=0$ QFI saturates immediately after the transient growth. This dynamics of QFI is aligned with the dynamics of charge and spin imbalances shown in the insets in Fig. 6.  For the non-interacting case ($\alpha=0$) imbalances do not decay because the system is in the regime of Anderson localization. As $\alpha$ increases the imbalances start to decay because of the additional charge/spin transport mediated by interactions (c.f. also Fig. 5).  One can argue that generally the fTWA should be more accurate in 2D than 1D because the system is closer to the mean field regime. So even for $\alpha=1$ we anticipate that the fTWA gives reliable results.

The qualitative and quantitative differences between dynamics in spin and charge sectors originates because the disorder potential in the Hamiltonian (\ref{eq: Hamiltonian - good}) directly couples only to charge degrees of freedom (see also discussion in Sec \ref{sec: Benchmark-of-fTWA}). In other words, there are perfect correlations between the disorder potential acting on both spin components manifested in the SU (2) symmetry of the model in the spin sector. However, by adding disorder also in the spin channel, e.g. by considering independent disorder potential for ``up'' and  ``down'' spins, dynamics of local charges and spins become equivalent as is demonstrated in Fig. \ref{fig: 2D - Z with uncorrelated disorder}  (for 1D system see also Ref. \cite{PhysRevB.94.241104}). For these simulations we used the independent disorder potential of the form
\be
 \sum_{i\sigma}\Delta_{i}\hat{n}_{i\sigma}\rightarrow\sum_{i\sigma}\Delta_{i\sigma}\hat{n}_{i\sigma}\,,
\ee
where $\Delta_{i\uparrow}$ and $\Delta_{i\downarrow}$ are independently distributed. This potential obviously breaks the SU(2) spin symmetry. The data presented in Fig. \ref{fig: 2D - Z with uncorrelated disorder}, was generated using the same initial conditions as before starting from either CDW or SDW stripe configurations and following the imbalance functions 
$\bar{z}_{C}(t)$ and $\bar{z}_{S}(t)$.  Apart for small differences at short times we see that the imbalance decay in both sectors is nearly identical, which is contrasted to the slower decay of the charge imbalance in the $SU(2)$ case discussed above and shown for completeness in the inset of Fig.~\ref{fig: 2D - Z with uncorrelated disorder}. This observation confirms that the difference of the dynamics in the charge and spin sectors in the system with the spin independent (correlated) disorder is not due to the difference in initial conditions but rather due to different thermalization mechanisms. One can also notice, that with the uncorrelated disorder, decay of the charge and the spin imbalances is still subdiffusive with the exponent, however, somewhat larger than for charge decay in the $SU(2)$ regime (c.f. the  fitting curves in Figs.~ \ref{fig: 2D - Z and QFI} and \ref{fig: 2D - Z with uncorrelated disorder} for $\Delta/J=8$).

Let us note that in all simulations shown in this section we used noise filtering to suppress the sampling noise, which is rather significant at long times. This noise goes down with the number of realizations of the initial conditions but the convergence of the results are rather slow. We checked that the filtering we use does not introduce any systematic error and that the filtered fTWA accurately describes all non-spurious short time oscillations of the observables. The effect of filtering on the charge imbalance together with the analysis of the finite size effects is shown in Appendix C.

\begin{figure*}[t]
\includegraphics[scale=0.4]{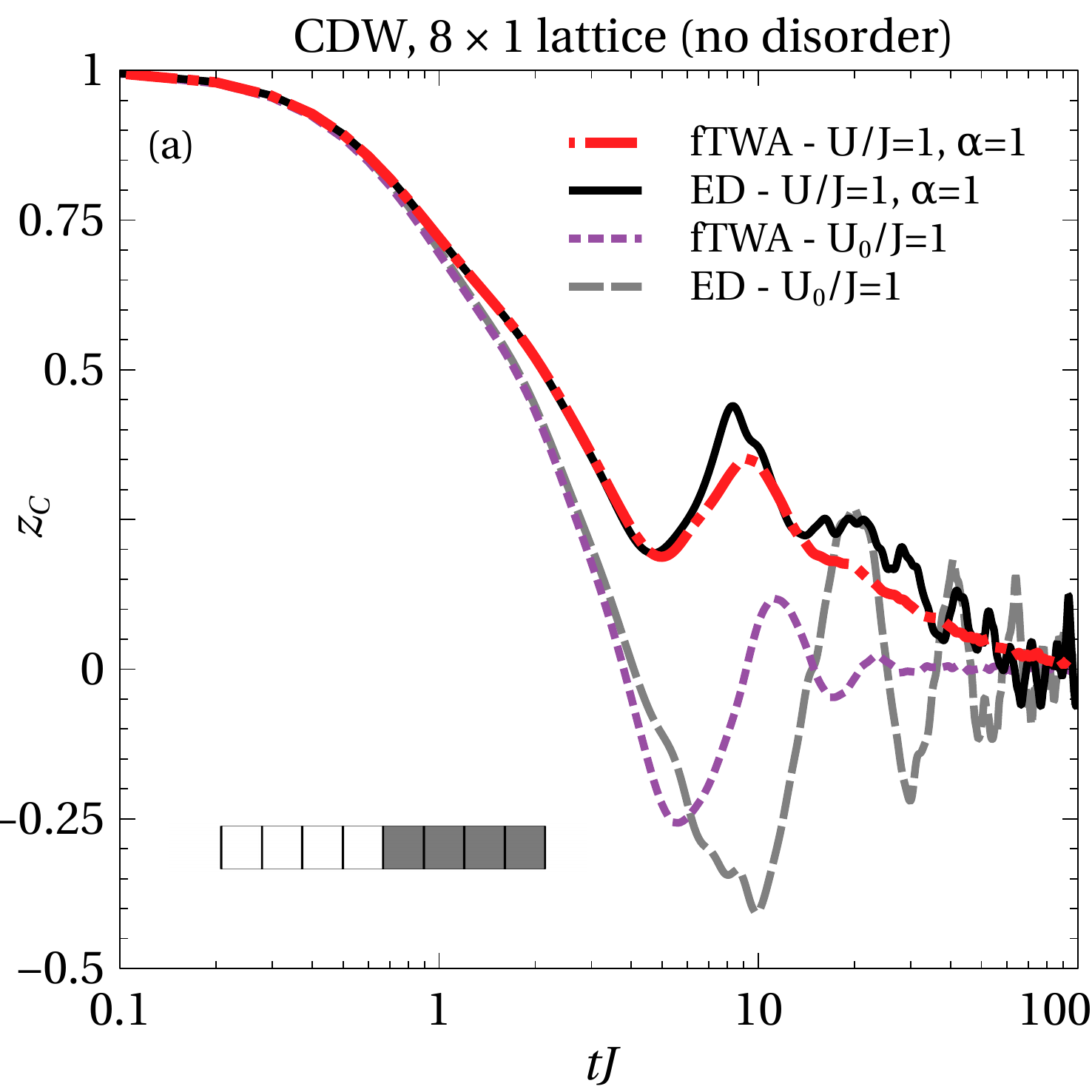}\includegraphics[scale=0.4]{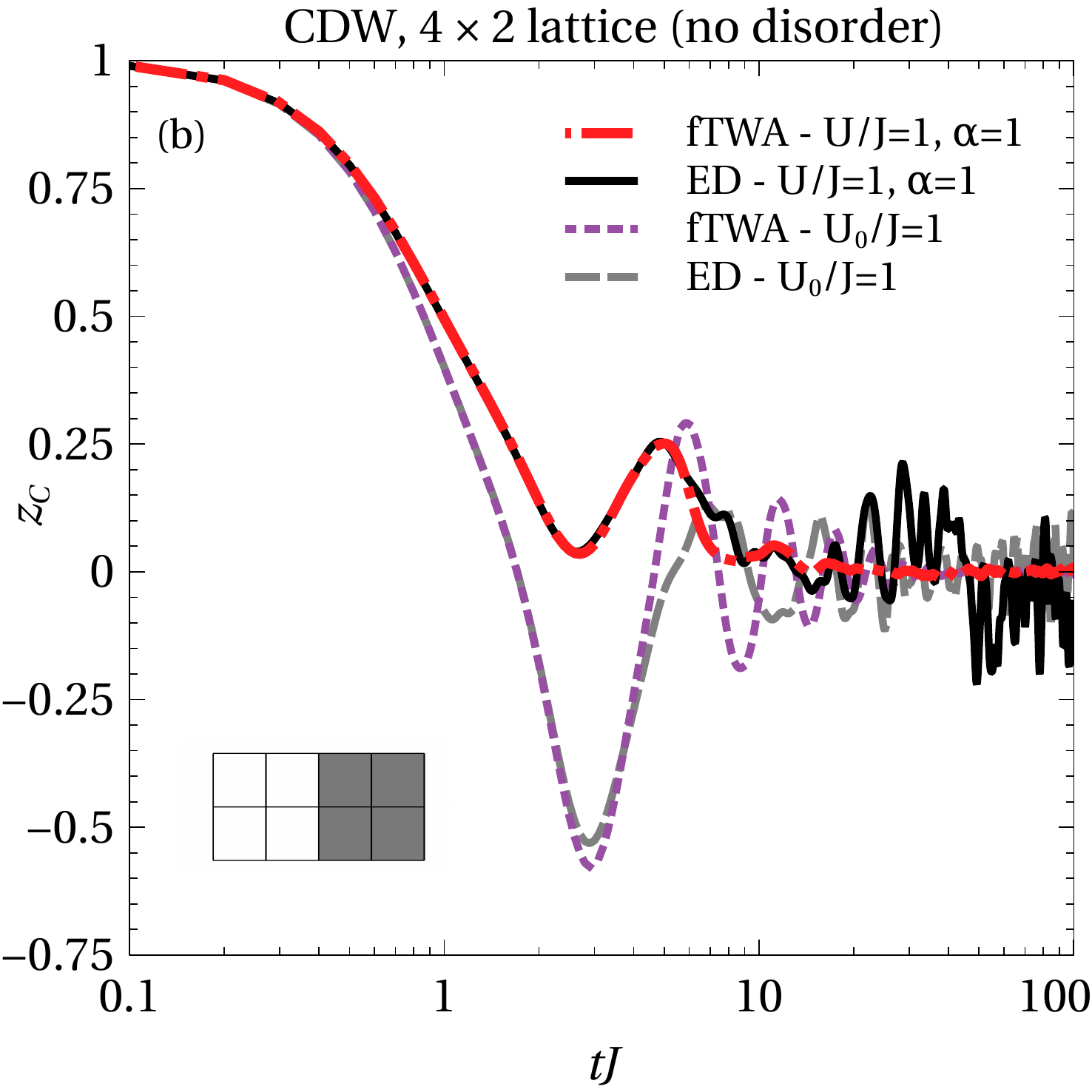}\includegraphics[scale=0.4]{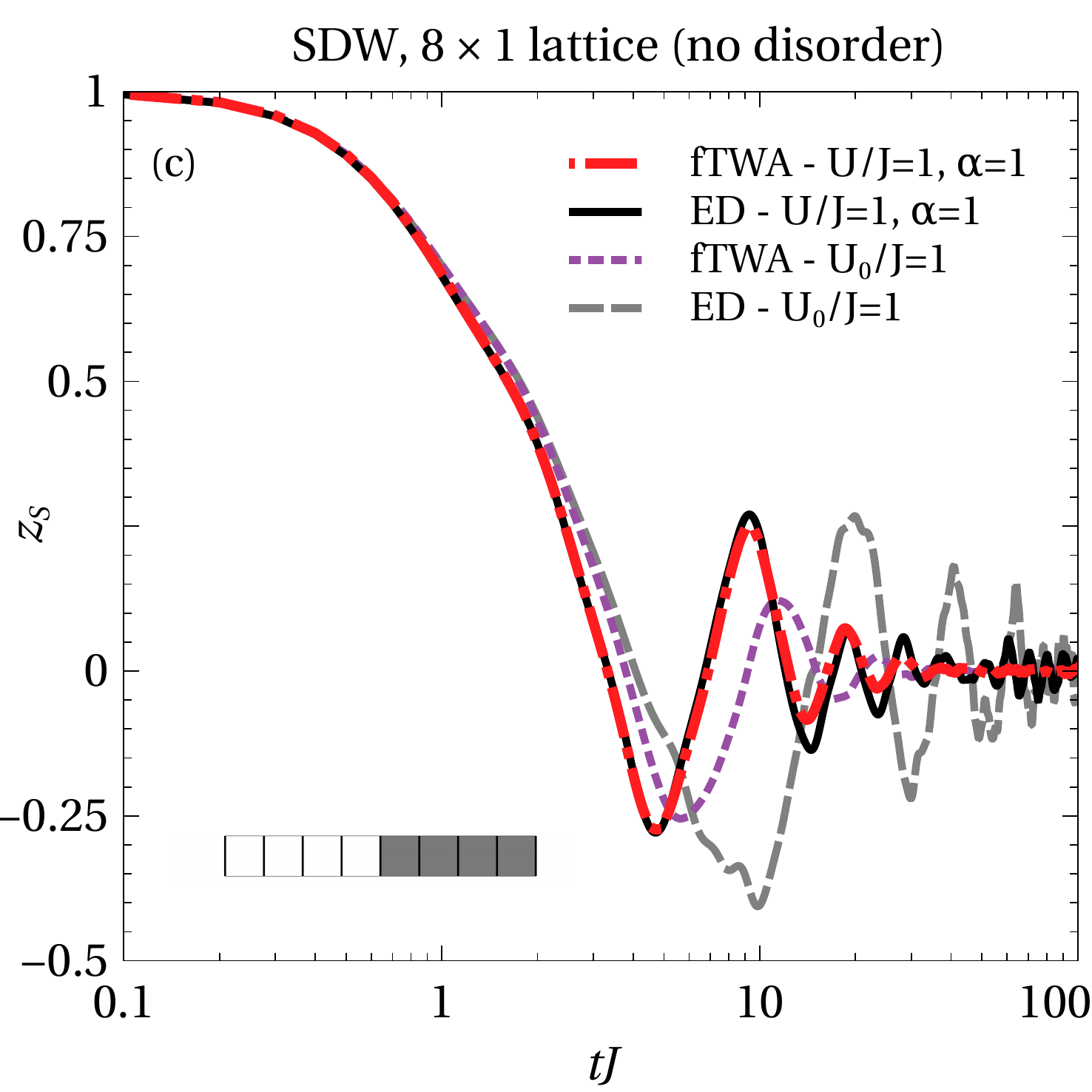}

\caption{Imbalance dynamics for small systems with short-range ($U_{0}/J=1)$ and long-range ($U/J=1$,
$\alpha=1$) interactions with domain wall initial conditions (see text for details). The simulations are performed for non-disordered $8\times1$ (a and c) and $4\times2$ (b) lattices. The a) and b) panels correspond to the initial CDW state and the c) panel does to the initial SDW state. 
\label{fig: no disorder ED and fTWA comparison}}
\end{figure*}

\section{Memory effects for different initial states}

Up to now we analyzed a particular stripe CDW/SDW initial states and saw how presence of disorder slows down dynamics in the system. In the absence of disorder the system is expected to quickly thermalize as illustrated  in Fig.~\ref{fig: no disorder 6x6} (a), (c). Interestingly in the long-range model the thermalization time strongly depends on the initial state. This is easily seen by changing the initial CDW/SDW stripe width from the unit length to the half of the system size and by analyzing the corresponding imbalance functions, which are adjusted according to the width of the initial configuration. For clarity, we represent these initial states in the insets of  Fig.~\ref{fig: no disorder 6x6}, i.e. doubly occupied or empty (up or down) sites are depicted as shadow or empty boxes for CDW (SDW) initial states, respectively. As we can see, for newly introduced initial conditions (panel (b) and (d)), the thermalization time for charge degrees of freedom becomes significantly longer than for the stripe initial configuration with the unit width (panels (a) and (c)). In particular, we observe that for the single domain wall initial state (i.e. for CDW/SDW stripe with the width of half of the system size)  the time scale at which the charge imbalance $z_{C}(t)$ decays to zero is around $tJ\approx700$ (panel (b)). A qualitatively similar slowing down in a fermionic system with long-range hopping was numerically observed in Ref.~\cite{Davidson2017}. We contrast the above results with those for the short-range interacting model with only on-site interactions between ``up'' and ``down'' species of strength $U_0$ (dashed purple lines), where the difference between the thermalization times for these two initial configurations is much less pronounced. Qualitatively this long-memory effect can be explained by a high energy released by the particles traveling from the filled to the empty part of the system in the presence of long-range interactions. This energy has to be redistributed among the other degrees of freedom resulting in a large kinematic barrier and hence in a smaller decay rate.  As one can see from the panels c) and d) the long memory effect is absent, or at least is much weaker, for the SDW initial state. 

\begin{figure}[tbh]
\includegraphics[scale=0.55]{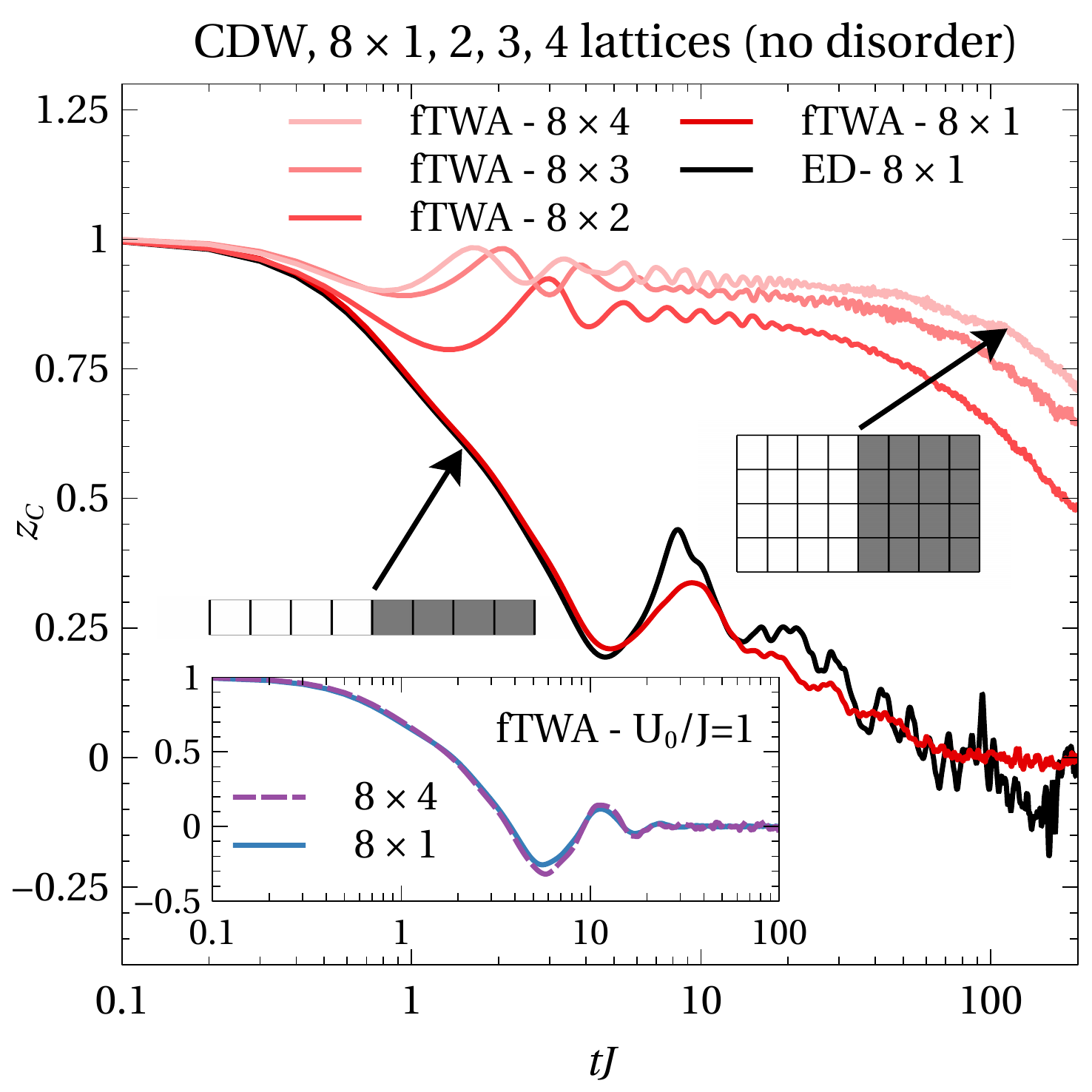}

\caption{The charge imbalance dynamics for non-disordered systems of a fixed length $L_x=8$ and different width $L_y=1,\,2,\,3,\,4$ with long-range interactions ($U/J=1$, $\alpha=1$) and the domain wall CDW initial condition. The inset shows imbalance decay for $8\times1$ and $8\times4$ lattices with on-site interactions strength $U_{0}/J=1$. \label{fig: no disorder fTWA 8 x 1, 2, 3, 4}}
\end{figure}

\begin{figure}[tbh]
\includegraphics[scale=0.55]{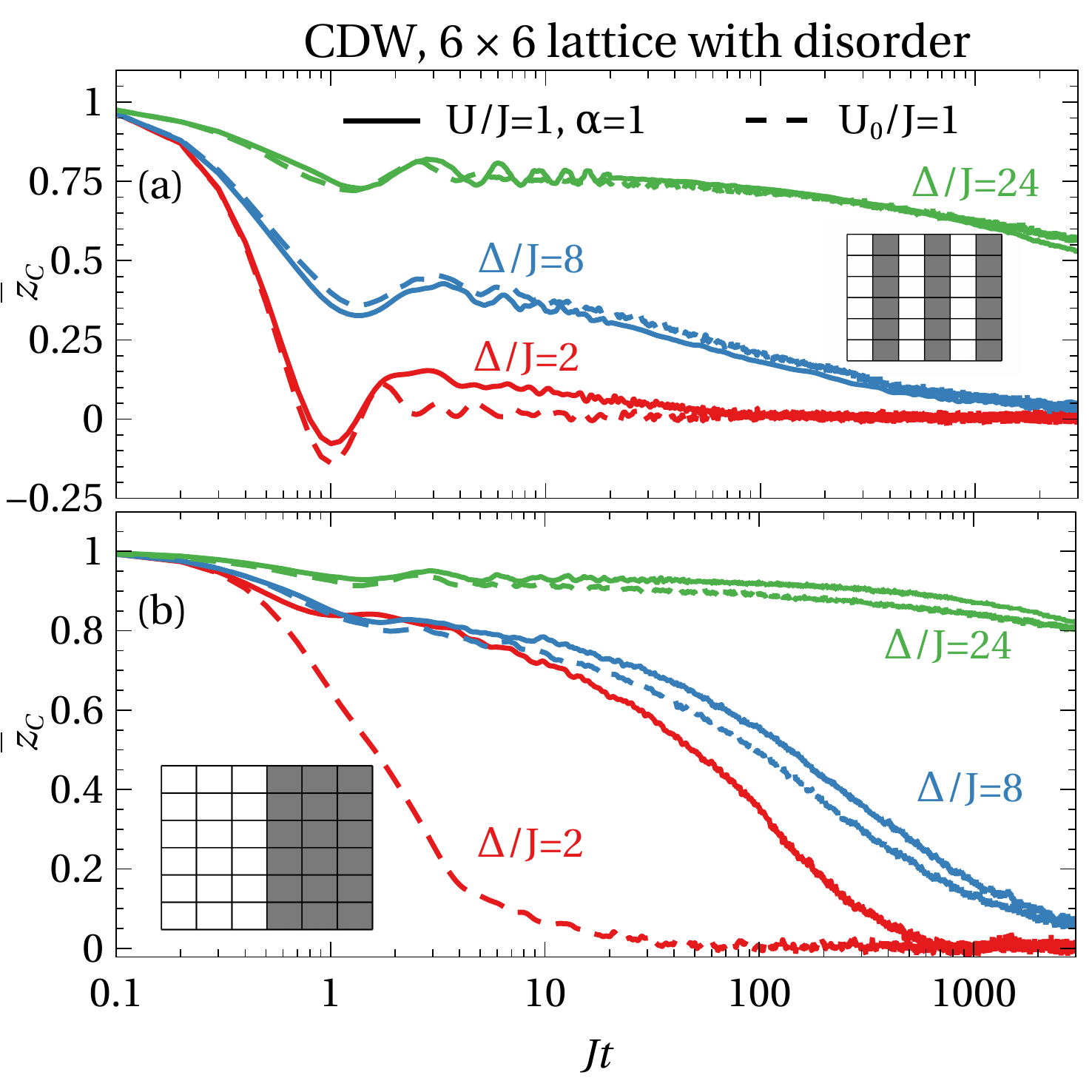}

\caption{Imbalance dynamics for different disorder strengths ($\Delta/J=2,\ 8,\ 24$)
and different interaction profiles: dashed lines - short-range interactions with $U_{0}/J=1$ and solid lines - long-range with $U/J=1$ and $\alpha=1$. Panel a) corresponds to the initial stripe CDW state (width of stripe is 1) and panel b) corresponds to the initial domain wall state (width of CDW stripe is 3). The system size is $6\times 6$.}
\label{disorder and domain wall}
\end{figure}

To confirm that the observed slow-down of thermalization of the charge sector in the long-range model we perform the ED simulations in small systems and contrast them with the fTWA simulations. In Fig. \ref{fig: no disorder ED and fTWA comparison} (a) and (b), we analyze the charge imbalance decay for the domain wall initial state for systems of  sizes $8\times1$ and $4\times2$, respectively. The initial state corresponds to the empty left half of the system and the fully occupied by doublons right half of the system. We see that long-range interactions lead to a slower thermalization rate of the charge imbalance compared to the short-range model, though the effect is not as strong as for larger system sizes analyzed in Fig.~\ref{fig: no disorder 6x6}. At the same time there is a little effect of the initial state on the decay of the SDW state (Fig. \ref{fig: no disorder ED and fTWA comparison}
c) again in agreement with the earlier fTWA results for larger system sizes. From the comparison of the fTWA and the ED predictions in small systems, we see that the agreement is very good especially for the long-range interactions and thus the fTWA simulations lead to reliable predictions. In Fig. \ref{fig: no disorder fTWA 8 x 1, 2, 3, 4}
we analyze how the decay of the charge imbalance depends on the width of the system, which is gradually increased from $L_y=1$ to $L_y=4$. We see that there is a dramatic jump in the relaxation time as the width increases from $1$ to $2$ followed by its more gradual dependence if the width increases further. In the inset we show that there is a very small effect of the width on the relaxation time for the short-range model. 

Interestingly, in the presence of stronger disorder, slowing down of the thermalization by long-range interactions gets smaller and the equilibration time scales for the models with short-range and long-range interactions become comparable. This is illustrated in Fig.~\ref{disorder and domain wall}, where we compare imbalance decay for short- and long-range interactions (dashed and solid lines respectively) for two different initial states (stripe CDW (a) and domain wall (b)) and different disorder strengths. We see that only for a small disorder $\Delta/J=2$ there is a very significant slowing down of the imbalance decay in the long-range model and for the domain wall initial state (two lowest lines in the panel (b)). Our findings suggest that extracting potential many-body localization transition in the system with long-range interactions in a 2D lattice using the experimental protocol proposed in Ref.~\citep{Choi2016} requires extra care with choosing a proper initial state.

\section{Summary}

In this work we developed an efficient semiclassical fTWA representation of dynamics in Hubbard model with long-range interactions. The method is based on the proper phase space representation of the Hubbard Hamiltonian. In particular, we resolved the ambiguity of finding the Weyl symbol of the Hamiltonian coming from the  operator identity $\hat n_\alpha^2=\hat n_\alpha$. We showed that this ambiguity can be eliminated by requiring that the fTWA becomes exact in the limit of the infinite range interactions. We showed that using the corresponding Weyl symbol of the Hamiltonian in the presence of algebraically decaying interactions significantly improves accuracy of the fTWA over a more naive choice of the Hamiltonian's phase space representation. 

Using the developed formalism we applied the fTWA to study quench dynamics in the fermionic Hubbard model with long-range interactions in the presence of disorder. We first benchmarked the method against the ED results in small one-dimensional systems and then applied the fTWA to a two-dimensional model, which is far beyond the reach of ED. In particular, motivated by recent experiments and theoretical works related to the Hubbard
model~\citep{mblschreiber, PhysRevLett.116.140401, PhysRevLett.119.260401, PhysRevX.7.041047, PhysRevLett.122.170403, PhysRevB.94.241104, PhysRevB.99.115111, PhysRevB.90.174204, pandeyMBL2019, PhysRevB.99.241114} we analyzed dynamics of charge and spin imbalance at half filling for different CDW/SDW type initial states and different disorder strengths. We showed that the fTWA can clearly distinguish different thermalization time scales of the charge and spin imbalance when the disorder potential is spin-independent. In particular, even for the weak or moderate disorder potentials we obtained subdiffusive charge transport
characterized by a power law decay of the charge imbalance with a disorder dependent exponent. At the same time the spin transport under the same conditions remained diffusive. This anisotropy between charge and spin transport is consistent with earlier studies in one-dimensional systems~\citep{PhysRevB.94.241104, PhysRevB.97.064204, PhysRevLett.120.246602, PhysRevB.98.115106, PhysRevB.98.014203, 1910.11889,PhysRevB.100.125132, PhysRevB.99.121110, PhysRevB.99.115111, pandeyMBL2019}. We also showed that fTWA can accurately reproduce the QFI and found that it grows logarithmically in time at strong disorder. Moreover we found that its growth rate is smaller for the QFI associated with the charge imbalance. This observation is consistent with slower charge transport and indicates that the information also spreads more slowly in the charge sector. We also investigated the role of different initial conditions and found that for long-range interactions there is an additional and very strong mechanism, which suppresses thermalization of the initial domain wall type CDW state, i.e. the state where all doubly occupied sites are initially clustered together, even at weak disorder. This effect can be important for properly designing experimental protocols, which aim to detect potential localization transition in the systems with long-range interactions.

\section*{Acknowledgments}

We would like to thank Jonathan Wurtz and Markus Schmitt for valuable discussions. A.S.S. acknowledges funding from the Polish Ministry of Science and Higher Education through a ``Mobilno\'{s}\'{c} Plus''
program nr 1651/MOB/V/2017/0. A.P.  were supported by NSF DMR-1813499 and AFOSR FA9550- 16-1-0334. We also thank the facilities of the Boston University Shared Computing Cluster, over which we run all the numerical simulations.

\section*{Appendix A. Fermionic truncated Wigner approximation (fTWA) \label{sec:Appendix-A.-Details}}

We will summarize here the main ideas of the fTWA formalism and its implementation. Additional details can be found in Refs.~\citep{Davidson2017,PhysRevB.99.134301}. The fTWA formalism is a direct generalization of the  standard TWA to fermionic systems where $\rho_{\alpha\beta}=\rho_{\beta\alpha}^\ast$, $\tau_{\alpha\beta}$, $\tau^\ast_{\alpha\beta}$ play the role of the complex phase space variables (for their definition see Sec. \ref{sec: Phase-space-representation of HH}). All observables including the Hamiltonian and the initial density matrix are represented by these phase space variables. However, as mentioned in Sec. \ref{sec: Phase-space-representation of HH}, one can consider Hamiltonian dynamics within $\rho$-representation which restrict phase space parametrization to $\rho_{\alpha\beta}$ variables only. Then, the expectation value of some time-dependent observable in the Heisenberg representation $\hat{\mathcal{O}}(t)$ within the fTWA is evaluated in the following way
\begin{equation}
\langle\hat{\mathcal{O}}(t)\rangle=\int d\boldsymbol{\rho}_{0}W(\boldsymbol{\rho}_{0})\mathcal{O}_{W}(\boldsymbol{\rho}(t))\label{eq: observable evolution in fTWA}
\end{equation}
where $W$ is the Wigner function, $\mathcal{O}_{W}$ is the Weyl symbol of the operator
$\hat{\mathcal{O}}$ and $\boldsymbol{\rho}=\left\{ \rho_{\alpha\beta}:\,\alpha,\,\beta\in\{1,\,...\,,\, N\}\right\} $.
Here the integration is performed over the initial conditions with the Wigner function $W(\boldsymbol{\rho}_{0})$ playing the role of their probability distribution. Following Refs.~\citep{Davidson2017,PhysRevB.99.134301} we approximate the Wigner function with a positive Gaussian distribution, which correctly reproduces both the expectation values of the phase space variables and their fluctuations in the initial state. Such positive representation is always possible for any Slater determinant type initial state and is likely possible for other states. In particular, the CDW/SDW initial states which are analyzed in this work and which are straightforward to realize in cold atoms~\citep{mblschreiber, Choi2016, PhysRevLett.116.140401, PhysRevLett.119.260401, PhysRevX.7.041047} belong to this category. We note that it is the presence of  fluctuations encoded in the Wigner function, which makes the fTWA fundamentally different from mean field and allows for extracting such purely quantum observables as the QFI. In order to find $\boldsymbol{\rho}(t)$ entering Eq.~\eqref{eq: observable evolution in fTWA} one has to solve deterministic classical and generally nonlinear equations of motion:
\be
\frac{\partial\rho_{\alpha\beta}}{\partial t}  =  \{\rho_{\alpha\beta}, H_W\}\equiv \sum_{\mu\nu\gamma\delta}f\left(\alpha,\beta,\mu,\nu,\gamma,\delta\right)\frac{\partial H_{W}}{\partial\rho_{\mu\nu}}\rho_{\gamma,\delta}\:,\label{eq: fTWA general equations for rho}
\ee
satisfying the randomly sampled initial conditions: $\boldsymbol{\rho(t=0)}=\boldsymbol{\rho}_0$. The evolution is governed by $H_W$, which plays the role of the classical Hamiltonian. We discuss the two possible choices for $H_W$ corresponding to the Hamiltonians $\hat{H}_{I}$ (Eq. (\ref{eq: Hamiltonian - operator form})) and $\hat{H}_{II}$ (Eq. (\ref{eq: Hamiltonian - good})) in the Appendix B (c.f. Eqs.~\eqref{eq: Weyl of H_I} and \eqref{eq: Weyl of H_II}). Finally $f$ in the equation of motion define the structure constants, which in turn define the Poisson brackets of the classical Hamiltonian evolution. These structure constants are found from the commutation relations:
\be
\left[\hat{E}_{\beta}^{\alpha},\,\hat{E}_{\nu}^{\mu}\right]=i\sum_{\gamma\delta}f\left(\alpha,\beta,\mu,\nu,\gamma,\delta\right)\hat{E}_{\delta}^{\gamma}
\ee
and are easy to compute~\cite{Davidson2017}. Instead of listing them here in the Appendix B we show the explicit form of Eq. (\ref{eq: fTWA general equations for rho}) for the two choices of the Hamiltonian $\hat{H}_{I}$ and $\hat{H}_{II}$.

.\begin{widetext}

\begin{figure*}
\includegraphics[scale=0.4]{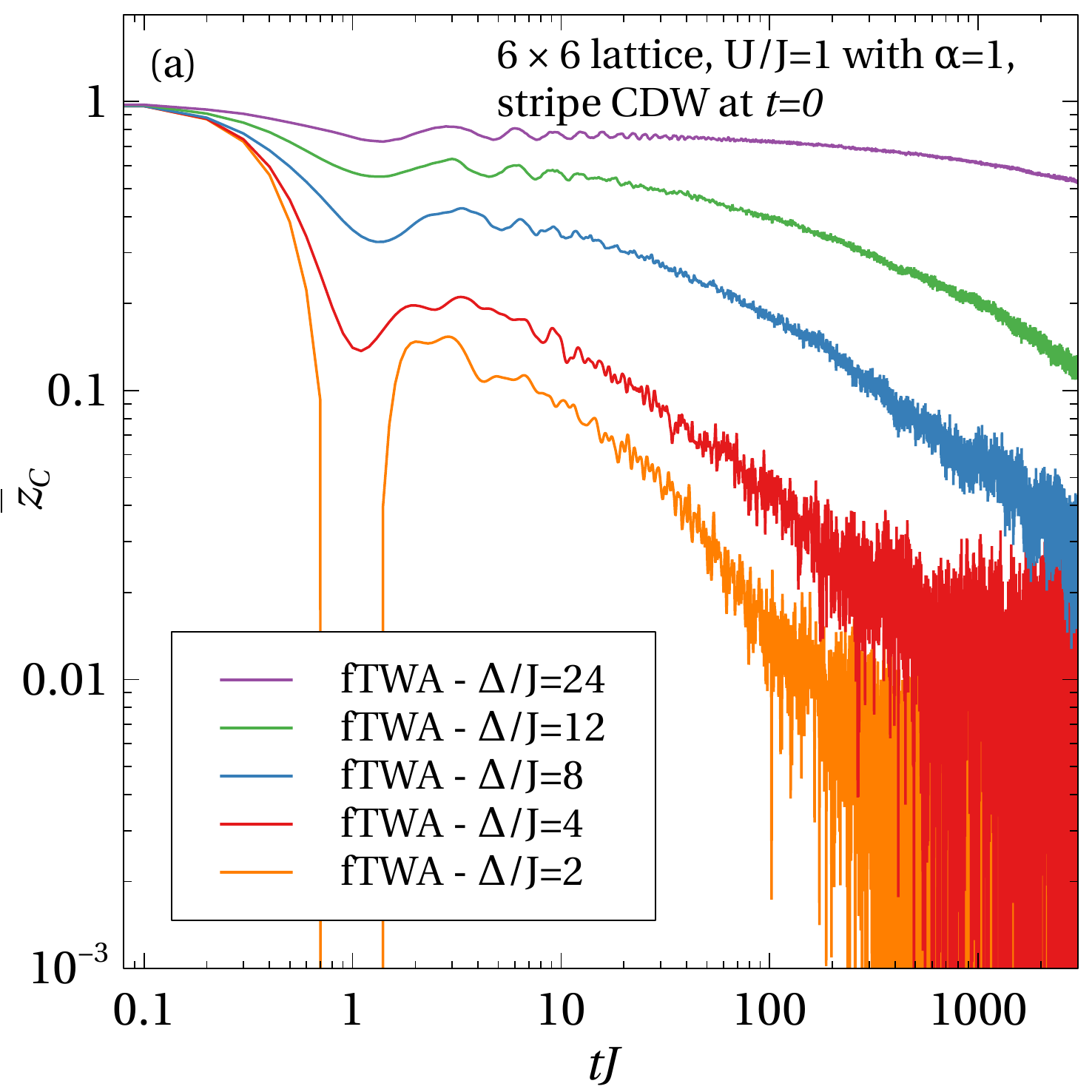}\includegraphics[scale=0.4]{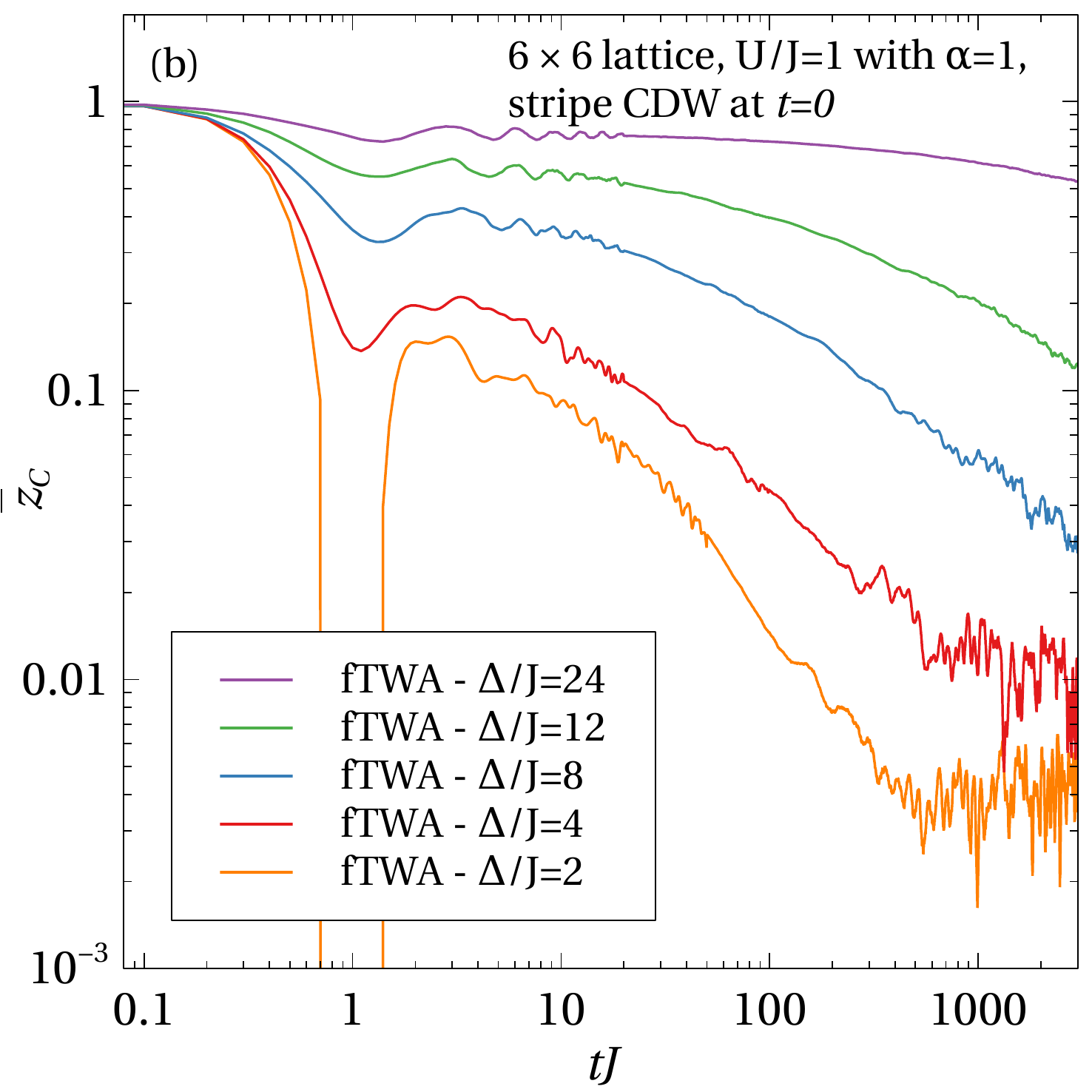}\includegraphics[scale=0.4]{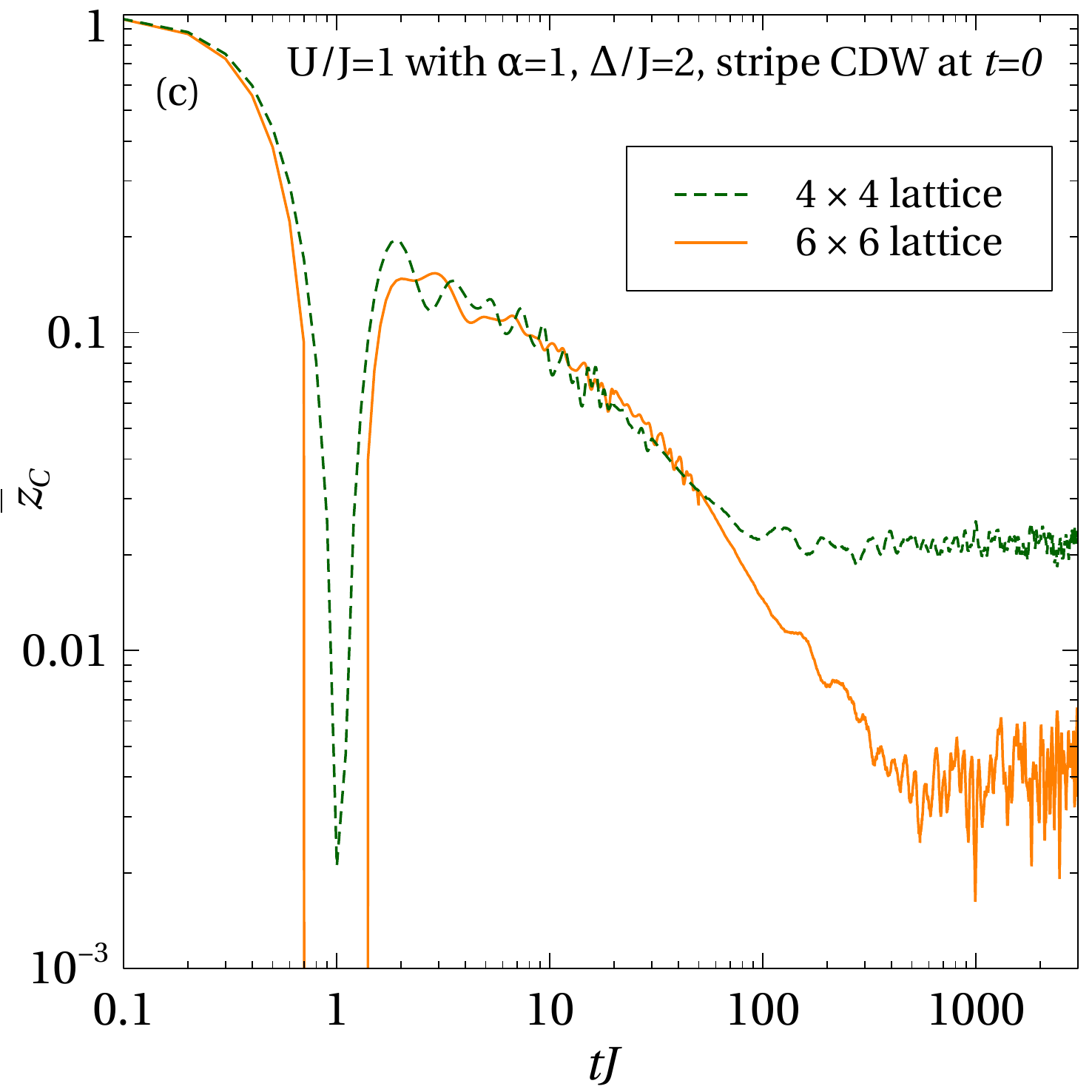}

\caption{(Color online) Long-time dynamics of the charge imbalance $\bar z_C(t)$. Panel (a) shows unfiltered data, while the panel (b)
presents the same data after filtering. The parameters used for the simulations are  $U/J=1$, $\alpha=1$, and the lattice size is $6\times6$. The results are averaged over at least 20 different disorder realizations. Different colors represent different disorder strengths:  $\Delta/J=2,4,8,12,24$ (from bottom to top (a,b)). Panel (c) shows 
dynamics of the charge imbalance for two different system sizes $4\times 4$ and $6\times 6$ suggesting that the long-time saturation of the imbalance at a small but nonzero value is a finite size effect.}
\label{signal filtering}
\end{figure*}

\section*{Appendix B. Semiclassical equations of motion for $\hat{H}_{I}$
and $\hat{H}_{II}$. \label{appendix - alpha=0 limit}}

The Weyl symbols of the Hamiltonians $\hat{H}_{I}$ and $\hat{H}_{II}$ can be found by standard means using the Bopp representation of the operators $\hat \rho_{\alpha\beta}$~\citep{Davidson2017}:
\be
H_{I,W} =  -J\sum_{\left\langle ij\right\rangle }\rho_{i\sigma j\sigma}+\sum_{i\sigma}\Delta_{i}\rho_{i\sigma i\sigma}+\sum_{ij}U_{ij}\left(\rho_{i\uparrow i\uparrow}+\frac{1}{2}\right)\left(\rho_{j\downarrow j\downarrow}+\frac{1}{2}\right)+\sum_{i<j,\sigma}V_{ij\sigma}\left(\rho_{i\sigma i\sigma}+\frac{1}{2}\right)\left(\rho_{j\sigma j\sigma}+\frac{1}{2}\right),\label{eq: Weyl of H_I}
\ee
\be
H_{II,W} =  -J\sum_{\left\langle ij\right\rangle }\rho_{i\sigma j\sigma}+\sum_{i\sigma}\Delta_{i}\rho_{i\sigma i\sigma}+\sum_{ij}U_{ij}\left(\rho_{i\uparrow i\uparrow}+\frac{1}{2}\right)\left(\rho_{j\downarrow j\downarrow}+\frac{1}{2}\right)+\frac{1}{2}\sum_{i,j,\sigma}V_{ij\sigma}\left(\rho_{i\sigma i\sigma}+\frac{1}{2}\right)\left(\rho_{j\sigma j\sigma}+\frac{1}{2}\right).\label{eq: Weyl of H_II}
\ee
Then the corresponding equations of motions for $\rho_{\alpha\beta}$ variables (see,
Eq.(\ref{eq: fTWA general equations for rho})) read:

\begin{eqnarray}
i\frac{\partial\rho_{m\sigma n\sigma}}{\partial t} & = & -J\sum_{\delta}\left(\rho_{m\sigma, n+\delta\;\sigma}-\rho_{m+\delta\;\sigma, n\sigma}\right)+\rho_{m\sigma n\sigma}\left[\left(\Delta_{n}-\Delta_{m}\right)+\sum_{r}\left(U_{nr}-U_{mr}\right)\left(\rho_{r-\sigma r-\sigma}+\frac{1}{2}\right)\right]\nonumber \\
 &  & +\rho_{m\sigma n\sigma}\sum_{r}\left((1-\delta_{nr})V_{nr\sigma}-(1-\delta_{mr})V_{mr\sigma}\right)\left(\rho_{r\sigma r\sigma}+\frac{1}{2}\right),\quad H=H_{I,W}\label{rho(t) equation FHM - simplified}
\end{eqnarray}
and
\begin{eqnarray}
i\frac{\partial\rho_{m\sigma n\sigma}}{\partial t} & = &-J\sum_{\delta}\left(\rho_{m\sigma, n+\delta\;\sigma}-\rho_{m+\delta\;\sigma, n\sigma}\right)+\rho_{m\sigma n\sigma}\left[\left(\Delta_{n}-\Delta_{m}\right)+\sum_{r}\left(U_{nr}-U_{mr}\right)\left(\rho_{r-\sigma r-\sigma}+\frac{1}{2}\right)\right]\nonumber \\
 &  & +\rho_{m\sigma n\sigma}\sum_{r}\left(V_{nr\sigma}-V_{mr\sigma}\right)\left(\rho_{r\sigma r\sigma}+\frac{1}{2}\right),\quad H=H_{II,W}.\label{eq: rho(t) - lepsze}
\end{eqnarray}
These equations are similar but not equivalent. In the limit of the infinite-range interactions, where $U_{ij}=U$ and
$V_{ij\sigma}=V$, Eq. (\ref{rho(t) equation FHM - simplified}) and
Eq. (\ref{eq: rho(t) - lepsze}) simplify to
\begin{eqnarray}
i\frac{\partial\rho_{m\sigma n\sigma}}{\partial t} & = & -J\sum_{\delta}\left(\rho_{m\sigma, n+\delta\;\sigma}-\rho_{m+\delta\;\sigma, n\sigma}\right)+\left(\Delta_{n}-\Delta_{m}\right)\rho_{m\sigma n\sigma}+V\rho_{m\sigma n\sigma}\sum_{r}\left(\delta_{mr}-\delta_{nr}\right)\left(\rho_{r\sigma r\sigma}+\frac{1}{2}\right).
\end{eqnarray}
and
\begin{eqnarray}
i\frac{\partial\rho_{m\sigma n\sigma}}{\partial t} & = & -J\sum_{\delta}\left(\rho_{m\sigma, n+\delta\;\sigma}-\rho_{m+\delta\;\sigma, n\sigma}\right)+\left(\Delta_{n}-\Delta_{m}\right)\rho_{m\sigma n\sigma}.\label{eq: linear fTWA}
\end{eqnarray}
We see that the first system of equations, which was obtained using $H_{I, W}$ still contains nonlinear terms, while the second system of equations based on $H_{II,W}$ representation of the Hamiltonian is linear, which has to be the case because the full quantum evolution is linear. Mathematically the origin of ambiguity comes from the fact that the operator identity for fermions $\hat n_\alpha^2=\hat n_\alpha$, which does not follow from the properties of the $U(N)$ algebra used to define the Poisson brackets, but rather from the fact that the operators $\hat E_\beta^\alpha$ form a particular fundamental representation of this algebra. At the same time the Weyl phase-space mapping of the operators is independent of the particular representation.

\end{widetext}

\section*{Appendix C Noise and finite size effects in fTWA simulations}

In Fig. \ref{signal filtering} a and b, we show the charge imbalance before and after removing substantial sampling noise from fTWA simulations.  For each disorder realization of fTWA simulations on a $6\times 6$ lattice we use at least 20 different trajectories corresponding to different random initial conditions. This number is to be contrasted with averaging over $400$ realization in Sec.~\ref{sec: Benchmark-of-fTWA} , where we benchmarked the fTWA against the exact results in small systems. While averaging over $20$ different initial conditions is sufficient to obtain a smooth short time behavior of the imbalance, more averaging is needed to eliminate the sampling noise at long times. Because even classical simulations in large systems are computationally costly we found that it is more efficient to apply filtering to the imbalance data to suppress this long-time spurious noise. As it is evident from comparing the curves shown in panels (a) and (b) the filtering does not introduce any systematic error.

In panel (c) of Fig.~\ref{signal filtering} we analyze finite size effects on the imbalance dynamics by comparing 
fTWA simulations for two different system sizes $4\times 4$ and $6\times 6$. We see that apart from a small difference in the first oscillation (enhanced in the plot because of using the logarithmic scale) the results for the two system sizes are very similar until the imbalance saturates at long times at a small positive value, which rapidly goes to zero with the system size. From this comparison we can conclude that the $6\times 6$ system size is sufficient to capture some key features of the imbalance dynamics in the thermodynamic limit.

\bibliographystyle{apsrev4-1}
\bibliography{library}

\begin{thebibliography}{54}%
\makeatletter
\providecommand \@ifxundefined [1]{%
 \@ifx{#1\undefined}
}%
\providecommand \@ifnum [1]{%
 \ifnum #1\expandafter \@firstoftwo
 \else \expandafter \@secondoftwo
 \fi
}%
\providecommand \@ifx [1]{%
 \ifx #1\expandafter \@firstoftwo
 \else \expandafter \@secondoftwo
 \fi
}%
\providecommand \natexlab [1]{#1}%
\providecommand \enquote  [1]{``#1''}%
\providecommand \bibnamefont  [1]{#1}%
\providecommand \bibfnamefont [1]{#1}%
\providecommand \citenamefont [1]{#1}%
\providecommand \href@noop [0]{\@secondoftwo}%
\providecommand \href [0]{\begingroup \@sanitize@url \@href}%
\providecommand \@href[1]{\@@startlink{#1}\@@href}%
\providecommand \@@href[1]{\endgroup#1\@@endlink}%
\providecommand \@sanitize@url [0]{\catcode `\\12\catcode `\$12\catcode
  `\&12\catcode `\#12\catcode `\^12\catcode `\_12\catcode `\%12\relax}%
\providecommand \@@startlink[1]{}%
\providecommand \@@endlink[0]{}%
\providecommand \url  [0]{\begingroup\@sanitize@url \@url }%
\providecommand \@url [1]{\endgroup\@href {#1}{\urlprefix }}%
\providecommand \urlprefix  [0]{URL }%
\providecommand \Eprint [0]{\href }%
\providecommand \doibase [0]{http://dx.doi.org/}%
\providecommand \selectlanguage [0]{\@gobble}%
\providecommand \bibinfo  [0]{\@secondoftwo}%
\providecommand \bibfield  [0]{\@secondoftwo}%
\providecommand \translation [1]{[#1]}%
\providecommand \BibitemOpen [0]{}%
\providecommand \bibitemStop [0]{}%
\providecommand \bibitemNoStop [0]{.\EOS\space}%
\providecommand \EOS [0]{\spacefactor3000\relax}%
\providecommand \BibitemShut  [1]{\csname bibitem#1\endcsname}%
\let\auto@bib@innerbib\@empty
\bibitem [{\citenamefont {Schreiber}\ \emph {et~al.}(2015)\citenamefont
  {Schreiber}, \citenamefont {Hodgman}, \citenamefont {Bordia}, \citenamefont
  {L{\"u}schen}, \citenamefont {Fischer}, \citenamefont {Vosk}, \citenamefont
  {Altman}, \citenamefont {Schneider},\ and\ \citenamefont
  {Bloch}}]{mblschreiber}%
  \BibitemOpen
  \bibfield  {author} {\bibinfo {author} {\bibfnamefont {M.}~\bibnamefont
  {Schreiber}}, \bibinfo {author} {\bibfnamefont {S.~S.}\ \bibnamefont
  {Hodgman}}, \bibinfo {author} {\bibfnamefont {P.}~\bibnamefont {Bordia}},
  \bibinfo {author} {\bibfnamefont {H.~P.}\ \bibnamefont {L{\"u}schen}},
  \bibinfo {author} {\bibfnamefont {M.~H.}\ \bibnamefont {Fischer}}, \bibinfo
  {author} {\bibfnamefont {R.}~\bibnamefont {Vosk}}, \bibinfo {author}
  {\bibfnamefont {E.}~\bibnamefont {Altman}}, \bibinfo {author} {\bibfnamefont
  {U.}~\bibnamefont {Schneider}}, \ and\ \bibinfo {author} {\bibfnamefont
  {I.}~\bibnamefont {Bloch}},\ }\href {\doibase
  doi.org/10.1126/science.aaa7432} {\bibfield  {journal} {\bibinfo  {journal}
  {Science}\ }\textbf {\bibinfo {volume} {349}},\ \bibinfo {pages} {842}
  (\bibinfo {year} {2015})}\BibitemShut {NoStop}%
\bibitem [{\citenamefont {y.~Choi}\ \emph {et~al.}(2016)\citenamefont
  {y.~Choi}, \citenamefont {Hild}, \citenamefont {Zeiher}, \citenamefont
  {Schauss}, \citenamefont {Rubio-Abadal}, \citenamefont {Yefsah},
  \citenamefont {Khemani}, \citenamefont {Huse}, \citenamefont {Bloch},\ and\
  \citenamefont {Gross}}]{Choi2016}%
  \BibitemOpen
  \bibfield  {author} {\bibinfo {author} {\bibfnamefont {J.}~\bibnamefont
  {y.~Choi}}, \bibinfo {author} {\bibfnamefont {S.}~\bibnamefont {Hild}},
  \bibinfo {author} {\bibfnamefont {J.}~\bibnamefont {Zeiher}}, \bibinfo
  {author} {\bibfnamefont {P.}~\bibnamefont {Schauss}}, \bibinfo {author}
  {\bibfnamefont {A.}~\bibnamefont {Rubio-Abadal}}, \bibinfo {author}
  {\bibfnamefont {T.}~\bibnamefont {Yefsah}}, \bibinfo {author} {\bibfnamefont
  {V.}~\bibnamefont {Khemani}}, \bibinfo {author} {\bibfnamefont {D.~A.}\
  \bibnamefont {Huse}}, \bibinfo {author} {\bibfnamefont {I.}~\bibnamefont
  {Bloch}}, \ and\ \bibinfo {author} {\bibfnamefont {C.}~\bibnamefont
  {Gross}},\ }\href {\doibase 10.1126/science.aaf8834} {\bibfield  {journal}
  {\bibinfo  {journal} {Science}\ }\textbf {\bibinfo {volume} {352}},\ \bibinfo
  {pages} {1547} (\bibinfo {year} {2016})}\BibitemShut {NoStop}%
\bibitem [{\citenamefont {Smith}\ \emph {et~al.}(2016)\citenamefont {Smith},
  \citenamefont {Lee}, \citenamefont {Richerme}, \citenamefont {Neyenhuis},
  \citenamefont {Hess}, \citenamefont {Hauke}, \citenamefont {Heyl},
  \citenamefont {Huse},\ and\ \citenamefont {Monroe}}]{Smith2016}%
  \BibitemOpen
  \bibfield  {author} {\bibinfo {author} {\bibfnamefont {J.}~\bibnamefont
  {Smith}}, \bibinfo {author} {\bibfnamefont {A.}~\bibnamefont {Lee}}, \bibinfo
  {author} {\bibfnamefont {P.}~\bibnamefont {Richerme}}, \bibinfo {author}
  {\bibfnamefont {B.}~\bibnamefont {Neyenhuis}}, \bibinfo {author}
  {\bibfnamefont {P.~W.}\ \bibnamefont {Hess}}, \bibinfo {author}
  {\bibfnamefont {P.}~\bibnamefont {Hauke}}, \bibinfo {author} {\bibfnamefont
  {M.}~\bibnamefont {Heyl}}, \bibinfo {author} {\bibfnamefont {D.~A.}\
  \bibnamefont {Huse}}, \ and\ \bibinfo {author} {\bibfnamefont
  {C.}~\bibnamefont {Monroe}},\ }\href {\doibase 10.1038/nphys3783} {\bibfield
  {journal} {\bibinfo  {journal} {Nature Physics}\ }\textbf {\bibinfo {volume}
  {12}},\ \bibinfo {pages} {907} (\bibinfo {year} {2016})}\BibitemShut
  {NoStop}%
\bibitem [{\citenamefont {Bordia}\ \emph {et~al.}(2016)\citenamefont {Bordia},
  \citenamefont {L\"uschen}, \citenamefont {Hodgman}, \citenamefont
  {Schreiber}, \citenamefont {Bloch},\ and\ \citenamefont
  {Schneider}}]{PhysRevLett.116.140401}%
  \BibitemOpen
  \bibfield  {author} {\bibinfo {author} {\bibfnamefont {P.}~\bibnamefont
  {Bordia}}, \bibinfo {author} {\bibfnamefont {H.~P.}\ \bibnamefont
  {L\"uschen}}, \bibinfo {author} {\bibfnamefont {S.~S.}\ \bibnamefont
  {Hodgman}}, \bibinfo {author} {\bibfnamefont {M.}~\bibnamefont {Schreiber}},
  \bibinfo {author} {\bibfnamefont {I.}~\bibnamefont {Bloch}}, \ and\ \bibinfo
  {author} {\bibfnamefont {U.}~\bibnamefont {Schneider}},\ }\href {\doibase
  10.1103/PhysRevLett.116.140401} {\bibfield  {journal} {\bibinfo  {journal}
  {Phys. Rev. Lett.}\ }\textbf {\bibinfo {volume} {116}},\ \bibinfo {pages}
  {140401} (\bibinfo {year} {2016})}\BibitemShut {NoStop}%
\bibitem [{\citenamefont {Zhang}\ \emph {et~al.}(2017)\citenamefont {Zhang},
  \citenamefont {Hess}, \citenamefont {Kyprianidis}, \citenamefont {Becker},
  \citenamefont {Lee}, \citenamefont {Smith}, \citenamefont {Pagano},
  \citenamefont {Potirniche}, \citenamefont {Potter}, \citenamefont
  {Vishwanath}, \citenamefont {Yao},\ and\ \citenamefont {Monroe}}]{Zhang2017}%
  \BibitemOpen
  \bibfield  {author} {\bibinfo {author} {\bibfnamefont {J.}~\bibnamefont
  {Zhang}}, \bibinfo {author} {\bibfnamefont {P.~W.}\ \bibnamefont {Hess}},
  \bibinfo {author} {\bibfnamefont {A.}~\bibnamefont {Kyprianidis}}, \bibinfo
  {author} {\bibfnamefont {P.}~\bibnamefont {Becker}}, \bibinfo {author}
  {\bibfnamefont {A.}~\bibnamefont {Lee}}, \bibinfo {author} {\bibfnamefont
  {J.}~\bibnamefont {Smith}}, \bibinfo {author} {\bibfnamefont
  {G.}~\bibnamefont {Pagano}}, \bibinfo {author} {\bibfnamefont {I.-D.}\
  \bibnamefont {Potirniche}}, \bibinfo {author} {\bibfnamefont {A.~C.}\
  \bibnamefont {Potter}}, \bibinfo {author} {\bibfnamefont {A.}~\bibnamefont
  {Vishwanath}}, \bibinfo {author} {\bibfnamefont {N.~Y.}\ \bibnamefont {Yao}},
  \ and\ \bibinfo {author} {\bibfnamefont {C.}~\bibnamefont {Monroe}},\ }\href
  {\doibase 10.1038/nature21413} {\bibfield  {journal} {\bibinfo  {journal}
  {Nature}\ }\textbf {\bibinfo {volume} {543}},\ \bibinfo {pages} {217}
  (\bibinfo {year} {2017})}\BibitemShut {NoStop}%
\bibitem [{\citenamefont {Choi}\ \emph {et~al.}(2017)\citenamefont {Choi},
  \citenamefont {Choi}, \citenamefont {Landig}, \citenamefont {Kucsko},
  \citenamefont {Zhou}, \citenamefont {Isoya}, \citenamefont {Jelezko},
  \citenamefont {Onoda}, \citenamefont {Sumiya}, \citenamefont {Khemani},
  \citenamefont {von Keyserlingk}, \citenamefont {Yao}, \citenamefont
  {Demler},\ and\ \citenamefont {Lukin}}]{Choi2017}%
  \BibitemOpen
  \bibfield  {author} {\bibinfo {author} {\bibfnamefont {S.}~\bibnamefont
  {Choi}}, \bibinfo {author} {\bibfnamefont {J.}~\bibnamefont {Choi}}, \bibinfo
  {author} {\bibfnamefont {R.}~\bibnamefont {Landig}}, \bibinfo {author}
  {\bibfnamefont {G.}~\bibnamefont {Kucsko}}, \bibinfo {author} {\bibfnamefont
  {H.}~\bibnamefont {Zhou}}, \bibinfo {author} {\bibfnamefont {J.}~\bibnamefont
  {Isoya}}, \bibinfo {author} {\bibfnamefont {F.}~\bibnamefont {Jelezko}},
  \bibinfo {author} {\bibfnamefont {S.}~\bibnamefont {Onoda}}, \bibinfo
  {author} {\bibfnamefont {H.}~\bibnamefont {Sumiya}}, \bibinfo {author}
  {\bibfnamefont {V.}~\bibnamefont {Khemani}}, \bibinfo {author} {\bibfnamefont
  {C.}~\bibnamefont {von Keyserlingk}}, \bibinfo {author} {\bibfnamefont
  {N.~Y.}\ \bibnamefont {Yao}}, \bibinfo {author} {\bibfnamefont
  {E.}~\bibnamefont {Demler}}, \ and\ \bibinfo {author} {\bibfnamefont {M.~D.}\
  \bibnamefont {Lukin}},\ }\href {\doibase 10.1038/nature21426} {\bibfield
  {journal} {\bibinfo  {journal} {Nature}\ }\textbf {\bibinfo {volume} {543}},\
  \bibinfo {pages} {221} (\bibinfo {year} {2017})}\BibitemShut {NoStop}%
\bibitem [{\citenamefont {L\"uschen}\ \emph
  {et~al.}(2017{\natexlab{a}})\citenamefont {L\"uschen}, \citenamefont
  {Bordia}, \citenamefont {Hodgman}, \citenamefont {Schreiber}, \citenamefont
  {Sarkar}, \citenamefont {Daley}, \citenamefont {Fischer}, \citenamefont
  {Altman}, \citenamefont {Bloch},\ and\ \citenamefont
  {Schneider}}]{Lschen2017}%
  \BibitemOpen
  \bibfield  {author} {\bibinfo {author} {\bibfnamefont {H.~P.}\ \bibnamefont
  {L\"uschen}}, \bibinfo {author} {\bibfnamefont {P.}~\bibnamefont {Bordia}},
  \bibinfo {author} {\bibfnamefont {S.~S.}\ \bibnamefont {Hodgman}}, \bibinfo
  {author} {\bibfnamefont {M.}~\bibnamefont {Schreiber}}, \bibinfo {author}
  {\bibfnamefont {S.}~\bibnamefont {Sarkar}}, \bibinfo {author} {\bibfnamefont
  {A.~J.}\ \bibnamefont {Daley}}, \bibinfo {author} {\bibfnamefont {M.~H.}\
  \bibnamefont {Fischer}}, \bibinfo {author} {\bibfnamefont {E.}~\bibnamefont
  {Altman}}, \bibinfo {author} {\bibfnamefont {I.}~\bibnamefont {Bloch}}, \
  and\ \bibinfo {author} {\bibfnamefont {U.}~\bibnamefont {Schneider}},\ }\href
  {\doibase 10.1103/PhysRevX.7.011034} {\bibfield  {journal} {\bibinfo
  {journal} {Phys. Rev. X}\ }\textbf {\bibinfo {volume} {7}},\ \bibinfo {pages}
  {011034} (\bibinfo {year} {2017}{\natexlab{a}})}\BibitemShut {NoStop}%
\bibitem [{\citenamefont {L\"uschen}\ \emph
  {et~al.}(2017{\natexlab{b}})\citenamefont {L\"uschen}, \citenamefont
  {Bordia}, \citenamefont {Scherg}, \citenamefont {Alet}, \citenamefont
  {Altman}, \citenamefont {Schneider},\ and\ \citenamefont
  {Bloch}}]{PhysRevLett.119.260401}%
  \BibitemOpen
  \bibfield  {author} {\bibinfo {author} {\bibfnamefont {H.~P.}\ \bibnamefont
  {L\"uschen}}, \bibinfo {author} {\bibfnamefont {P.}~\bibnamefont {Bordia}},
  \bibinfo {author} {\bibfnamefont {S.}~\bibnamefont {Scherg}}, \bibinfo
  {author} {\bibfnamefont {F.}~\bibnamefont {Alet}}, \bibinfo {author}
  {\bibfnamefont {E.}~\bibnamefont {Altman}}, \bibinfo {author} {\bibfnamefont
  {U.}~\bibnamefont {Schneider}}, \ and\ \bibinfo {author} {\bibfnamefont
  {I.}~\bibnamefont {Bloch}},\ }\href {\doibase 10.1103/PhysRevLett.119.260401}
  {\bibfield  {journal} {\bibinfo  {journal} {Phys. Rev. Lett.}\ }\textbf
  {\bibinfo {volume} {119}},\ \bibinfo {pages} {260401} (\bibinfo {year}
  {2017}{\natexlab{b}})}\BibitemShut {NoStop}%
\bibitem [{\citenamefont {Bordia}\ \emph {et~al.}(2017)\citenamefont {Bordia},
  \citenamefont {L\"uschen}, \citenamefont {Scherg}, \citenamefont
  {Gopalakrishnan}, \citenamefont {Knap}, \citenamefont {Schneider},\ and\
  \citenamefont {Bloch}}]{PhysRevX.7.041047}%
  \BibitemOpen
  \bibfield  {author} {\bibinfo {author} {\bibfnamefont {P.}~\bibnamefont
  {Bordia}}, \bibinfo {author} {\bibfnamefont {H.}~\bibnamefont {L\"uschen}},
  \bibinfo {author} {\bibfnamefont {S.}~\bibnamefont {Scherg}}, \bibinfo
  {author} {\bibfnamefont {S.}~\bibnamefont {Gopalakrishnan}}, \bibinfo
  {author} {\bibfnamefont {M.}~\bibnamefont {Knap}}, \bibinfo {author}
  {\bibfnamefont {U.}~\bibnamefont {Schneider}}, \ and\ \bibinfo {author}
  {\bibfnamefont {I.}~\bibnamefont {Bloch}},\ }\href {\doibase
  10.1103/PhysRevX.7.041047} {\bibfield  {journal} {\bibinfo  {journal} {Phys.
  Rev. X}\ }\textbf {\bibinfo {volume} {7}},\ \bibinfo {pages} {041047}
  (\bibinfo {year} {2017})}\BibitemShut {NoStop}%
\bibitem [{\citenamefont {Kucsko}\ \emph {et~al.}(2018)\citenamefont {Kucsko},
  \citenamefont {Choi}, \citenamefont {Choi}, \citenamefont {Maurer},
  \citenamefont {Zhou}, \citenamefont {Landig}, \citenamefont {Sumiya},
  \citenamefont {Onoda}, \citenamefont {Isoya}, \citenamefont {Jelezko},
  \citenamefont {Demler}, \citenamefont {Yao},\ and\ \citenamefont
  {Lukin}}]{Kucsko2018}%
  \BibitemOpen
  \bibfield  {author} {\bibinfo {author} {\bibfnamefont {G.}~\bibnamefont
  {Kucsko}}, \bibinfo {author} {\bibfnamefont {S.}~\bibnamefont {Choi}},
  \bibinfo {author} {\bibfnamefont {J.}~\bibnamefont {Choi}}, \bibinfo {author}
  {\bibfnamefont {P.~C.}\ \bibnamefont {Maurer}}, \bibinfo {author}
  {\bibfnamefont {H.}~\bibnamefont {Zhou}}, \bibinfo {author} {\bibfnamefont
  {R.}~\bibnamefont {Landig}}, \bibinfo {author} {\bibfnamefont
  {H.}~\bibnamefont {Sumiya}}, \bibinfo {author} {\bibfnamefont
  {S.}~\bibnamefont {Onoda}}, \bibinfo {author} {\bibfnamefont
  {J.}~\bibnamefont {Isoya}}, \bibinfo {author} {\bibfnamefont
  {F.}~\bibnamefont {Jelezko}}, \bibinfo {author} {\bibfnamefont
  {E.}~\bibnamefont {Demler}}, \bibinfo {author} {\bibfnamefont {N.~Y.}\
  \bibnamefont {Yao}}, \ and\ \bibinfo {author} {\bibfnamefont {M.~D.}\
  \bibnamefont {Lukin}},\ }\href {\doibase 10.1103/PhysRevLett.121.023601}
  {\bibfield  {journal} {\bibinfo  {journal} {Phys. Rev. Lett.}\ }\textbf
  {\bibinfo {volume} {121}},\ \bibinfo {pages} {023601} (\bibinfo {year}
  {2018})}\BibitemShut {NoStop}%
\bibitem [{\citenamefont {Lukin}\ \emph {et~al.}(2019)\citenamefont {Lukin},
  \citenamefont {Rispoli}, \citenamefont {Schittko}, \citenamefont {Tai},
  \citenamefont {Kaufman}, \citenamefont {Choi}, \citenamefont {Khemani},
  \citenamefont {L{\'e}onard},\ and\ \citenamefont {Greiner}}]{Lukin256}%
  \BibitemOpen
  \bibfield  {author} {\bibinfo {author} {\bibfnamefont {A.}~\bibnamefont
  {Lukin}}, \bibinfo {author} {\bibfnamefont {M.}~\bibnamefont {Rispoli}},
  \bibinfo {author} {\bibfnamefont {R.}~\bibnamefont {Schittko}}, \bibinfo
  {author} {\bibfnamefont {M.~E.}\ \bibnamefont {Tai}}, \bibinfo {author}
  {\bibfnamefont {A.~M.}\ \bibnamefont {Kaufman}}, \bibinfo {author}
  {\bibfnamefont {S.}~\bibnamefont {Choi}}, \bibinfo {author} {\bibfnamefont
  {V.}~\bibnamefont {Khemani}}, \bibinfo {author} {\bibfnamefont
  {J.}~\bibnamefont {L{\'e}onard}}, \ and\ \bibinfo {author} {\bibfnamefont
  {M.}~\bibnamefont {Greiner}},\ }\href {\doibase 10.1126/science.aau0818}
  {\bibfield  {journal} {\bibinfo  {journal} {Science}\ }\textbf {\bibinfo
  {volume} {364}},\ \bibinfo {pages} {256} (\bibinfo {year} {2019})},\ \Eprint
  {http://arxiv.org/abs/1805.09819} {1805.09819} \BibitemShut {NoStop}%
\bibitem [{\citenamefont {Chiaro}\ \emph {et~al.}(2019)\citenamefont {Chiaro},
  \citenamefont {Neill}, \citenamefont {Bohrdt}, \citenamefont {Filippone},
  \citenamefont {Arute}, \citenamefont {Arya}, \citenamefont {Babbush},
  \citenamefont {Bacon}, \citenamefont {Bardin}, \citenamefont {Barends},
  \citenamefont {Boixo}, \citenamefont {Buell}, \citenamefont {Burkett},
  \citenamefont {Chen}, \citenamefont {Chen}, \citenamefont {Collins},
  \citenamefont {Dunsworth}, \citenamefont {Farhi}, \citenamefont {Fowler},
  \citenamefont {Foxen}, \citenamefont {Gidney}, \citenamefont {Giustina},
  \citenamefont {Harrigan}, \citenamefont {Huang}, \citenamefont {Isakov},
  \citenamefont {Jeffrey}, \citenamefont {Jiang}, \citenamefont {Kafri},
  \citenamefont {Kechedzhi}, \citenamefont {Kelly}, \citenamefont {Klimov},
  \citenamefont {Korotkov}, \citenamefont {Kostritsa}, \citenamefont
  {Landhuis}, \citenamefont {Lucero}, \citenamefont {McClean}, \citenamefont
  {Mi}, \citenamefont {Megrant}, \citenamefont {Mohseni}, \citenamefont
  {Mutus}, \citenamefont {McEwen}, \citenamefont {Naaman}, \citenamefont
  {Neeley}, \citenamefont {Niu}, \citenamefont {Petukhov}, \citenamefont
  {Quintana}, \citenamefont {Rubin}, \citenamefont {Sank}, \citenamefont
  {Satzinger}, \citenamefont {Vainsencher}, \citenamefont {White},
  \citenamefont {Yao}, \citenamefont {Yeh}, \citenamefont {Zalcman},
  \citenamefont {Smelyanskiy}, \citenamefont {Neven}, \citenamefont
  {Gopalakrishnan}, \citenamefont {Abanin}, \citenamefont {Knap}, \citenamefont
  {Martinis},\ and\ \citenamefont {Roushan}}]{1910.06024}%
  \BibitemOpen
  \bibfield  {author} {\bibinfo {author} {\bibfnamefont {B.}~\bibnamefont
  {Chiaro}}, \bibinfo {author} {\bibfnamefont {C.}~\bibnamefont {Neill}},
  \bibinfo {author} {\bibfnamefont {A.}~\bibnamefont {Bohrdt}}, \bibinfo
  {author} {\bibfnamefont {M.}~\bibnamefont {Filippone}}, \bibinfo {author}
  {\bibfnamefont {F.}~\bibnamefont {Arute}}, \bibinfo {author} {\bibfnamefont
  {K.}~\bibnamefont {Arya}}, \bibinfo {author} {\bibfnamefont {R.}~\bibnamefont
  {Babbush}}, \bibinfo {author} {\bibfnamefont {D.}~\bibnamefont {Bacon}},
  \bibinfo {author} {\bibfnamefont {J.}~\bibnamefont {Bardin}}, \bibinfo
  {author} {\bibfnamefont {R.}~\bibnamefont {Barends}}, \bibinfo {author}
  {\bibfnamefont {S.}~\bibnamefont {Boixo}}, \bibinfo {author} {\bibfnamefont
  {D.}~\bibnamefont {Buell}}, \bibinfo {author} {\bibfnamefont
  {B.}~\bibnamefont {Burkett}}, \bibinfo {author} {\bibfnamefont
  {Y.}~\bibnamefont {Chen}}, \bibinfo {author} {\bibfnamefont {Z.}~\bibnamefont
  {Chen}}, \bibinfo {author} {\bibfnamefont {R.}~\bibnamefont {Collins}},
  \bibinfo {author} {\bibfnamefont {A.}~\bibnamefont {Dunsworth}}, \bibinfo
  {author} {\bibfnamefont {E.}~\bibnamefont {Farhi}}, \bibinfo {author}
  {\bibfnamefont {A.}~\bibnamefont {Fowler}}, \bibinfo {author} {\bibfnamefont
  {B.}~\bibnamefont {Foxen}}, \bibinfo {author} {\bibfnamefont
  {C.}~\bibnamefont {Gidney}}, \bibinfo {author} {\bibfnamefont
  {M.}~\bibnamefont {Giustina}}, \bibinfo {author} {\bibfnamefont
  {M.}~\bibnamefont {Harrigan}}, \bibinfo {author} {\bibfnamefont
  {T.}~\bibnamefont {Huang}}, \bibinfo {author} {\bibfnamefont
  {S.}~\bibnamefont {Isakov}}, \bibinfo {author} {\bibfnamefont
  {E.}~\bibnamefont {Jeffrey}}, \bibinfo {author} {\bibfnamefont
  {Z.}~\bibnamefont {Jiang}}, \bibinfo {author} {\bibfnamefont
  {D.}~\bibnamefont {Kafri}}, \bibinfo {author} {\bibfnamefont
  {K.}~\bibnamefont {Kechedzhi}}, \bibinfo {author} {\bibfnamefont
  {J.}~\bibnamefont {Kelly}}, \bibinfo {author} {\bibfnamefont
  {P.}~\bibnamefont {Klimov}}, \bibinfo {author} {\bibfnamefont
  {A.}~\bibnamefont {Korotkov}}, \bibinfo {author} {\bibfnamefont
  {F.}~\bibnamefont {Kostritsa}}, \bibinfo {author} {\bibfnamefont
  {D.}~\bibnamefont {Landhuis}}, \bibinfo {author} {\bibfnamefont
  {E.}~\bibnamefont {Lucero}}, \bibinfo {author} {\bibfnamefont
  {J.}~\bibnamefont {McClean}}, \bibinfo {author} {\bibfnamefont
  {X.}~\bibnamefont {Mi}}, \bibinfo {author} {\bibfnamefont {A.}~\bibnamefont
  {Megrant}}, \bibinfo {author} {\bibfnamefont {M.}~\bibnamefont {Mohseni}},
  \bibinfo {author} {\bibfnamefont {J.}~\bibnamefont {Mutus}}, \bibinfo
  {author} {\bibfnamefont {M.}~\bibnamefont {McEwen}}, \bibinfo {author}
  {\bibfnamefont {O.}~\bibnamefont {Naaman}}, \bibinfo {author} {\bibfnamefont
  {M.}~\bibnamefont {Neeley}}, \bibinfo {author} {\bibfnamefont
  {M.}~\bibnamefont {Niu}}, \bibinfo {author} {\bibfnamefont {A.}~\bibnamefont
  {Petukhov}}, \bibinfo {author} {\bibfnamefont {C.}~\bibnamefont {Quintana}},
  \bibinfo {author} {\bibfnamefont {N.}~\bibnamefont {Rubin}}, \bibinfo
  {author} {\bibfnamefont {D.}~\bibnamefont {Sank}}, \bibinfo {author}
  {\bibfnamefont {K.}~\bibnamefont {Satzinger}}, \bibinfo {author}
  {\bibfnamefont {A.}~\bibnamefont {Vainsencher}}, \bibinfo {author}
  {\bibfnamefont {T.}~\bibnamefont {White}}, \bibinfo {author} {\bibfnamefont
  {Z.}~\bibnamefont {Yao}}, \bibinfo {author} {\bibfnamefont {P.}~\bibnamefont
  {Yeh}}, \bibinfo {author} {\bibfnamefont {A.}~\bibnamefont {Zalcman}},
  \bibinfo {author} {\bibfnamefont {V.}~\bibnamefont {Smelyanskiy}}, \bibinfo
  {author} {\bibfnamefont {H.}~\bibnamefont {Neven}}, \bibinfo {author}
  {\bibfnamefont {S.}~\bibnamefont {Gopalakrishnan}}, \bibinfo {author}
  {\bibfnamefont {D.}~\bibnamefont {Abanin}}, \bibinfo {author} {\bibfnamefont
  {M.}~\bibnamefont {Knap}}, \bibinfo {author} {\bibfnamefont {J.}~\bibnamefont
  {Martinis}}, \ and\ \bibinfo {author} {\bibfnamefont {P.}~\bibnamefont
  {Roushan}},\ }\href@noop {} {\  (\bibinfo {year} {2019})},\ \Eprint
  {http://arxiv.org/abs/arXiv:1910.06024} {arXiv:1910.06024} \BibitemShut
  {NoStop}%
\bibitem [{\citenamefont {Nandkishore}\ and\ \citenamefont
  {Huse}(2015)}]{Nandkishore2015}%
  \BibitemOpen
  \bibfield  {author} {\bibinfo {author} {\bibfnamefont {R.}~\bibnamefont
  {Nandkishore}}\ and\ \bibinfo {author} {\bibfnamefont {D.~A.}\ \bibnamefont
  {Huse}},\ }\href {\doibase 10.1146/annurev-conmatphys-031214-014726}
  {\bibfield  {journal} {\bibinfo  {journal} {Annual Review of Condensed Matter
  Physics}\ }\textbf {\bibinfo {volume} {6}},\ \bibinfo {pages} {15} (\bibinfo
  {year} {2015})}\BibitemShut {NoStop}%
\bibitem [{\citenamefont {Abanin}\ \emph {et~al.}(2019)\citenamefont {Abanin},
  \citenamefont {Altman}, \citenamefont {Bloch},\ and\ \citenamefont
  {Serbyn}}]{RevModPhys.91.021001}%
  \BibitemOpen
  \bibfield  {author} {\bibinfo {author} {\bibfnamefont {D.~A.}\ \bibnamefont
  {Abanin}}, \bibinfo {author} {\bibfnamefont {E.}~\bibnamefont {Altman}},
  \bibinfo {author} {\bibfnamefont {I.}~\bibnamefont {Bloch}}, \ and\ \bibinfo
  {author} {\bibfnamefont {M.}~\bibnamefont {Serbyn}},\ }\href@noop {}
  {\bibfield  {journal} {\bibinfo  {journal} {Rev. Mod. Phys.}\ }\textbf
  {\bibinfo {volume} {91}},\ \bibinfo {pages} {021001} (\bibinfo {year}
  {2019})}\BibitemShut {NoStop}%
\bibitem [{\citenamefont {Kohlert}\ \emph {et~al.}(2019)\citenamefont
  {Kohlert}, \citenamefont {Scherg}, \citenamefont {Li}, \citenamefont
  {L\"uschen}, \citenamefont {Das~Sarma}, \citenamefont {Bloch},\ and\
  \citenamefont {Aidelsburger}}]{PhysRevLett.122.170403}%
  \BibitemOpen
  \bibfield  {author} {\bibinfo {author} {\bibfnamefont {T.}~\bibnamefont
  {Kohlert}}, \bibinfo {author} {\bibfnamefont {S.}~\bibnamefont {Scherg}},
  \bibinfo {author} {\bibfnamefont {X.}~\bibnamefont {Li}}, \bibinfo {author}
  {\bibfnamefont {H.~P.}\ \bibnamefont {L\"uschen}}, \bibinfo {author}
  {\bibfnamefont {S.}~\bibnamefont {Das~Sarma}}, \bibinfo {author}
  {\bibfnamefont {I.}~\bibnamefont {Bloch}}, \ and\ \bibinfo {author}
  {\bibfnamefont {M.}~\bibnamefont {Aidelsburger}},\ }\href {\doibase
  10.1103/PhysRevLett.122.170403} {\bibfield  {journal} {\bibinfo  {journal}
  {Phys. Rev. Lett.}\ }\textbf {\bibinfo {volume} {122}},\ \bibinfo {pages}
  {170403} (\bibinfo {year} {2019})}\BibitemShut {NoStop}%
\bibitem [{\citenamefont {Luitz}\ and\ \citenamefont {Lev}(2017)}]{Luitz2017}%
  \BibitemOpen
  \bibfield  {author} {\bibinfo {author} {\bibfnamefont {D.~J.}\ \bibnamefont
  {Luitz}}\ and\ \bibinfo {author} {\bibfnamefont {Y.~B.}\ \bibnamefont
  {Lev}},\ }\href {\doibase 10.1002/andp.201600350} {\bibfield  {journal}
  {\bibinfo  {journal} {Annalen der Physik}\ }\textbf {\bibinfo {volume}
  {529}},\ \bibinfo {pages} {1600350} (\bibinfo {year} {2017})}\BibitemShut
  {NoStop}%
\bibitem [{\citenamefont {\ifmmode \check{Z}\else
  \v{Z}\fi{}nidari\ifmmode~\check{c}\else \v{c}\fi{}}\ \emph
  {et~al.}(2008)\citenamefont {\ifmmode \check{Z}\else
  \v{Z}\fi{}nidari\ifmmode~\check{c}\else \v{c}\fi{}}, \citenamefont {Prosen},\
  and\ \citenamefont {Prelov\ifmmode~\check{s}\else
  \v{s}\fi{}ek}}]{PhysRevB.77.064426}%
  \BibitemOpen
  \bibfield  {author} {\bibinfo {author} {\bibfnamefont {M.}~\bibnamefont
  {\ifmmode \check{Z}\else \v{Z}\fi{}nidari\ifmmode~\check{c}\else
  \v{c}\fi{}}}, \bibinfo {author} {\bibfnamefont {T.}~\bibnamefont {Prosen}}, \
  and\ \bibinfo {author} {\bibfnamefont {P.}~\bibnamefont
  {Prelov\ifmmode~\check{s}\else \v{s}\fi{}ek}},\ }\href {\doibase
  10.1103/PhysRevB.77.064426} {\bibfield  {journal} {\bibinfo  {journal} {Phys.
  Rev. B}\ }\textbf {\bibinfo {volume} {77}},\ \bibinfo {pages} {064426}
  (\bibinfo {year} {2008})}\BibitemShut {NoStop}%
\bibitem [{\citenamefont {Bardarson}\ \emph {et~al.}(2012)\citenamefont
  {Bardarson}, \citenamefont {Pollmann},\ and\ \citenamefont
  {Moore}}]{PhysRevLett.109.017202}%
  \BibitemOpen
  \bibfield  {author} {\bibinfo {author} {\bibfnamefont {J.~H.}\ \bibnamefont
  {Bardarson}}, \bibinfo {author} {\bibfnamefont {F.}~\bibnamefont {Pollmann}},
  \ and\ \bibinfo {author} {\bibfnamefont {J.~E.}\ \bibnamefont {Moore}},\
  }\href {\doibase 10.1103/PhysRevLett.109.017202} {\bibfield  {journal}
  {\bibinfo  {journal} {Phys. Rev. Lett.}\ }\textbf {\bibinfo {volume} {109}},\
  \bibinfo {pages} {017202} (\bibinfo {year} {2012})}\BibitemShut {NoStop}%
\bibitem [{\citenamefont {Serbyn}\ \emph {et~al.}(2013)\citenamefont {Serbyn},
  \citenamefont {Papi\ifmmode~\acute{c}\else \'{c}\fi{}},\ and\ \citenamefont
  {Abanin}}]{PhysRevLett.110.260601}%
  \BibitemOpen
  \bibfield  {author} {\bibinfo {author} {\bibfnamefont {M.}~\bibnamefont
  {Serbyn}}, \bibinfo {author} {\bibfnamefont {Z.}~\bibnamefont
  {Papi\ifmmode~\acute{c}\else \'{c}\fi{}}}, \ and\ \bibinfo {author}
  {\bibfnamefont {D.~A.}\ \bibnamefont {Abanin}},\ }\href {\doibase
  10.1103/PhysRevLett.110.260601} {\bibfield  {journal} {\bibinfo  {journal}
  {Phys. Rev. Lett.}\ }\textbf {\bibinfo {volume} {110}},\ \bibinfo {pages}
  {260601} (\bibinfo {year} {2013})}\BibitemShut {NoStop}%
\bibitem [{\citenamefont {De~Tomasi}\ \emph {et~al.}(2019)\citenamefont
  {De~Tomasi}, \citenamefont {Pollmann},\ and\ \citenamefont
  {Heyl}}]{PhysRevB.99.241114}%
  \BibitemOpen
  \bibfield  {author} {\bibinfo {author} {\bibfnamefont {G.}~\bibnamefont
  {De~Tomasi}}, \bibinfo {author} {\bibfnamefont {F.}~\bibnamefont {Pollmann}},
  \ and\ \bibinfo {author} {\bibfnamefont {M.}~\bibnamefont {Heyl}},\ }\href
  {\doibase 10.1103/PhysRevB.99.241114} {\bibfield  {journal} {\bibinfo
  {journal} {Phys. Rev. B}\ }\textbf {\bibinfo {volume} {99}},\ \bibinfo
  {pages} {241114(R)} (\bibinfo {year} {2019})}\BibitemShut {NoStop}%
\bibitem [{\citenamefont {Pino}(2014)}]{PhysRevB.90.174204}%
  \BibitemOpen
  \bibfield  {author} {\bibinfo {author} {\bibfnamefont {M.}~\bibnamefont
  {Pino}},\ }\href {\doibase 10.1103/PhysRevB.90.174204} {\bibfield  {journal}
  {\bibinfo  {journal} {Phys. Rev. B}\ }\textbf {\bibinfo {volume} {90}},\
  \bibinfo {pages} {174204} (\bibinfo {year} {2014})}\BibitemShut {NoStop}%
\bibitem [{\citenamefont {Singh}\ \emph {et~al.}(2017)\citenamefont {Singh},
  \citenamefont {Moessner},\ and\ \citenamefont {Roy}}]{PhysRevB.95.094205}%
  \BibitemOpen
  \bibfield  {author} {\bibinfo {author} {\bibfnamefont {R.}~\bibnamefont
  {Singh}}, \bibinfo {author} {\bibfnamefont {R.}~\bibnamefont {Moessner}}, \
  and\ \bibinfo {author} {\bibfnamefont {D.}~\bibnamefont {Roy}},\ }\href
  {\doibase 10.1103/PhysRevB.95.094205} {\bibfield  {journal} {\bibinfo
  {journal} {Phys. Rev. B}\ }\textbf {\bibinfo {volume} {95}},\ \bibinfo
  {pages} {094205} (\bibinfo {year} {2017})}\BibitemShut {NoStop}%
\bibitem [{\citenamefont {Pandey}\ and\ \citenamefont
  {Pati}(2019)}]{pandeyMBL2019}%
  \BibitemOpen
  \bibfield  {author} {\bibinfo {author} {\bibfnamefont {B.}~\bibnamefont
  {Pandey}}\ and\ \bibinfo {author} {\bibfnamefont {S.~K.}\ \bibnamefont
  {Pati}},\ }\href@noop {} {\  (\bibinfo {year} {2019})},\ \Eprint
  {http://arxiv.org/abs/arXiv:1905.06677} {arXiv:1905.06677} \BibitemShut
  {NoStop}%
\bibitem [{\citenamefont {De~Tomasi}(2019)}]{PhysRevB.99.054204}%
  \BibitemOpen
  \bibfield  {author} {\bibinfo {author} {\bibfnamefont {G.}~\bibnamefont
  {De~Tomasi}},\ }\href {\doibase 10.1103/PhysRevB.99.054204} {\bibfield
  {journal} {\bibinfo  {journal} {Phys. Rev. B}\ }\textbf {\bibinfo {volume}
  {99}},\ \bibinfo {pages} {054204} (\bibinfo {year} {2019})}\BibitemShut
  {NoStop}%
\bibitem [{\citenamefont {Safavi-Naini}\ \emph {et~al.}(2019)\citenamefont
  {Safavi-Naini}, \citenamefont {Wall}, \citenamefont {Acevedo}, \citenamefont
  {Rey},\ and\ \citenamefont {Nandkishore}}]{PhysRevA.99.033610}%
  \BibitemOpen
  \bibfield  {author} {\bibinfo {author} {\bibfnamefont {A.}~\bibnamefont
  {Safavi-Naini}}, \bibinfo {author} {\bibfnamefont {M.~L.}\ \bibnamefont
  {Wall}}, \bibinfo {author} {\bibfnamefont {O.~L.}\ \bibnamefont {Acevedo}},
  \bibinfo {author} {\bibfnamefont {A.~M.}\ \bibnamefont {Rey}}, \ and\
  \bibinfo {author} {\bibfnamefont {R.~M.}\ \bibnamefont {Nandkishore}},\
  }\href {\doibase 10.1103/PhysRevA.99.033610} {\bibfield  {journal} {\bibinfo
  {journal} {Phys. Rev. A}\ }\textbf {\bibinfo {volume} {99}},\ \bibinfo
  {pages} {033610} (\bibinfo {year} {2019})}\BibitemShut {NoStop}%
\bibitem [{\citenamefont {Pollak}(2013)}]{pollak2013}%
  \BibitemOpen
  \bibfield  {author} {\bibinfo {author} {\bibfnamefont {M.}~\bibnamefont
  {Pollak}},\ }\href {https://doi.org/10.1017/CBO9780511978999} {\emph
  {\bibinfo {title} {The Electron Glass}}}\ (\bibinfo  {publisher} {Cambridge
  University Press},\ \bibinfo {year} {2013})\BibitemShut {NoStop}%
\bibitem [{\citenamefont {Kotov}\ \emph {et~al.}(2012)\citenamefont {Kotov},
  \citenamefont {Uchoa}, \citenamefont {Pereira}, \citenamefont {Guinea},\ and\
  \citenamefont {Castro~Neto}}]{RevModPhys.84.1067}%
  \BibitemOpen
  \bibfield  {author} {\bibinfo {author} {\bibfnamefont {V.~N.}\ \bibnamefont
  {Kotov}}, \bibinfo {author} {\bibfnamefont {B.}~\bibnamefont {Uchoa}},
  \bibinfo {author} {\bibfnamefont {V.~M.}\ \bibnamefont {Pereira}}, \bibinfo
  {author} {\bibfnamefont {F.}~\bibnamefont {Guinea}}, \ and\ \bibinfo {author}
  {\bibfnamefont {A.~H.}\ \bibnamefont {Castro~Neto}},\ }\href {\doibase
  10.1103/RevModPhys.84.1067} {\bibfield  {journal} {\bibinfo  {journal} {Rev.
  Mod. Phys.}\ }\textbf {\bibinfo {volume} {84}},\ \bibinfo {pages} {1067}
  (\bibinfo {year} {2012})}\BibitemShut {NoStop}%
\bibitem [{\citenamefont {Altland}\ and\ \citenamefont
  {Simons}(2010)}]{Altland}%
  \BibitemOpen
  \bibfield  {author} {\bibinfo {author} {\bibfnamefont {A.}~\bibnamefont
  {Altland}}\ and\ \bibinfo {author} {\bibfnamefont {B.~D.}\ \bibnamefont
  {Simons}},\ }\href@noop {} {\emph {\bibinfo {title} {Condensed matter field
  theory}}}\ (\bibinfo  {publisher} {Cambridge University Press},\ \bibinfo
  {year} {2010})\BibitemShut {NoStop}%
\bibitem [{\citenamefont {Prelov\ifmmode~\check{s}\else \v{s}\fi{}ek}\ \emph
  {et~al.}(2016)\citenamefont {Prelov\ifmmode~\check{s}\else \v{s}\fi{}ek},
  \citenamefont {Bari\ifmmode \check{s}\else \v{s}\fi{}i\ifmmode~\acute{c}\else
  \'{c}\fi{}},\ and\ \citenamefont {\ifmmode \check{Z}\else
  \v{Z}\fi{}nidari\ifmmode~\check{c}\else \v{c}\fi{}}}]{PhysRevB.94.241104}%
  \BibitemOpen
  \bibfield  {author} {\bibinfo {author} {\bibfnamefont {P.}~\bibnamefont
  {Prelov\ifmmode~\check{s}\else \v{s}\fi{}ek}}, \bibinfo {author}
  {\bibfnamefont {O.~S.}\ \bibnamefont {Bari\ifmmode \check{s}\else
  \v{s}\fi{}i\ifmmode~\acute{c}\else \'{c}\fi{}}}, \ and\ \bibinfo {author}
  {\bibfnamefont {M.}~\bibnamefont {\ifmmode \check{Z}\else
  \v{Z}\fi{}nidari\ifmmode~\check{c}\else \v{c}\fi{}}},\ }\href {\doibase
  10.1103/PhysRevB.94.241104} {\bibfield  {journal} {\bibinfo  {journal} {Phys.
  Rev. B}\ }\textbf {\bibinfo {volume} {94}},\ \bibinfo {pages} {241104(R)}
  (\bibinfo {year} {2016})}\BibitemShut {NoStop}%
\bibitem [{\citenamefont {Mierzejewski}\ \emph {et~al.}(2018)\citenamefont
  {Mierzejewski}, \citenamefont {Kozarzewski},\ and\ \citenamefont
  {Prelov\ifmmode~\check{s}\else \v{s}\fi{}ek}}]{PhysRevB.97.064204}%
  \BibitemOpen
  \bibfield  {author} {\bibinfo {author} {\bibfnamefont {M.}~\bibnamefont
  {Mierzejewski}}, \bibinfo {author} {\bibfnamefont {M.}~\bibnamefont
  {Kozarzewski}}, \ and\ \bibinfo {author} {\bibfnamefont {P.}~\bibnamefont
  {Prelov\ifmmode~\check{s}\else \v{s}\fi{}ek}},\ }\href {\doibase
  10.1103/PhysRevB.97.064204} {\bibfield  {journal} {\bibinfo  {journal} {Phys.
  Rev. B}\ }\textbf {\bibinfo {volume} {97}},\ \bibinfo {pages} {064204}
  (\bibinfo {year} {2018})}\BibitemShut {NoStop}%
\bibitem [{\citenamefont {Kozarzewski}\ \emph {et~al.}(2018)\citenamefont
  {Kozarzewski}, \citenamefont {Prelov\ifmmode~\check{s}\else \v{s}\fi{}ek},\
  and\ \citenamefont {Mierzejewski}}]{PhysRevLett.120.246602}%
  \BibitemOpen
  \bibfield  {author} {\bibinfo {author} {\bibfnamefont {M.}~\bibnamefont
  {Kozarzewski}}, \bibinfo {author} {\bibfnamefont {P.}~\bibnamefont
  {Prelov\ifmmode~\check{s}\else \v{s}\fi{}ek}}, \ and\ \bibinfo {author}
  {\bibfnamefont {M.}~\bibnamefont {Mierzejewski}},\ }\href {\doibase
  10.1103/PhysRevLett.120.246602} {\bibfield  {journal} {\bibinfo  {journal}
  {Phys. Rev. Lett.}\ }\textbf {\bibinfo {volume} {120}},\ \bibinfo {pages}
  {246602} (\bibinfo {year} {2018})}\BibitemShut {NoStop}%
\bibitem [{\citenamefont {Yu}\ \emph {et~al.}(2018)\citenamefont {Yu},
  \citenamefont {Luo},\ and\ \citenamefont {Clark}}]{PhysRevB.98.115106}%
  \BibitemOpen
  \bibfield  {author} {\bibinfo {author} {\bibfnamefont {X.}~\bibnamefont
  {Yu}}, \bibinfo {author} {\bibfnamefont {D.}~\bibnamefont {Luo}}, \ and\
  \bibinfo {author} {\bibfnamefont {B.~K.}\ \bibnamefont {Clark}},\ }\href
  {\doibase 10.1103/PhysRevB.98.115106} {\bibfield  {journal} {\bibinfo
  {journal} {Phys. Rev. B}\ }\textbf {\bibinfo {volume} {98}},\ \bibinfo
  {pages} {115106} (\bibinfo {year} {2018})}\BibitemShut {NoStop}%
\bibitem [{\citenamefont {Wurtz}\ \emph {et~al.}(2020)\citenamefont {Wurtz},
  \citenamefont {Claeys},\ and\ \citenamefont {Polkovnikov}}]{1910.11889}%
  \BibitemOpen
  \bibfield  {author} {\bibinfo {author} {\bibfnamefont {J.}~\bibnamefont
  {Wurtz}}, \bibinfo {author} {\bibfnamefont {P.~W.}\ \bibnamefont {Claeys}}, \
  and\ \bibinfo {author} {\bibfnamefont {A.}~\bibnamefont {Polkovnikov}},\
  }\href {\doibase 10.1103/PhysRevB.101.014302} {\bibfield  {journal} {\bibinfo
   {journal} {Phys. Rev. B}\ }\textbf {\bibinfo {volume} {101}},\ \bibinfo
  {pages} {014302} (\bibinfo {year} {2020})}\BibitemShut {NoStop}%
\bibitem [{\citenamefont {Krause}\ \emph {et~al.}(2019)\citenamefont {Krause},
  \citenamefont {Pellegrin}, \citenamefont {Brouwer}, \citenamefont {Abanin},\
  and\ \citenamefont {Filippone}}]{1911.11711}%
  \BibitemOpen
  \bibfield  {author} {\bibinfo {author} {\bibfnamefont {U.}~\bibnamefont
  {Krause}}, \bibinfo {author} {\bibfnamefont {T.}~\bibnamefont {Pellegrin}},
  \bibinfo {author} {\bibfnamefont {P.~W.}\ \bibnamefont {Brouwer}}, \bibinfo
  {author} {\bibfnamefont {D.~A.}\ \bibnamefont {Abanin}}, \ and\ \bibinfo
  {author} {\bibfnamefont {M.}~\bibnamefont {Filippone}},\ }\href@noop {} {\
  (\bibinfo {year} {2019})},\ \Eprint {http://arxiv.org/abs/arXiv:1911.11711}
  {arXiv:1911.11711} \BibitemShut {NoStop}%
\bibitem [{\citenamefont {Leipner-Johns}\ and\ \citenamefont
  {Wortis}(2019)}]{PhysRevB.100.125132}%
  \BibitemOpen
  \bibfield  {author} {\bibinfo {author} {\bibfnamefont {B.}~\bibnamefont
  {Leipner-Johns}}\ and\ \bibinfo {author} {\bibfnamefont {R.}~\bibnamefont
  {Wortis}},\ }\href {\doibase 10.1103/PhysRevB.100.125132} {\bibfield
  {journal} {\bibinfo  {journal} {Phys. Rev. B}\ }\textbf {\bibinfo {volume}
  {100}},\ \bibinfo {pages} {125132} (\bibinfo {year} {2019})}\BibitemShut
  {NoStop}%
\bibitem [{\citenamefont {\ifmmode~\acute{S}\else \'{S}\fi{}roda}\ \emph
  {et~al.}(2019)\citenamefont {\ifmmode~\acute{S}\else \'{S}\fi{}roda},
  \citenamefont {Prelov\ifmmode~\check{s}\else \v{s}\fi{}ek},\ and\
  \citenamefont {Mierzejewski}}]{PhysRevB.99.121110}%
  \BibitemOpen
  \bibfield  {author} {\bibinfo {author} {\bibfnamefont {M.}~\bibnamefont
  {\ifmmode~\acute{S}\else \'{S}\fi{}roda}}, \bibinfo {author} {\bibfnamefont
  {P.}~\bibnamefont {Prelov\ifmmode~\check{s}\else \v{s}\fi{}ek}}, \ and\
  \bibinfo {author} {\bibfnamefont {M.}~\bibnamefont {Mierzejewski}},\ }\href
  {\doibase 10.1103/PhysRevB.99.121110} {\bibfield  {journal} {\bibinfo
  {journal} {Phys. Rev. B}\ }\textbf {\bibinfo {volume} {99}},\ \bibinfo
  {pages} {121110(R)} (\bibinfo {year} {2019})}\BibitemShut {NoStop}%
\bibitem [{\citenamefont {Zakrzewski}\ and\ \citenamefont
  {Delande}(2018)}]{PhysRevB.98.014203}%
  \BibitemOpen
  \bibfield  {author} {\bibinfo {author} {\bibfnamefont {J.}~\bibnamefont
  {Zakrzewski}}\ and\ \bibinfo {author} {\bibfnamefont {D.}~\bibnamefont
  {Delande}},\ }\href {\doibase 10.1103/PhysRevB.98.014203} {\bibfield
  {journal} {\bibinfo  {journal} {Phys. Rev. B}\ }\textbf {\bibinfo {volume}
  {98}},\ \bibinfo {pages} {014203} (\bibinfo {year} {2018})}\BibitemShut
  {NoStop}%
\bibitem [{\citenamefont {Protopopov}\ and\ \citenamefont
  {Abanin}(2019)}]{PhysRevB.99.115111}%
  \BibitemOpen
  \bibfield  {author} {\bibinfo {author} {\bibfnamefont {I.~V.}\ \bibnamefont
  {Protopopov}}\ and\ \bibinfo {author} {\bibfnamefont {D.~A.}\ \bibnamefont
  {Abanin}},\ }\href {\doibase 10.1103/PhysRevB.99.115111} {\bibfield
  {journal} {\bibinfo  {journal} {Phys. Rev. B}\ }\textbf {\bibinfo {volume}
  {99}},\ \bibinfo {pages} {115111} (\bibinfo {year} {2019})}\BibitemShut
  {NoStop}%
\bibitem [{\citenamefont {Davidson}\ \emph {et~al.}(2017)\citenamefont
  {Davidson}, \citenamefont {Sels},\ and\ \citenamefont
  {Polkovnikov}}]{Davidson2017}%
  \BibitemOpen
  \bibfield  {author} {\bibinfo {author} {\bibfnamefont {S.~M.}\ \bibnamefont
  {Davidson}}, \bibinfo {author} {\bibfnamefont {D.}~\bibnamefont {Sels}}, \
  and\ \bibinfo {author} {\bibfnamefont {A.}~\bibnamefont {Polkovnikov}},\
  }\href {\doibase 10.1016/j.aop.2017.07.003} {\bibfield  {journal} {\bibinfo
  {journal} {Annals of Physics}\ }\textbf {\bibinfo {volume} {384}},\ \bibinfo
  {pages} {128} (\bibinfo {year} {2017})}\BibitemShut {NoStop}%
\bibitem [{\citenamefont {Schmitt}\ \emph {et~al.}(2019)\citenamefont
  {Schmitt}, \citenamefont {Sels}, \citenamefont {Kehrein},\ and\ \citenamefont
  {Polkovnikov}}]{PhysRevB.99.134301}%
  \BibitemOpen
  \bibfield  {author} {\bibinfo {author} {\bibfnamefont {M.}~\bibnamefont
  {Schmitt}}, \bibinfo {author} {\bibfnamefont {D.}~\bibnamefont {Sels}},
  \bibinfo {author} {\bibfnamefont {S.}~\bibnamefont {Kehrein}}, \ and\
  \bibinfo {author} {\bibfnamefont {A.}~\bibnamefont {Polkovnikov}},\ }\href
  {\doibase 10.1103/PhysRevB.99.134301} {\bibfield  {journal} {\bibinfo
  {journal} {Phys. Rev. B}\ }\textbf {\bibinfo {volume} {99}},\ \bibinfo
  {pages} {134301} (\bibinfo {year} {2019})}\BibitemShut {NoStop}%
\bibitem [{\citenamefont {Japaridze}\ and\ \citenamefont
  {Kampf}(1999)}]{PhysRevB.59.12822}%
  \BibitemOpen
  \bibfield  {author} {\bibinfo {author} {\bibfnamefont {G.~I.}\ \bibnamefont
  {Japaridze}}\ and\ \bibinfo {author} {\bibfnamefont {A.~P.}\ \bibnamefont
  {Kampf}},\ }\href {\doibase 10.1103/PhysRevB.59.12822} {\bibfield  {journal}
  {\bibinfo  {journal} {Phys. Rev. B}\ }\textbf {\bibinfo {volume} {59}},\
  \bibinfo {pages} {12822} (\bibinfo {year} {1999})}\BibitemShut {NoStop}%
\bibitem [{\citenamefont {Garrison}\ \emph {et~al.}(2017)\citenamefont
  {Garrison}, \citenamefont {Mishmash},\ and\ \citenamefont
  {Fisher}}]{PhysRevB.95.054204}%
  \BibitemOpen
  \bibfield  {author} {\bibinfo {author} {\bibfnamefont {J.~R.}\ \bibnamefont
  {Garrison}}, \bibinfo {author} {\bibfnamefont {R.~V.}\ \bibnamefont
  {Mishmash}}, \ and\ \bibinfo {author} {\bibfnamefont {M.~P.~A.}\ \bibnamefont
  {Fisher}},\ }\href {\doibase 10.1103/PhysRevB.95.054204} {\bibfield
  {journal} {\bibinfo  {journal} {Phys. Rev. B}\ }\textbf {\bibinfo {volume}
  {95}},\ \bibinfo {pages} {054204} (\bibinfo {year} {2017})}\BibitemShut
  {NoStop}%
\bibitem [{\citenamefont {Anderson}(1958)}]{PhysRev.109.1492}%
  \BibitemOpen
  \bibfield  {author} {\bibinfo {author} {\bibfnamefont {P.~W.}\ \bibnamefont
  {Anderson}},\ }\href {\doibase 10.1103/PhysRev.109.1492} {\bibfield
  {journal} {\bibinfo  {journal} {Phys. Rev.}\ }\textbf {\bibinfo {volume}
  {109}},\ \bibinfo {pages} {1492} (\bibinfo {year} {1958})}\BibitemShut
  {NoStop}%
\bibitem [{\citenamefont {Basko}\ \emph {et~al.}(2006)\citenamefont {Basko},
  \citenamefont {Aleiner},\ and\ \citenamefont {Altshuler}}]{Basko2006}%
  \BibitemOpen
  \bibfield  {author} {\bibinfo {author} {\bibfnamefont {D.}~\bibnamefont
  {Basko}}, \bibinfo {author} {\bibfnamefont {I.}~\bibnamefont {Aleiner}}, \
  and\ \bibinfo {author} {\bibfnamefont {B.}~\bibnamefont {Altshuler}},\ }\href
  {\doibase 10.1016/j.aop.2005.11.014} {\bibfield  {journal} {\bibinfo
  {journal} {Annals of Physics}\ }\textbf {\bibinfo {volume} {321}},\ \bibinfo
  {pages} {1126} (\bibinfo {year} {2006})}\BibitemShut {NoStop}%
\bibitem [{\citenamefont {Pal}\ and\ \citenamefont
  {Huse}(2010)}]{PhysRevB.82.174411}%
  \BibitemOpen
  \bibfield  {author} {\bibinfo {author} {\bibfnamefont {A.}~\bibnamefont
  {Pal}}\ and\ \bibinfo {author} {\bibfnamefont {D.~A.}\ \bibnamefont {Huse}},\
  }\href {\doibase 10.1103/PhysRevB.82.174411} {\bibfield  {journal} {\bibinfo
  {journal} {Phys. Rev. B}\ }\textbf {\bibinfo {volume} {82}},\ \bibinfo
  {pages} {174411} (\bibinfo {year} {2010})}\BibitemShut {NoStop}%
\bibitem [{\citenamefont {Braunstein}\ and\ \citenamefont
  {Caves}(1994)}]{PhysRevLett.72.3439}%
  \BibitemOpen
  \bibfield  {author} {\bibinfo {author} {\bibfnamefont {S.~L.}\ \bibnamefont
  {Braunstein}}\ and\ \bibinfo {author} {\bibfnamefont {C.~M.}\ \bibnamefont
  {Caves}},\ }\href {\doibase 10.1103/PhysRevLett.72.3439} {\bibfield
  {journal} {\bibinfo  {journal} {Phys. Rev. Lett.}\ }\textbf {\bibinfo
  {volume} {72}},\ \bibinfo {pages} {3439} (\bibinfo {year}
  {1994})}\BibitemShut {NoStop}%
\bibitem [{\citenamefont {Hyllus}\ \emph {et~al.}(2012)\citenamefont {Hyllus},
  \citenamefont {Laskowski}, \citenamefont {Krischek}, \citenamefont
  {Schwemmer}, \citenamefont {Wieczorek}, \citenamefont {Weinfurter},
  \citenamefont {Pezz\'e},\ and\ \citenamefont {Smerzi}}]{PhysRevA.85.022321}%
  \BibitemOpen
  \bibfield  {author} {\bibinfo {author} {\bibfnamefont {P.}~\bibnamefont
  {Hyllus}}, \bibinfo {author} {\bibfnamefont {W.}~\bibnamefont {Laskowski}},
  \bibinfo {author} {\bibfnamefont {R.}~\bibnamefont {Krischek}}, \bibinfo
  {author} {\bibfnamefont {C.}~\bibnamefont {Schwemmer}}, \bibinfo {author}
  {\bibfnamefont {W.}~\bibnamefont {Wieczorek}}, \bibinfo {author}
  {\bibfnamefont {H.}~\bibnamefont {Weinfurter}}, \bibinfo {author}
  {\bibfnamefont {L.}~\bibnamefont {Pezz\'e}}, \ and\ \bibinfo {author}
  {\bibfnamefont {A.}~\bibnamefont {Smerzi}},\ }\href {\doibase
  10.1103/PhysRevA.85.022321} {\bibfield  {journal} {\bibinfo  {journal} {Phys.
  Rev. A}\ }\textbf {\bibinfo {volume} {85}},\ \bibinfo {pages} {022321}
  (\bibinfo {year} {2012})}\BibitemShut {NoStop}%
\bibitem [{\citenamefont {T\'oth}(2012)}]{PhysRevA.85.022322}%
  \BibitemOpen
  \bibfield  {author} {\bibinfo {author} {\bibfnamefont {G.}~\bibnamefont
  {T\'oth}},\ }\href {\doibase 10.1103/PhysRevA.85.022322} {\bibfield
  {journal} {\bibinfo  {journal} {Phys. Rev. A}\ }\textbf {\bibinfo {volume}
  {85}},\ \bibinfo {pages} {022322} (\bibinfo {year} {2012})}\BibitemShut
  {NoStop}%
\bibitem [{\citenamefont {Hauke}\ \emph {et~al.}(2016)\citenamefont {Hauke},
  \citenamefont {Heyl}, \citenamefont {Tagliacozzo},\ and\ \citenamefont
  {Zoller}}]{Hauke2016}%
  \BibitemOpen
  \bibfield  {author} {\bibinfo {author} {\bibfnamefont {P.}~\bibnamefont
  {Hauke}}, \bibinfo {author} {\bibfnamefont {M.}~\bibnamefont {Heyl}},
  \bibinfo {author} {\bibfnamefont {L.}~\bibnamefont {Tagliacozzo}}, \ and\
  \bibinfo {author} {\bibfnamefont {P.}~\bibnamefont {Zoller}},\ }\href
  {\doibase 10.1038/nphys3700} {\bibfield  {journal} {\bibinfo  {journal}
  {Nature Physics}\ }\textbf {\bibinfo {volume} {12}},\ \bibinfo {pages} {778}
  (\bibinfo {year} {2016})}\BibitemShut {NoStop}%
\bibitem [{\citenamefont {Acevedo}\ \emph {et~al.}(2017)\citenamefont
  {Acevedo}, \citenamefont {Safavi-Naini}, \citenamefont {Schachenmayer},
  \citenamefont {Wall}, \citenamefont {Nandkishore},\ and\ \citenamefont
  {Rey}}]{PhysRevA.96.033604}%
  \BibitemOpen
  \bibfield  {author} {\bibinfo {author} {\bibfnamefont {O.~L.}\ \bibnamefont
  {Acevedo}}, \bibinfo {author} {\bibfnamefont {A.}~\bibnamefont
  {Safavi-Naini}}, \bibinfo {author} {\bibfnamefont {J.}~\bibnamefont
  {Schachenmayer}}, \bibinfo {author} {\bibfnamefont {M.~L.}\ \bibnamefont
  {Wall}}, \bibinfo {author} {\bibfnamefont {R.}~\bibnamefont {Nandkishore}}, \
  and\ \bibinfo {author} {\bibfnamefont {A.~M.}\ \bibnamefont {Rey}},\ }\href
  {\doibase 10.1103/PhysRevA.96.033604} {\bibfield  {journal} {\bibinfo
  {journal} {Phys. Rev. A}\ }\textbf {\bibinfo {volume} {96}},\ \bibinfo
  {pages} {033604} (\bibinfo {year} {2017})}\BibitemShut {NoStop}%
\bibitem [{\citenamefont {Wurtz}\ \emph {et~al.}(2018)\citenamefont {Wurtz},
  \citenamefont {Polkovnikov},\ and\ \citenamefont {Sels}}]{WURTZ2018341}%
  \BibitemOpen
  \bibfield  {author} {\bibinfo {author} {\bibfnamefont {J.}~\bibnamefont
  {Wurtz}}, \bibinfo {author} {\bibfnamefont {A.}~\bibnamefont {Polkovnikov}},
  \ and\ \bibinfo {author} {\bibfnamefont {D.}~\bibnamefont {Sels}},\ }\href
  {\doibase https://doi.org/10.1016/j.aop.2018.06.001} {\bibfield  {journal}
  {\bibinfo  {journal} {Annals of Physics}\ }\textbf {\bibinfo {volume}
  {395}},\ \bibinfo {pages} {341 } (\bibinfo {year} {2018})}\BibitemShut
  {NoStop}%
\bibitem [{\citenamefont {Oganesyan}\ \emph {et~al.}(2009)\citenamefont
  {Oganesyan}, \citenamefont {Pal},\ and\ \citenamefont
  {Huse}}]{Oganesyan_2009}%
  \BibitemOpen
  \bibfield  {author} {\bibinfo {author} {\bibfnamefont {V.}~\bibnamefont
  {Oganesyan}}, \bibinfo {author} {\bibfnamefont {A.}~\bibnamefont {Pal}}, \
  and\ \bibinfo {author} {\bibfnamefont {D.~A.}\ \bibnamefont {Huse}},\ }\href
  {\doibase 10.1103/PhysRevB.80.115104} {\bibfield  {journal} {\bibinfo
  {journal} {Phys. Rev. B}\ }\textbf {\bibinfo {volume} {80}},\ \bibinfo
  {pages} {115104} (\bibinfo {year} {2009})}\BibitemShut {NoStop}%
\bibitem [{\citenamefont {Kulshreshtha}\ \emph {et~al.}(2018)\citenamefont
  {Kulshreshtha}, \citenamefont {Pal}, \citenamefont {Wahl},\ and\
  \citenamefont {Simon}}]{PhysRevB.98.184201}%
  \BibitemOpen
  \bibfield  {author} {\bibinfo {author} {\bibfnamefont {A.~K.}\ \bibnamefont
  {Kulshreshtha}}, \bibinfo {author} {\bibfnamefont {A.}~\bibnamefont {Pal}},
  \bibinfo {author} {\bibfnamefont {T.~B.}\ \bibnamefont {Wahl}}, \ and\
  \bibinfo {author} {\bibfnamefont {S.~H.}\ \bibnamefont {Simon}},\ }\href
  {\doibase 10.1103/PhysRevB.98.184201} {\bibfield  {journal} {\bibinfo
  {journal} {Phys. Rev. B}\ }\textbf {\bibinfo {volume} {98}},\ \bibinfo
  {pages} {184201} (\bibinfo {year} {2018})}\BibitemShut {NoStop}%
\bibitem [{\citenamefont {Lev}\ and\ \citenamefont
  {Reichman}(2016)}]{BarLev2016}%
  \BibitemOpen
  \bibfield  {author} {\bibinfo {author} {\bibfnamefont {Y.~B.}\ \bibnamefont
  {Lev}}\ and\ \bibinfo {author} {\bibfnamefont {D.~R.}\ \bibnamefont
  {Reichman}},\ }\href {\doibase 10.1209/0295-5075/113/46001} {\bibfield
  {journal} {\bibinfo  {journal} {{EPL} (Europhysics Letters)}\ }\textbf
  {\bibinfo {volume} {113}},\ \bibinfo {pages} {46001} (\bibinfo {year}
  {2016})}\BibitemShut {NoStop}%
\end{thebibliography}%

\end{document}